\documentclass[fleqn,usenatbib]{mnras}

\usepackage{mathptmx}

\usepackage[T1]{fontenc}

\DeclareRobustCommand{\VAN}[3]{#2}
\let\VANthebibliography\thebibliography
\def\thebibliography{\DeclareRobustCommand{\VAN}[3]{##3}\VANthebibliography}

\usepackage{graphicx}	
\usepackage{amsmath}	
\usepackage{amssymb}	

\title[Gravitationally lensed Pop III stars]{The detection and characterization of highly magnified stars with JWST: Prospects of finding Population III}

\author[E. Zackrisson et al.]{
Erik Zackrisson,$^{1,10}$\thanks{E-mail: erik.zackrisson@physics.uu.se}
Adam Hultquist,$^{1}$
Aron Kordt,$^{1}$
Jose M. Diego,$^{2}$
Armin Nabizadeh,$^{1}$
Anton Vikaeus,$^{1}$\newauthor
Ashish Kumar Meena,$^{3}$
Adi Zitrin,$^{3}$
Guglielmo Volpato,$^{4,5}$
Emma Lundqvist,$^{1}$
Brian Welch,$^{6,7}$\newauthor
Guglielmo Costa,$^{8,5}$ 
and Rogier A. Windhorst,$^{9}$
\\
$^{1}$Observational Astrophysics, Department of Physics and Astronomy, Uppsala University, Box 516, SE-751 20 Uppsala, Sweden\\
$^{2}$ Instituto de F\'isica de Cantabria (CSIC-UC). Avda. Los Castros s/n. 39005 Santander, Spain\\
$^{3}$ Physics Department, Ben-Gurion University of the Negev, P.O. Box 653, Be'er-Sheva 8410501, Israel\\
$^{4}$ Dipartimento di Fisica e Astronomia Galileo Galilei, Università di Padova, Vicolo dell’Osservatorio 3, I–35122 Padova, Italy\\
$^{5}$ INAF–Padova, Vicolo dell’Osservatorio 5, I–35122 Padova, Italy\\
$^{6}$ Observational Cosmology Lab, NASA Goddard Space Flight Center, Greenbelt, MD 20771, USA\\
$^7$ Department of Astronomy, University of Maryland, College Park, 20742, USA\\
$^{8}$ Univ Lyon, Univ Lyon, Ens de Lyon, CNRS, Centre de Recherche Astrophysique de Lyon UMR5574, F-69230 Saint-Genis-Laval, France
\\
$^{9}$School of Earth and Space Exploration, Arizona State University, Tempe, AZ 85287-1404\\
$^{10}$ Swedish Collegium for Advanced Study, Linneanum, Thunbergsv\"a{}gen 2
SE-752 38 Uppsala, Sweden}

\date{Accepted XXX. Received YYY; in original form ZZZ}

\pubyear{2021}

\begin{document}
\label{firstpage}
\pagerange{\pageref{firstpage}--\pageref{lastpage}}
\maketitle

\begin{abstract}
Gravitational lensing may render individual high-mass stars detectable out to cosmological distances, and several extremely magnified stars have in recent years been detected out to redshifts $z\approx 6$. Here, we present Muspelheim, a model for the evolving spectral energy distributions of both metal-enriched and metal-free stars at high redshifts. Using this model, we argue that lensed stars will form a highly biased sample of the intrinsic distribution of stars across the Hertzsprung-Russell diagram, and that this bias will typically tend to favour the detection of lensed stars in evolved stages characterized by low effective temperatures, even though stars only spend a minor fraction of their lifetimes in such states. We also explore the prospects of detecting individual, lensed metal-free (Population III) stars  at high redshifts using the James Webb Space Telescope (JWST). We find that very massive ($\gtrsim 100\ M_\odot$) Population III stars at $z\gtrsim 6$ may potentially be detected by JWST in surveys covering large numbers of strong lensing clusters, provided that the Population III stellar initial mass function is sufficiently top-heavy, that these stars evolve to effective temperatures $\leq 15000$ K, and that the cosmic star formation rate density of Pop III stars reaches $\gtrsim 10^{-4}\ M_\odot$ cMpc$^{-3}$ yr$^{-1}$ at $z\approx 6$--10. Various ways to distinguish metal-free lensed stars from metal-enriched ones are also discussed.
\end{abstract}




\section{Introduction}
The very first stars, created from gas with primordial composition, likely started forming at a time when the Universe was $\approx 50-100$ Myr old (cosmic redshift $z\approx 30-50$), thereby bringing the cosmic ``dark ages'' to an end.  Due to the lack of efficient coolants and fragmentation in the chemically unenriched gas at these early epochs, the resulting metal-free (a.k.a. Population III, hereafter Pop III) stars are expected to exhibit a stellar initial mass function (IMF) which differs significantly from that of metal-enriched stars. Current simulations predict this IMF to be top-heavy and possibly logarithmically flat across a very wide mass range \citep[from $\lesssim 1\ M_\odot$ and up to several hundred $M_\odot$; for a review, see][]{Klessen23}. A substantial Pop III mass fraction may hence be locked up in the form of massive stars ($\gtrsim 10\ M_\odot$), with short lifetimes \citep[$\approx 2-40$ Myr for stars in the 10--1000 $M_\odot$ mass range;][]{Yoon12}. As some of these stars exploded as supernovae, the ambient gas was enriched with heavy elements which initiated the transition to the normal mode of star formation (population II and I), with much lower characteristic stellar masses as a result. While the Pop III star formation mode likely became cosmologically subdominant already at $z\approx 15$--25 (less than 200 Myr after the Big Bang), these stars may have continued forming at a low rate for several billion years within rare pockets of gas that remained unpolluted by metals \citep{Magg18,Mebane18,Liu20}. However, no massive Pop III stars have ever been directly detected, and -- with the exception of supermassive Pop III stars \citep[$\sim 10^5\ M_\odot$; e.g.,][]{Haemmerle18,Surace18,Surace19} or Pop III stars at the lowest redshifts ($z\lesssim 1.5$ for $M\leq 1000\ M_\odot$) -- their intrinsic fluxes place them far below the detection limits of both current and upcoming telescopes \citep{Rydberg13,Schauer20}.

Gravitational lensing can boost the fluxes of individual stars beyond the local Universe to detectable levels, and dozens of such stars have already been detected at $z\approx 1$--6 \citep[e.g.][]{Kelly18,Welch22,Fudamoto24}. These highly magnified stars have all been found through observations targeting foreground strong-lensing galaxy cluster fields. In this situation, a background, high-redshift star drifts into the region close to the cluster macrolensing caustic, where it may be subjected to magnifications $\mu>1000$ due to the combined effects of the macrolensing from the galaxy cluster itself, microlensing from intracluster stars, and potentially also millilensing from globular clusters and dark matter substructures \citep[e.g.][]{Diego24}. The magnification is time dependent, both because of the transit of the star across the macrocaustic (timescale of months to years) and the movement of the star though the corrugated network of microcaustics (timescale of days to weeks). Most of the lensed stars detected so far have been detected because of their lensing-induced variability \citep[e.g.][]{Kelly22,Yan23,Fudamoto24}, but Earendel, the highest-redshift lensed-star candidate currently known, has so far not exhibited any substantial variability \citep{Welch22b}, possibly because of a very high microlensing optical depth which would result in small magnification fluctuations \citep{Welch22}.

Very high levels of gravitational magnification ($\mu\gtrsim 10^3-10^4$) would also be able to lift massive Pop III stars at redshifts as high as $z\approx 20$ into the detection range of the James Webb Space Telescope (JWST), but the probability for such detections is sensitive to the detailed cosmic star formation history and stellar initial mass function of the Pop III stars \citep{Rydberg13,Windhorst18}.

The question of how to distinguish between a lensed, metal-enriched star and a bona fide Pop III object using JWST photometry or spectroscopy, and what stellar properties one can realistically hope to derive from such data, remains open.

While many papers have been devoted to the spectral signatures of Pop III galaxies or star clusters \citep[e.g.][]{Tumlinson00,Tumlinson01,Schaerer02,Schaerer03,Tumlinson03,Inoue11,Zackrisson11,Mas-Ribas16,Nakajima22,Trussler22}, not much work has been devoted to the spectral signatures of individual lensed stars. Here, we present the Muspelheim\footnote{In Norse mythology, the stars of the night sky are interpreted as glowing sparks from Muspelheim, the realm of fire.} models for the spectral energy distributions of individual, lensed stars at high redshifts, derive the required conditions for detecting highly magnified Pop III stars at $z\gtrsim 6$ and explore the prospects of distinguishing such stars from metal-enriched counterparts using JWST.

Section~\ref{sec:muspelheim} includes an introduction to the computational machinery of the Muspelheim models. In Section~\ref{sec:HR_bias}, we use Muspelheim to explain why lensed, high-$z$ samples of stars obtained using JWST photometry will typically result in a highly biased sample of high-mass stars, favouring specific regions in the Hertzsprung-Russell (HR) diagram over others. This turns out to have a large impact on our forecasts for lensed Pop III star detections in Section~\ref{sec:PopIII_detection_probability}, as some evolutionary pathways predicted for Pop III stars can significantly boost the detection probabilities of such stars in lensing surveys. In Section~\ref{sec:PopIII features} we explore a number of features that can potentially be used to distinguish Pop III stars from metal-enriched lensed stars. Various uncertainties in the modelling of lensed stars are discussed in Section~\ref{sec:discussion}, and  section~\ref{sec:summary} summarizes our findings.

\section{The Muspelheim models}
\label{sec:muspelheim}
The Muspelheim model predicts the spectral energy distributions (SEDs) of stars as a function of age and initial (Zero-Age Main Sequence, ZAMS) mass $M_\mathrm{ZAMS}$ at different metallicities. The model start from a set of stellar evolutionary tracks (section~\ref{subsec:tracks}) which describe the bolometric luminosity $L$, effective temperature $T_\mathrm{eff}$, surface gravity $\log(g)$ and radius $R$ of stars as a function of age $t$, $M_\mathrm{ZAMS}$,  metallicity ([M/H]) and rotation. SEDs are generated for positions along the tracks separated by $\Delta(t)\approx 10^5$ yr\footnote{ provided that this is possible given the time sampling of the original track data points}, or whenever $T_\mathrm{eff}$ has changed by $\Delta\log_{10} T_\mathrm{eff}\geq 0.05$ compared to the last saved SED, by matching the $T_\mathrm{eff}$ and $\log(g)$ to a suitable stellar atmosphere spectrum from one of several pre-calculated grids (Section~\ref{subsec:atmos}). This SED is then integrated and rescaled to match the bolometric luminosity of the star. As small mismatches between the $T_\mathrm{eff}$ and surface gravity $\log(g)$ of the track data point and the available pre-calculated stellar atmospheres are inevitable, this rescaling is required to minimize the impact of such defects on the predicted brightness of the star. This results, for every ZAMS mass featured in the stellar evolutionary tracks, in an age sequence of stellar spectra with relatively low spectral resolution (SEDs) that cover the rest-frame ultraviolet to mid-infrared wavelength range.
Using these spectra, we derive the ionizing luminosities and redshift-dependent AB magnitudes in a set of JWST broad- and medium band filters of these stars, assuming a $\Omega_\Lambda=0.7$, $\Omega_\mathrm{M}=0.3$, $H_0=70$ km s$^{-1}$ Mpc$^{-1}$ cosmology. The Muspelheim model grids (spectra, ionizing fluxes, AB magnitudes in various filter sets) are publicly available (see Data Availability section for access link). 

\subsection{Stellar evolutionary tracks}
\label{subsec:tracks}
In this paper, we use the Bonn Optimized Stellar Tracks (BoOST, v.~1.3.; \citealt{Szecsi20}) for low-metallicity stars (SMC metallicity; $Z\sim 0.1Z_\odot$) in the $M_\mathrm{ZAMS} \approx 9$--575 $M_\odot$ range. For Pop III stars, several sets of alternative tracks are used: tracks in the $M_\mathrm{ZAMS}=10$--1000 $M_\odot$ range from \citet{Yoon12}, $M_\mathrm{ZAMS}=10$--100 $M_\odot$ tracks by \citet{Windhorst18}, $M_\mathrm{ZAMS}=9$--120 $M_\odot$ tracks by \citet{Murphy21a} and $M_\mathrm{ZAMS} = 100$--1000 $M_\odot$ tracks by \citet{Volpato23}. These Pop III tracks differ in their assumptions on e.g., stellar rotation, magnetic torques, mass loss and overshooting, and consequently result in different predictions for mass, surface temperature, luminosity as a function on stellar age even for Pop III stars with identical ZAMS masses. We refer to the original papers for detailed descriptions on the input physics of these models, but note that whereas the  \citet{Windhorst18} and \citet{Volpato23} models are for non-rotating stars only, \citet{Yoon12} and \citet{Murphy21a} also consider models with various levels of rotation. The \citet{Yoon12} models cover initial rotational velocities $v_\mathrm{init}=0.0$--0.6 $v_\mathrm{K}$ for 10--250 $M_\odot$ and $v_\mathrm{init}=0.0$--0.4 $v_\mathrm{K}$ at 300--1000 $M_\odot$, where $v_\mathrm{K}$ is the Keplerian velocity at the equatorial surface. The \citet{Murphy21a} models instead quantify the initial rotation of their models as the fraction of the critical break-up velocity $v_\mathrm{crit}$, covering $v_\mathrm{init} = 0.0$--0.4 $v_\mathrm{crit}$ throughout the 9--120 $M_\odot$ mass interval. 
 
\subsection{Stellar atmospheres}
\label{subsec:atmos}
To produce stellar atmosphere spectra for metal-poor stars, we combine the O- and B-star TLUSTY grids of \citet{Lanz03,Lanz07} for $T_\mathrm{eff} = 15000$--55000 K with the grid of \citet{Lejeune97} at $T_\mathrm{eff}<15000$ K, both at metallicity $[\mathrm{M}/\mathrm{H}]=-1$ and scaled solar abundances. For our Pop III models, we use the TLUSTY stellar atmosphere code \citep{Hubeny95} to generate a grid of stellar atmosphere models with primordial chemical composition for effective surface temperatures $T_\mathrm{eff}$ in the range from 15000 K to $\approx 3\times 10^5$ K, and augment this with $T_\mathrm{eff}<15000$ K with the lowest metallicity models ($[\mathrm{M}/\mathrm{H}]=-5$) models from \citet{Lejeune97} set. The wavelength coverage of these spectra is $\approx 45$ to $\approx 1.6\times 10^6$ \AA{}  for the low-metallicity \citet{Lanz03,Lanz07} TLUSTY grids, from $\approx 12$--114 \AA{} (depending on $T_\mathrm{eff}$) to $\approx 3\times 10^6$ \AA{} for the Pop III TLUSTY grids and  $\approx 91$ to $\approx 3\times 10^6$ \AA{} for the 
\citet{Lejeune97} set. While the high-$T_\mathrm{eff}$ part of the grid (TLUSTY) does take the effects of non-local thermal thermodynamic equilibrium into account, the low-$T_\mathrm{eff}$
part (the \citealt{Lejeune97} set) is based on the assumption of local thermodynamic equilibrium. While this will affect the strengths and shapes of absorption lines \citep[e.g.][]{Bergemann14}, the impact on the overall shape of the SEDs is likely minor, as the continuum of the stellar atmosphere spectra used in the Lejeune have already been subjected to empirical corrections. 

To derive a suitable SED for each extracted data point along the stellar evolutionary tracks, these grids of stellar atmosphere spectra are interpolated in $T_\mathrm{eff}$ and $\log(g)$. Details on this procedure can be found in Appendix~\ref{sec:appendix_stellar_atmospheres}.

While our models assume an approximately primordial chemical composition for all Pop III stellar atmospheres, one should note that scenarios involving chemically homogeneous evolution and dredge-ups \citep[e.g.][]{Yoon12,Song20,Liu21,Volpato23,Volpato24} are expected to alter the surface composition of the star during their lifetimes (see Section~\ref{subsec:surface_enrichment} for a discussion). While this will affect individual spectral features, the impact on the overall photospheric SED as measured by broadband filter fluxes is likely to be negligible.

\section{An observational bias in the Hertzsprung-Russell diagram}
\label{sec:HR_bias}
In this section, we use the Muspelheim models for the JWST broadband filter fluxes of high-redshift stars, as a function of initial mass and age, to demonstrate the existence of a strong selection bias of where in the HR diagram lensed stars are expected to turn up. 

Massive, metal-enriched stars typically spend most of their lifetimes at relatively high $T_\mathrm{eff}$ ($\gtrsim 25000$ K; \citealt{Szecsi20}), but many of the currently known, lensed $z\gtrsim 1$ stars \citep[e.g.][]{Kelly18,Meena23,Diego22b} are found at considerably lower temperatures ($T_\mathrm{eff}\lesssim 15000$). We argue that this is due to a strong selection bias for lensed stars, caused by the combined effects of the detection limits (as a function of wavelength) and the lensing magnification distribution. Because of this bias, JWST surveys for lensed, high-$z$ stars will not be able to populate the Hertzsprung-Russell diagram uniformly, but will instead preferentially pick up stars in certain regions of the diagram \citep[see also][for a discussion on this]{Diego23}. Consequently, stars that evolve into such regions will be easier to detect, and this has direct consequences for the detection probabilities of Pop III stars, since not all proposed evolutionary paths for such stars will venture into the HR diagram regions where the detection prospects are optimized. 
 
In general, the lensed-star selection function (i.e. the bias that determines what stars are lifted above the observational detection threshold through a combination of macro- and microlensing) is expected to be highly complex. If star formation were unclustered, we argue that this selection function should depend on the wavelength-dependent detection limits, on the shape of the magnification probability distribution, on the shape of the stellar IMF, on the evolution of stars of a given initial mass across the HR diagram, and the dust obscuration and reddening along the line of sight (with potential contributions from the circumstellar medium of the host star, the ISM at its location of birth, the ISM elsewhere in its host galaxy and within the cluster lens). The path that a star of a given initial mass follows across the HR diagram in turn depends on metallicity, rotation rate, the impact of binary evolution and the potential mergers of stars. The fact that most stars are likely to be born in star clusters brings additional complications, since the difference in magnification between one star and its immediate neighbours dictates whether the observer detects a single lensed star or the blended light from many. IMF sampling effects in small star clusters \citep[e.g.][]{Yan23} and the potential ejection of some massive stars from their birth clusters may also play a role. Here, we make no attempt to provide a definitive model for all of these effects, but simply use Muspelheim to demonstrate the likely biases in the highly simplified situation of single-metallicity stars obeying a specific IMF, while neglecting binary evolution, clustering and dust obscuration effects.

As stars age, they typically evolve from a long-lasting high $T_\mathrm{eff}$ state on the main sequence, to more short-lived states at lower $T_\mathrm{eff}$ (although rapid rotation, binary evolution and mergers can alter this behaviour; e.g., \citealt{deMink09,Yoon12,Szecsi20,Wang22}). This is typically accompanied by an increase in $L_\mathrm{bol}$, although this increase can be rather modest for the most massive stars, which are predicted to evolve almost horizontally across the Hertzsprung-Russell diagram \citep[e.g.][]{Yoon12,Windhorst18,Volpato23}. 

\begin{figure*}
\includegraphics[width=\columnwidth]{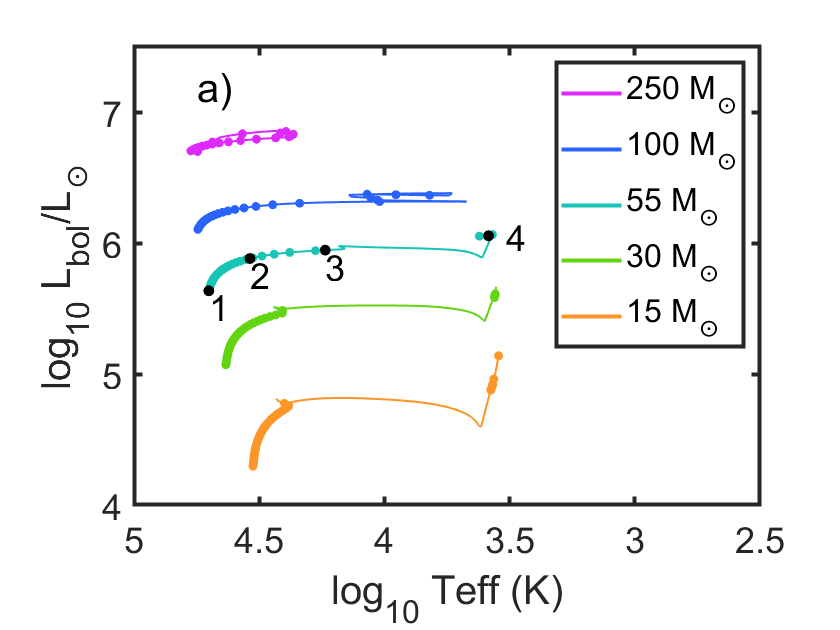}
\includegraphics[width=\columnwidth]{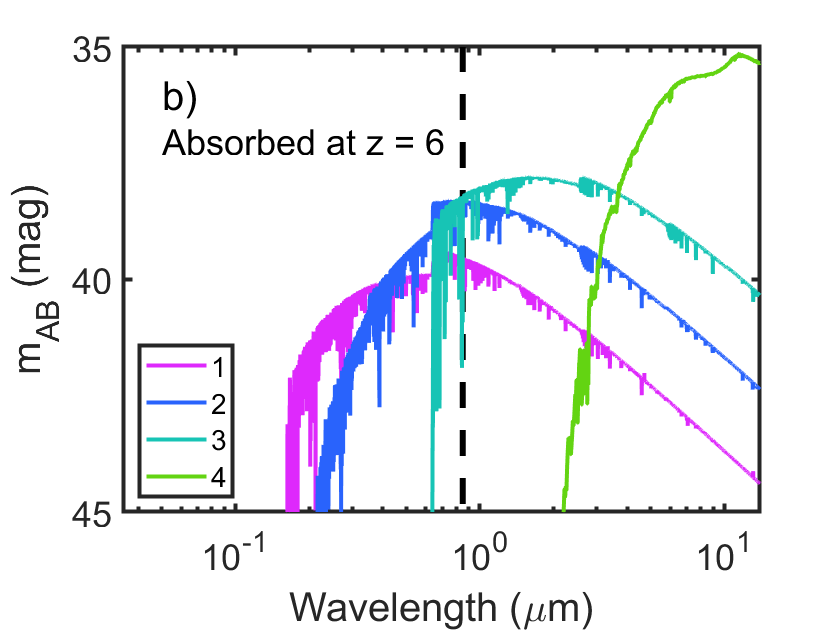}
\includegraphics[width=\columnwidth]{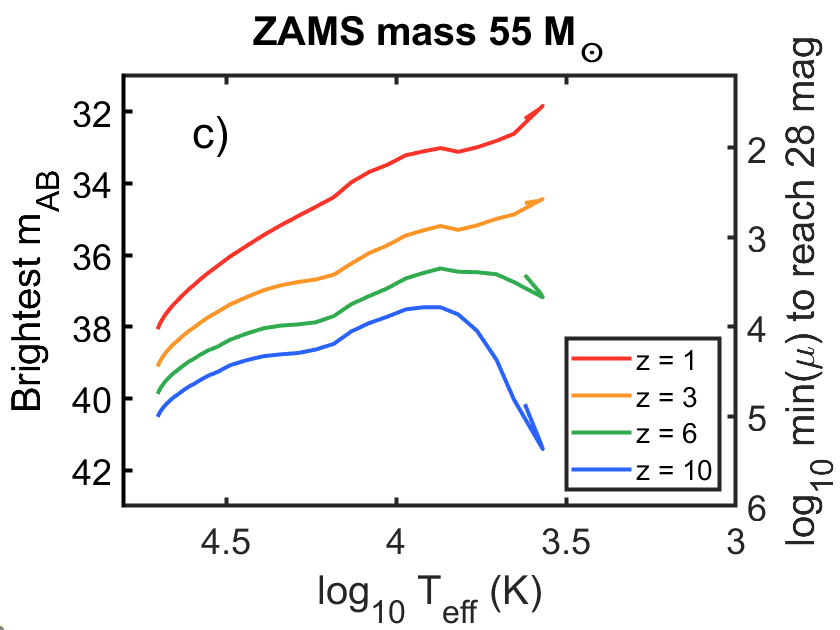}
\includegraphics[width=\columnwidth]{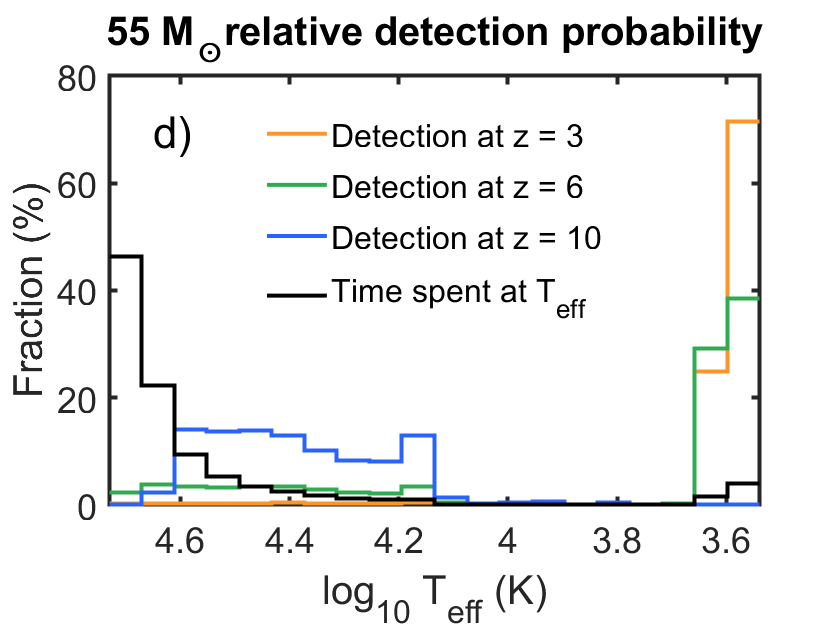}
\caption{The reason why lensed stars form a biased sample of the the intrinsic distribution in the HR diagram. {\bf a)} A selection of evolutionary tracks at different ZAMS masses from the BoOST SMC set at $Z/Z_\odot \sim 0.1$. Coloured markers along each track indicate time steps of $\approx 10^5$ yr. Please note how, in the case of the 15 and 30 $M_\odot$ tracks, there are no age markers between $\log T_\mathrm{eff}\approx 4.4$ and $\approx 3.6$, as the evolution between these states is very rapid. {\bf b)} Redshift $z=6$ SEDs corresponding to the evolutionary stages labeled 1-4 along the 55 $M_\odot$ track in panel a. The vertical dashed line indicates the Lyman-$\alpha$ limit at this redshift, shortward of which the flux is expected to be absorbed by the neutral IGM. As seen, the SEDs become considerably brighter with decreasing $T_\mathrm{eff}$ throughout the $\approx 1$--5 $\mu$m range of JWST/NIRCam, even though the evolution in $L_\mathrm{bol}$ is modest. 
{\bf c)} The brightest NIRCam wide-band magnitude attained as a function of $T_\mathrm{eff}$ along the 55 $M_\odot$ track in panel a, at redshifts $z=1$, 3, 6 and 10. While the brightness increases with decreasing $T_\mathrm{eff}$ at $z=1$--3, this trend starts to turn over at $z=6$, and at $z=10$, the lowest-$T_\mathrm{eff}$ point at the end is no longer the brightest, since the peak of the $\approx 4000$ K ($\log T_\mathrm{eff}\approx 3.6$) SED no longer falls within the NIRCam bands at $z=10$. {\bf d)} The relative probability of detection as a function of $T_\mathrm{eff}$ along the 55 $M_\odot$ track for a detection threshold of 28 AB mag across the JWST/NIRCam broadband filters. In the case where a 55 $M_\odot$ star is detected through lensing (and a $P(>\mu)\propto \mu^{-2}$ magnification probability distribution is assumed in the source plane), the coloured orange, green and blue lines represent the observed probability distribution for different $T_\mathrm{eff}$ at $z=3$, 6 and 10. The black line instead describes the intrinsic distribution of $T_\mathrm{eff}$ across the lifespan of the star, without any observational bias. While most of the lifetime of this star is spent at $\log T_\mathrm{eff}\gtrsim 4.6$ (black line; this means that there are more stars with this initial mass at $\log T_\mathrm{eff}\gtrsim 4.6$ than at lower $T_\mathrm{eff}$ in any source plane with active star formation), such high temperatures are very unlikely to be detected. Instead, it is the red supergiant state at $\log T_\mathrm{eff}\approx 3.6$ that is the most likely to be detected at $z=3$ and $z=6$.}
\label{HR_evol}
\end{figure*}

One may perhaps naively expect that most lensed stars would be detected in their more-long lasting main-sequence states, but there are observational biases that favour the detection of more evolved stars. At fixed $L_\mathrm{bol}$, a low-$T_\mathrm{eff}$ star attains a higher $f_\nu$ flux at the peak of its SED compared to that of a high-$T_\mathrm{eff}$ star\footnote{This is because the luminosity in a fixed wavelength interval $\Delta\lambda$ goes as $L(\Delta\lambda)\propto f_\nu \lambda^{-2} \Delta\lambda$, so that $f_\nu$ fluxes at long wavelengths contribute less to $L_\mathrm{bol}$ than a similar $f_\nu$ at short wavelengths. A low-$T_\mathrm{eff}$ star, for which the SED peaks at long wavelengths, will therefore need a higher peak $f_\nu$ flux to reach the same $L_\mathrm{bol}$ as a high-$T_\mathrm{eff}$ star, for which the SED peaks at short wavelengths. The increase in peak brightness with decreasing $T_\mathrm{eff}$ at fixed $L_\mathrm{bol}$ holds for both stellar atmosphere spectra and black body spectra if flux is expressed in $f_\nu$ units, but not if it's expressed in $f_\lambda$ units.} Since the JWST/NIRCam detection limits are fairly constant in $f_\nu$ units across the $\approx 1-5\ \mu$m range\footnote{Across the wide NIRCam filters F090W, F115W, F150W, F200W, F277W, F356W and F444W filters used in this paper, the $f_\nu$ detection limits are within a factor of 3 of each other, at fixed signal-to-noise and exposure time \citep{Rieke23}. When the peak $f_\nu$ flux of a stellar SED is redshifted into the wavelength range of this filter set, the difference between the maximum continuum $f_\nu$ seen in the SED and that captured photometrically is moreover typically minor, due to the shape of stellar SEDs.}, this means that a high-$z$, low-$T_\mathrm{eff}$ star requires a lower gravitational magnification $\mu$ to be rendered detectable compared to a high-$T_\mathrm{eff}$ star of the same $L_\mathrm{bol}$ and at the same redshift. Since the probability $P$ to attain a certain magnification approximately goes as $P(>\mu)\propto \mu^{-2}$ in the source plane (up to some limit; see Section~\ref{sec:selection function}), short-lived, low-$T_\mathrm{eff}$ evolutionary states that require low $\mu$ to be detected can in lensed samples trump long-lived, high-$T_\mathrm{eff}$ states that require high $\mu$ for detection (see Appendix~\ref{sec:Teff dependence} for further discussion on this).

We demonstrate this effect for a 55 $M_\odot$ star  at $z=1$--10 in Fig.~\ref{HR_evol}. In general, this bias will favour detection of the lowest-$T_\mathrm{eff}$ states along the stellar evolutionary tracks, as long as these are not too short-lived, and as long as the peak of the SEDs of these low-$T_\mathrm{eff}$ stars falls within the wavelength range of JWST/NIRCam at the redshift of the star. As the redshift is increased, the most favoured $T_\mathrm{eff}$ gradually shifts towards higher $T_\mathrm{eff}$, since the $f_\nu$ peak of lower-$T_\mathrm{eff}$ stars redshifts out of NIRCam range. However, the selection function still looks very different from one based simply on how long stars intrinsically occupy different $T_\mathrm{eff}$ states.

\subsection{Modelling the selection function of lensed stars}
\label{sec:selection function}
Here, we use Muspelheim to simulate the expected distribution of low-metallicity ($Z\sim 0.1\ Z_\odot$) lensed stars at $z=1$--10 in the HR diagram for a JWST/NIRCam survey of cluster-lens fields. To do this, we assume a fixed detection limit (in the 27--30 AB magnitude range) in the NIRCam filters F090W, F115W, F150W, F200W, F277W, F356W and F444W. 

For plotting purposes, we at each redshift interpolate the Muspelheim models for the AB magnitudes of stars along the \citet{Szecsi20} evolutionary tracks to a finer mass resolution, and derive the brightest AB magnitudes in the JWST/NIRCam bands as a function of redshift at each data point along these interpolated tracks.

To model the intrinsic distribution of stellar masses, we adopt a stellar initial mass function (IMF) with the same high-mass slope as the \citet{Kroupa01} universal IMF ($\mathrm{d}N/\mathrm{d}M\propto M^{-2.3}$) throughout the 10--150 $M_\odot$ mass range. 

We then run a Monte Carlo simulation which draws a sample of stars that obey the adopted IMF, randomly assigns ages up to 30 Myr and rejects stars which, based on their initial mass, have expired at their selected age. This corresponds to a situation where $M_\mathrm{ZAMS}\geq 10 M_\odot$ stars of any age may drift across the source-plane caustic of the clusters surveyed. This is a fair assumption in a case where the sample of lensed stars comes from many highly magnified galaxies with different star formation histories. However, this may not accurately reflect a situation in which the sample of lensed stars comes from a single, highly magnified galaxy, since the actual age distribution of stars would then be dictated by the detailed star formation history of that particular object. 

For every Monte Carlo generated star, a random magnification $\mu$ is assigned under the assumption of a source-plane probability distribution function $P(\mu)\propto \mu^{-3}$ (or equivalently $P(>\mu)\propto \mu^{-2}$; see \citealt{Schneider92} for an analytical derivation of this distribution), and the brightest NIRCam wide-band flux is shifted accordingly to predict the brightest apparent magnitude across the NIRCam bands of the star. Stars that do not reach the adopted JWST detection threshold in at least one of the NIRCam filters assumed are rejected from the simulated sample.

\begin{figure*}
\includegraphics[scale=0.4]{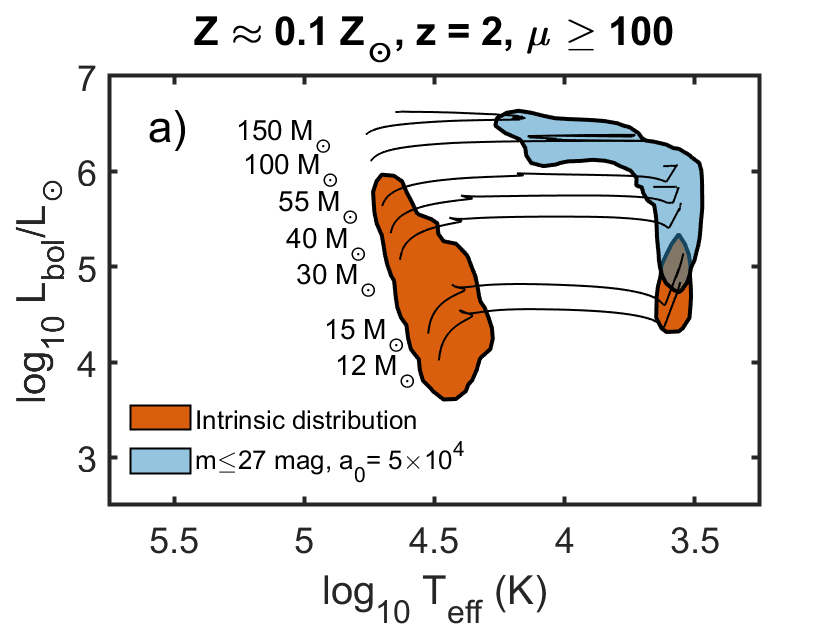}
\includegraphics[scale=0.4]{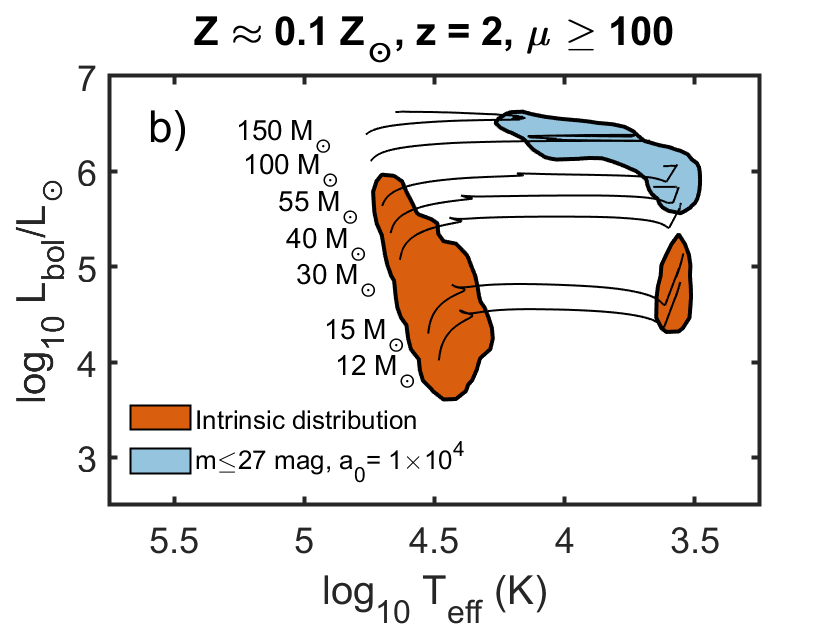}
\includegraphics[scale=0.4]{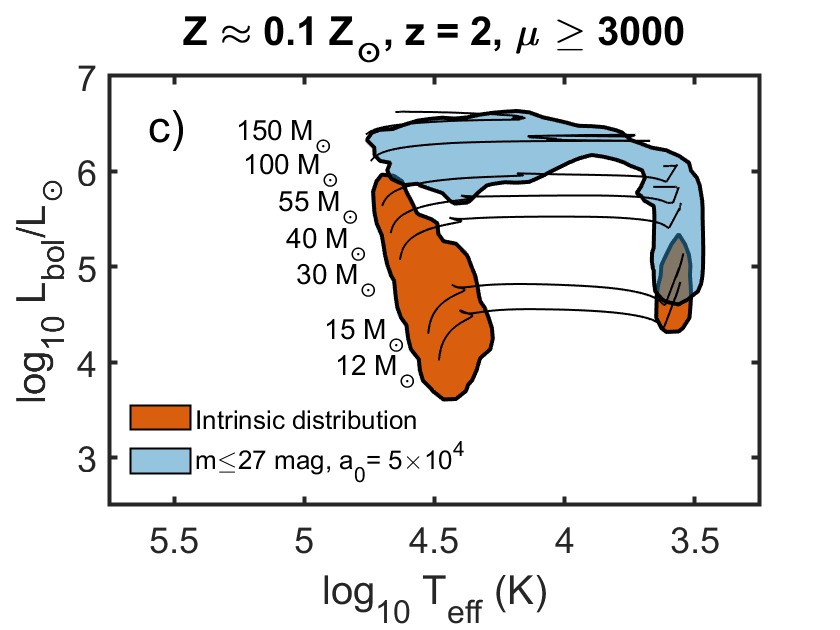}
\includegraphics[scale=0.4]{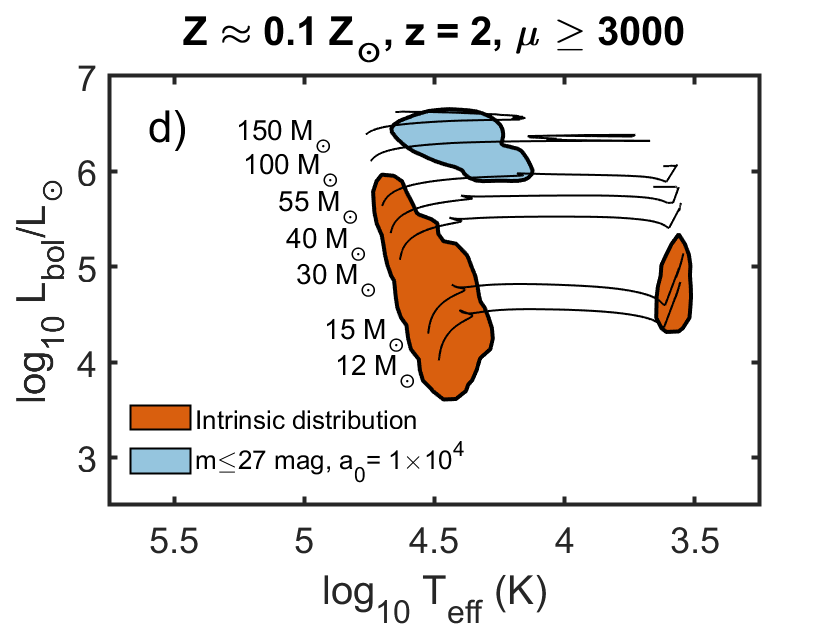}
\caption{The selection bias of lensed, low-metallicity ($Z\approx 0.1\ Z_\odot$), $M_\mathrm{ZAMS}=10$--150 $M_\odot$ stars in the Hertzsprung-Russell diagram at $z=2$, for two different microlensing size-magnitude relations, and two different lower limits on the magnification. The coloured, filled contours indicate the regions in which $\approx 95\%$ of the stars will fall, under different assumptions on the selection function. In all panels, the dark orange regions marks the intrinsic distribution of the $M_\mathrm{ZAMS}=10$--150 $M_\odot$ stars (i.e., only considering the stellar IMF and the time spend in different evolutionary stages, without considering any brightness or lensing bias). The light blue region represents the regions occupied by the stars most likely to end up in JWST/NIRCam surveys reaching $\leq 27.0$ AB mag in lensed fields in the case of a source-plane $P(>\mu)\propto \mu^{-2}$ magnification distribution throughout the range $\mu=1\times 10^2$--$5\times 10^4$ (upper panels) or $\mu=3\times 10^3$--$5\times 10^4$ (lower panels), with an additional microlensing size-magnification limit (equation~\ref{eq:mu-size}) with $A_0=1\times 10^4$ (left panels) and $A_0=5\times 10^4$ (right panels). In all panels, lensed samples produce a strong selection bias, as seen in the difference between the dark orange and light blue regions. This bias typically favours stars that have evolved significantly away from their ZAMS effective temperatures.
Only the most massive stars are likely to be detected at $\log T_\mathrm{eff}>4.0$ whereas less massive stars may be detected on the red supergiant branch $\log T_\mathrm{eff}\approx 3.6$. The prospects of detecting red supergiants in the lensed sample are, however, sensitive to the adopted microlensing size-magnification constraint, since a lower $A_0$ (as in panel b) makes it harder for stars with large radii to reach the magnification required for detection. The subset of high-magnification stars ($\mu\geq 3000$; panels c and d) displays a wider range in effective temperature that extends to higher $\log T_\mathrm{eff}$, since very high magnifcaitons are required to detect the hottest stars. In the case of the most restrictive microlensing size-magnification relation considered ($A_0=1\times 10^4$), the $\mu\geq 3000$ sample (panel d) rejects low-$T_\mathrm{eff}$ stars due to their large radii, leaving only the $\log T_\mathrm{eff}>4.2$ stars detectable.}
\label{HR_bias_low_metallicity_z2}
\end{figure*}

We note that in the resulting sample of the lensed stars that make it above the JWST detection threshold, the $P(>\mu)\propto \mu^{-2}$ distribution will no longer be obeyed, even though stars were initially assigned magnifications from this distribution. This happens because some stars can be rendered detectable at lower magnifications than others. The reason for this bias can be understood through the following thought experiment: Let's assume that there are only two types of stars, A and B, which both retain fixed positions in the HR diagram until they expire. Due to having different intrinsic brightness, star A may be detected when $\mu>1000$, whereas star B requires $\mu>10000$ to be lifted above the detection threshold. If these stars were equally common, then one would expect to detect B one hundred times less often than A ($P_\mathrm{B}(\mu>10000)/P_\mathrm{A}(\mu>1000) = 10^{-2}$). However, if B is $\sim 100$ times more common than A (due to the IMF or a longer lifetime), then the probability to detect A and B become comparable, and hence magnifications of $\mu>1000$ and $\mu>10000$ turn out about equally likely in the observed sample of lensed stars, which violates the $P(>\mu) \propto \mu^{-2}$ distribution. In practice, however, the magnification distributions of our simulated samples are never flat, and the highest magnifications are still the rarest, but the mean or median magnifications may vary substantially with the adopted detection limit and choice of stellar evolutionary tracks. 

Throughout this paper, we only consider stars with magnifications in the $\mu=1\times 10^2$--$5\times 10^4$ range. The lower limit has been set in order not to exclude cases which may resemble some of the faintest lensed stars detected at low redshifts with JWST through microlensing-induced variability \citep[e.g][]{Fudamoto24}. The upper limit is set by the fact that higher magnifications are highly unlikely in realistic settings, as argued by \citet{Diego18}. In some figures, we also present results exclusively for the high-magnification regime of $\mu=3\times 10^3$--$5\times 10^4$, which corresponds to a range of magnifications that  {$\gtrsim 1$ pc} star clusters cannot reach (see eq.~\ref{mumax_macro} and Figure~\ref{lightcurve} and the associated discussion). This higher-magnification interval may be more appropiate in situations where lensed-star candidates do not show significant variability, and there may be some ambiguity whether these objects are high-magnification stars or lower-magnficication star clusters. In such cases, the distinction can be made on the likely macromagnification at the position of the object. We stress, however, that many of the variability-selected lensed stars detected at low redshifts by JWST may also be in this high-magnification range \citep{Broadhurst24}. 

While the absolute scaling of the $P(>\mu) \propto \mu^{-2}$ magnification distribution is important for the {\it absolute} number of lensed stars expected to be detected in a given survey, we are in this section only concerned with the {\it relative} distribution across the HR diagram of the lensed stars that are detected (i.e. {\it where} in the HR diagram lensed stars are the most likely to turn up, not the number of stars that turn up in those regions), for which the scaling factor does not matter.   

\begin{figure*}
\includegraphics[scale=0.4]{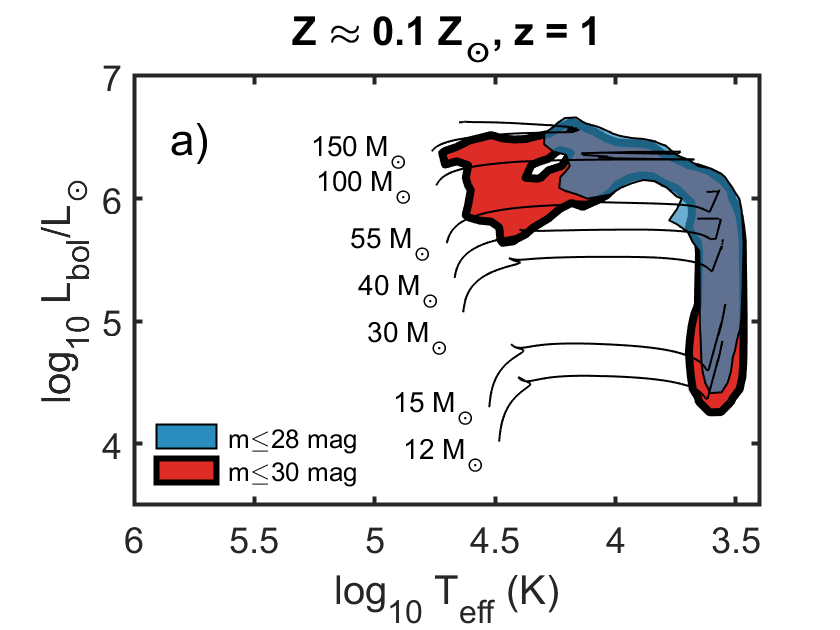} 
\includegraphics[scale=0.4]{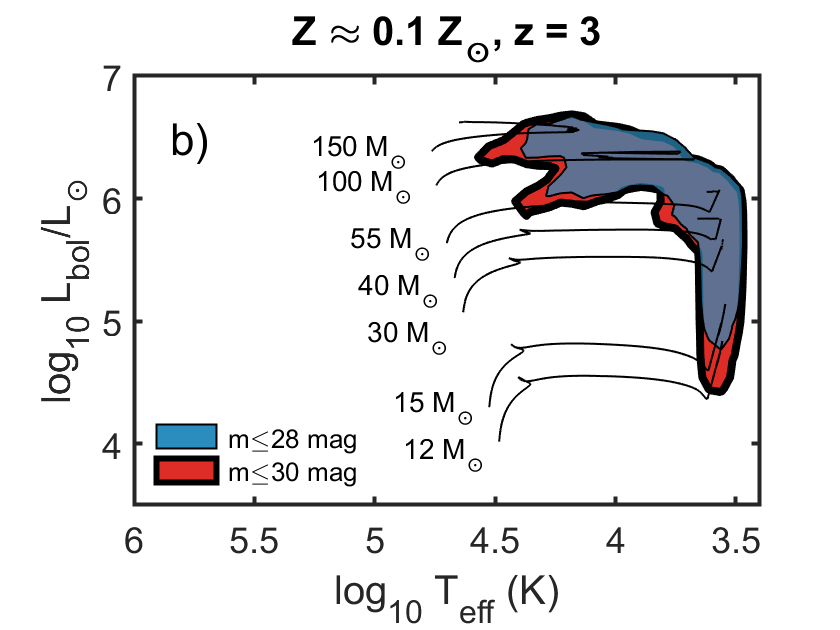}
\includegraphics[scale=0.4]{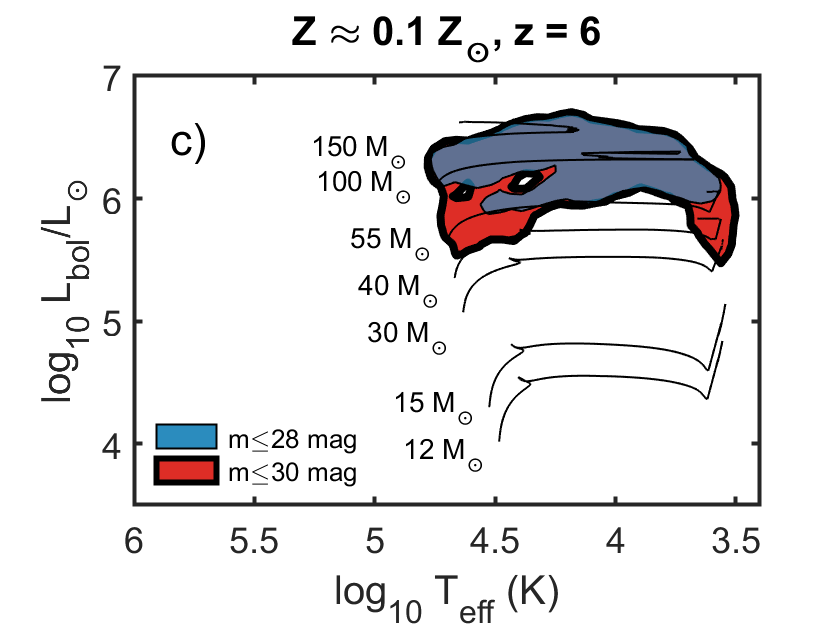}
\includegraphics[scale=0.4]{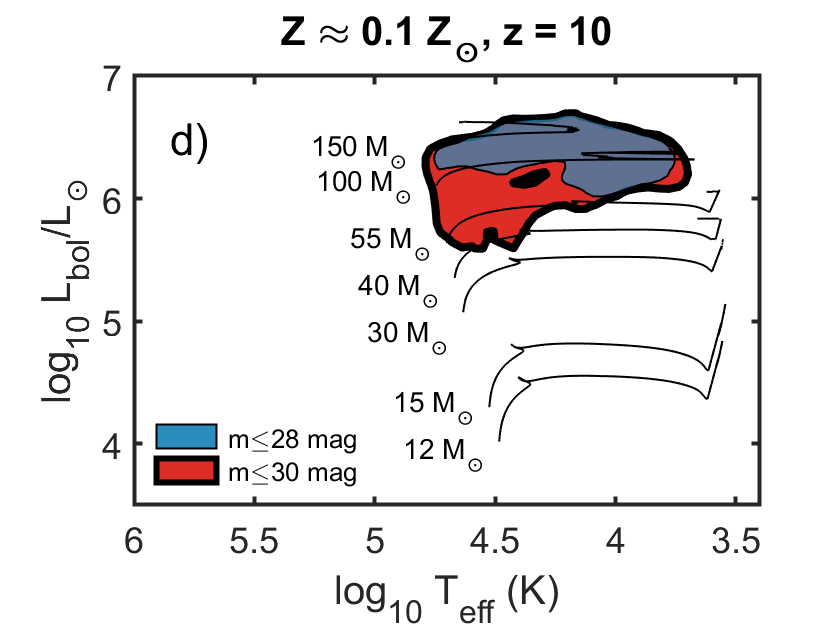}
\caption{The selection bias of lensed, low-metallicity ($Z\approx 0.1\ Z_\odot$) $M_\mathrm{ZAMS}=10$--150 $M_\odot$ stars in the Hertzsprung-Russell diagram at {\bf a)} $z=1$, {\bf b)} $z=3$, {\bf c)} $z=6$ and {\bf d)} $z=10$ under the assumption of a source-plane $P(>\mu)\propto \mu^{-2}$ magnification distribution with $\mu=1\times 10^2$--$5\times 10^4$ and an additional size-magnification limit (equation~\ref{eq:mu-size}) with $A_0=5\times 10^4$. The coloured, filled contours indicate the regions in which $\approx 95\%$ of the lensed stars will appear, in the case of a 28.0 AB mag (blue patch, thin contour) and 30.0 AB mag (red patch, thick contour) survey limit respectively. The prospects of detecting red supergiants drops drops at the highest redshifts, since the SED peak of these stars moves out of JWST/NIRCam range, and at $z=10$ (panel d), no such stars are expected to be seen.}
\label{HR_bias_low_metallicity_z1to10}
\end{figure*}

Stars that attain their high magnifications due to microlensing by stars in a foreground galaxy cluster, for instance because they were selected based on variability criteria in multi-epoch observations, may show a slightly different HR diagram bias compared to stars selected due to their close proximity of a cluster macrolensing caustic. In general, the size of a star sets an upper limit on its microlensing-induced magnification, which would make microlensing disfavour detection of stars with the largest radii. 

While the exact size-magnification limit depends on the gradient of the lensing potential at the position of the event, and the surface mass density in stars at this location in the cluster lens \citep[e.g.][]{Venumadhav17,Meena23}, we here adopt an approximate size-magnification limit of the form
\begin{equation}
\max(\mu_\mathrm{micro})\approx A_0 (D_\mathrm{S}/1Gpc)^{1/2}(R/10R_\odot)^{-1/2},
\label{eq:mu-size}
\end{equation}
with $A_0$ either $1\times 10^4$ or $5\times 10^4$, where the higher value corresponds to a lower contribution from microlenses to the surface mass density. Here, $D_\mathrm{S}$ is the angular size distance to the source at redshift $z$, and $R$ is the radius of the star. The $A_0=1\times 10^4$ value is similar to the lowest value considered by \citet{Meena23}, while the $A_0=5\times 10^4$ is the highest we can consider given our overall cut-off in overall the magnification distribution. This range of $A_0$ corresponds to a change in the surface mass density of microlenses by a factor of $\approx 8$ \citep{Venumadhav17}.

\subsection{Selection bias for metal-enriched stars obeying a standard IMF}
To illustrate the effects of the resulting selection biases, we in Fig.~\ref{HR_bias_low_metallicity_z2} compare the intrinsic distribution of stars with initial masses 10--150 $M_\odot$ in the HR diagram with the most likely regions for lensed stars to appear at $z=2$ in a JWST survey reaching 27.0 AB magnitudes, given our $Z/Z_\odot \sim 0.1$ models. The intrinsic distribution is dominated by $M_\mathrm{ZAMS} \lesssim 55 M_\odot$ stars on the main sequence at $\log T_\mathrm{eff}>4.3$, and by $M_\mathrm{ZAMS} \lesssim 20 M_\odot$ stars that have evolved to red supergiants at $\log T_\mathrm{eff} \approx 3.6$. 

In a lensed sample with $\mu=1\times 10^2$ -- $5\times 10^4$ and a microlensing size-magnification limit with $A_0=5\times 10^4$ (Figure~\ref{HR_bias_low_metallicity_z2}a), the observed distribution instead gets dominated by cooler stars in the $T_\mathrm{eff}<20000$ K ($\log T_\mathrm{eff}<4.3$) range, and with initial masses from $M_\mathrm{ZAMS}\gtrsim 10 \ M_\odot$ all the way up to $\approx 150\ M_\odot$. Adopting a more restrictive microlensing size-magnification limit (equation~\ref{eq:mu-size}) with $A_0=1\times 10^4$ (Figure~\ref{HR_bias_low_metallicity_z2}b) removes some of the lower-mass red supergiants with initial masses $<30\ M_\odot$, which have large radii ($\approx 300$--$1000\ R_\odot$) and require $\mu>1000$ to be rendered detectable. 

In Figure~\ref{HR_bias_low_metallicity_z2}c and d, we show the corresponding distributions for the highest-magnification stars with $\mu\geq 3000$. In this subset, one finds an increase in the fraction of 
$T_\mathrm{eff}\gtrsim 20000$ K ($\log T_\mathrm{eff}\gtrsim 4.3$ stars, which require these very high magnifications to get boosted above the detection threshold. In the $A_0=5\times 10^4$ (Figure~\ref{HR_bias_low_metallicity_z2}c) case, stars with initial masses $\approx 100$--150 $M_\odot$ can in fact be detected at their ZAMS temperatures of $T_\mathrm{eff}\gtrsim 50000$ K ($\log T_\mathrm{eff}\gtrsim 4.7$). In the case of $A_0=1\times 10^4$ (Figure~\ref{HR_bias_low_metallicity_z2}d), the $T_\mathrm{eff}<12000$ K stars fall out of the sample because of their large radii and the size-magnification limit set by equation~\ref{eq:mu-size}. Some of the high-mass, highest-$\log T_\mathrm{eff}$ stars also get removed by this criterion, since these stars have radii $\sim 10\ R_\odot$ already on the ZAMS and require $\mu>10^4$ to be detectable because of their high $T_\mathrm{eff}$.

Using $A_0=5\times 10^4$, and $\mu=1\times 10^2$--$5\times 10^4$, we in Fig.~\ref{HR_bias_low_metallicity_z1to10} indicate the most likely regions in the HR diagram for $Z/Z_\odot \sim 0.1$ stars to appear at $z=1$, 3, 6 and 10, in the case of two deeper JWST photometric surveys that reach 28.0 and 30.0 AB magnitudes. 

At $z=1$ and $z=3$, the resulting distributions are similar to those at $z=2$ in Figure~\ref{HR_bias_low_metallicity_z2}a, but extend to higher $T_\mathrm{eff}$ at these fainter detection thresholds.
At $z=6$, the chance of detecting red supergiants from stars with initial masses $M<55 \ M_\odot$ (28 AB mag survey limit) or 
$M<30 \ M_\odot$ (30 AB mag survey limit) diminishes, and stars with $T_\mathrm{eff}$ up to their ZAMS values enter the sample for initial masses $\gtrsim 100 \ M_\odot$ (28 AB mag) or $\gtrsim 40 \ M_\odot$ (30 AB mag). 
At $z=10$ the chance of detecting red supergiants at $\log T_\mathrm{eff}<3.6$ is gone, since the SED peaks of these stars have redshifted far outside the JWST/NIRcam wavelength range. 

In the redshift range $z=1$--6, it is interesting to note that for the brighter detection limits of 27--28 AB mag (Fig.~ \ref{HR_bias_low_metallicity_z2}abc and \ref{HR_bias_low_metallicity_z1to10}abc), the relative probability of lensed stars to turn up in the 4000--15000 K range ($\log(T_\mathrm{eff})\approx 3.6$--4.2) becomes substantial, in agreement with recent detections of lensed stars \citep{Kelly18,Chen22,Welch22b,Meena23,Furtak23,Diego23b,Diego23c} at these redshifts, even though stars in this temperature range are exceedingly rare.

\begin{figure*}
\includegraphics[width=\columnwidth]{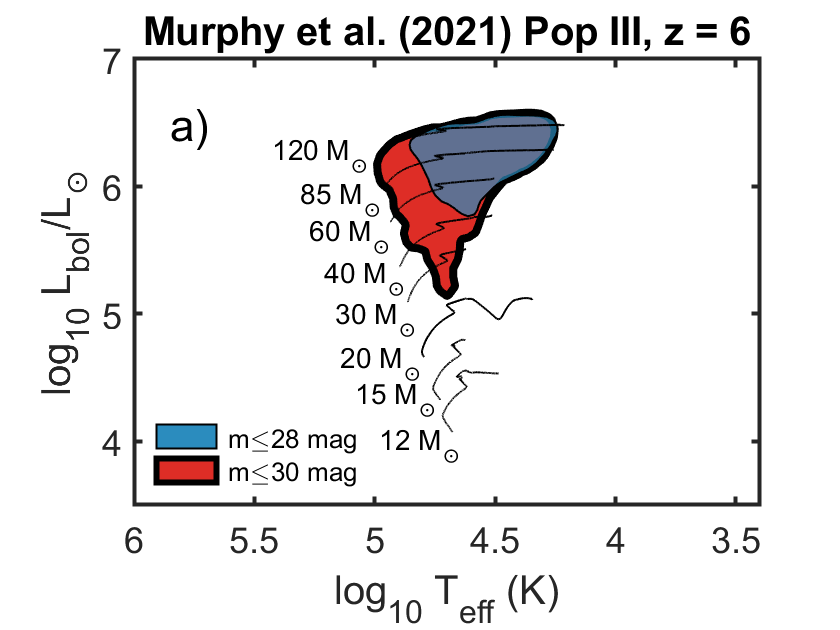}
\includegraphics[width=\columnwidth]{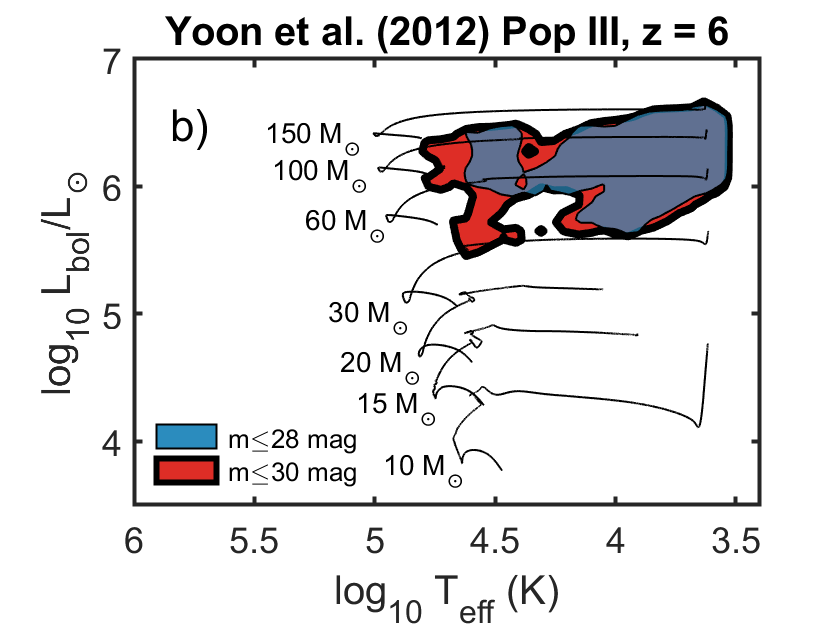}
\caption{The most likely positions in the HR diagram of lensed $z=6$, 10-120 $M_\odot$ Pop III stars following the {\bf a)} \citet{Murphy21a} and {\bf b)} \citet{Yoon12} stellar evolutionary tracks for non-rotating stars, under the assumption of a source-plane $P(>\mu)\propto \mu^{-2}$ magnification distribution with $\mu=1\times 10^2$--$5\times 10^4$ and a size-magnification limit (equation~\ref{eq:mu-size}) with $A_0=5\times 10^4$. The coloured, filled contours indicate the regions in which $\approx 95\%$ of the lensed stars will appear, in the case of a 28.0 AB mag (blue patch, thin contour) and 30.0 AB mag (red patch, thick contour) survey limit respectively. A top-heavy, log-normal IMF \citep[][with characteristic mass $M_\mathrm{c}=10\ M_\odot$ and $\sigma =1$]{Tumlinson06} is assumed for all scenarios depicted. In both panels, stars captured by a 30 AB mag sample extend to somewhat higher $T_\mathrm{eff}$ than those captured in the 28 AB mag sample. It is only in the 30 AB mag case, and for 
the \citet{Murphy21a} tracks (panel a), that lensed stars at the characteristically high main-sequence temperatures ($T_\mathrm{eff}\sim 10^5$ K) of $\gtrsim 100\ M_\odot$ Pop III stars make up any signficant fraction of those rendered detectable. Because the \citet{Murphy21a} tracks do not extend to $T\mathrm{eff}<15000$ K for Pop III masses in the range plotted, whereas the \citet{Yoon12} ones do, the contours in panel a appear more compact along the $T_\mathrm{eff}$ directions than in panel b.}
\label{HR_bias_PopIII} 
\end{figure*}

We stress that we in Figure ~\ref{HR_bias_low_metallicity_z1to10} only show the expected HR diagram distribution of stars with ZAMS mass $\geq 10\ M_\odot$. However, at the lowest redshifts (e.g., $z=1$--3), it is possible that stars with ZAMS mass $<10\ M_\odot$ could also enter surveys reaching 28--30 AB mag, since the contours extend down to the supergiant branch of the lowest-mass tracks plotted.

\subsection{Selection bias for Pop III stars with top-heavy IMFs}
In Figure~\ref{HR_bias_PopIII}, we depict the same selection bias (here with $A_0=5\times 10^4$, 28.0 and 30.0 AB mag survey limits) as in Figure~\ref{HR_bias_low_metallicity_z1to10}, but now in the case of Pop III tracks from \citet{Murphy21a} and \citet{Yoon12} with no initial rotation, under the assumption of a top-heavy, log-normal IMF \citep[][with characteristic mass $M_\mathrm{c}=10\ M_\odot$ and $\sigma =1$]{Tumlinson06}, for $M=10$--$120\ M_\odot$ stars at $z=6$. 

A fundamental difference between the \citet{Murphy21a} and \citet{Yoon12} tracks is that the latter predict non-rotating Pop III stars to reach a $\approx 4000$ K stage for $M_{ZAMS}\geq 30\ M_\odot$ whereas the former stay above 15000 K until the end of the tracks. As we have previously shown that low-$T_\mathrm{eff}$ stages, even if very short-lived, can have a significant impact on the detectability of lensed stars, this difference leads to very different predictions on where in the HR diagram lensed Pop III stars are the most likely to be detected. 

As a result, the \citet{Yoon12} tracks favour the appearance of red supergiants in samples with detection limits at both 28.0 and 30.0 AB magnitudes, whereas the \citet{Murphy21a} tracks produce distributions that are limited to the $T_\mathrm{eff}\gtrsim 15000$ K range. 

It is worth noting that lensed stars at the peak temperatures of Pop III stars during early stages of evolution ($\sim 10^5$ K) only make up a signficant fraction of the simulated sample in the case of the \citet{Murphy21a} tracks and a survey limit of 30.0 AB magnitudes. This extension towards very high $T_\mathrm{eff}$ represents an important difference between the predicted HR-diagram distribution of lensed 10--120 $M_\odot$ Pop III stars (Figure~\ref{HR_bias_PopIII}) and the corresponding distribution for lensed metal-enriched stars at $z=6$ (Figure~\ref{HR_bias_low_metallicity_z1to10}c), as the latter do not appear at $T_\mathrm{eff}>60000$ K ($\log T_\mathrm{eff}> 4.8$) given our chosen set of tracks. However, even if $\sim 10^5$ K objects were to appear in a sample of lensed stars, measuring their temperatures sufficiently well to constrain their location to this part of the HR diagram may be extremely challenging at faint magnitudes, as discussed in Sect.~\ref{sec:PopIII features}. Moreover, Wolf-Rayet stars can also attain temperatures of $\sim 10^5K$, and alternative sets of tracks for metal-enriched stars that include this phase \citep[e.g.][]{Eldridge06} may shift the predictions to higher $T_\mathrm{eff}$ than shown in 
Figure~\ref{HR_bias_low_metallicity_z1to10}c.

\begin{figure}
\includegraphics[width=\columnwidth]{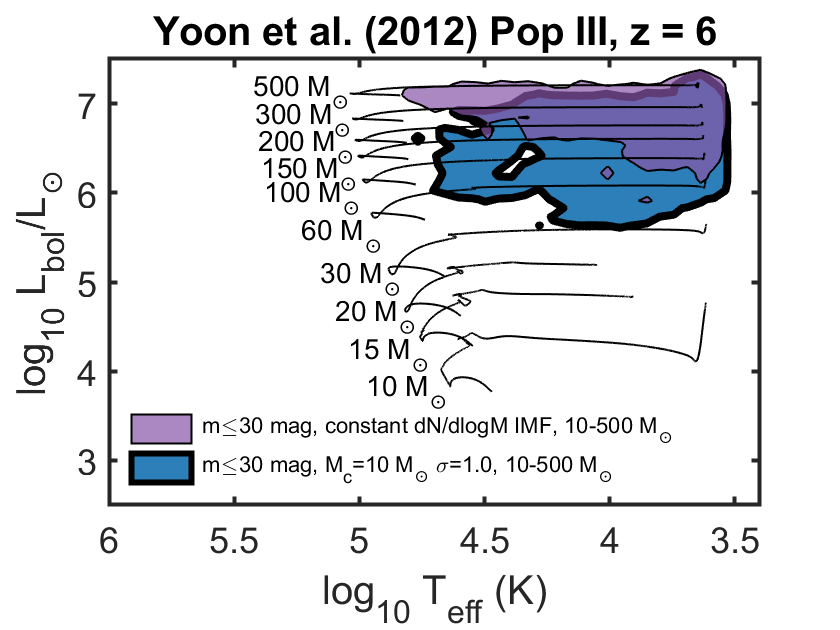}
\caption{IMF effects on the HR diagram distribution of lensed Pop III stars at $z=6$. The coloured, filled contours indicate the regions in which $\approx 95\%$ of the lensed stars are predicted appear, in the case of a $\mathrm{d}N/\log\mathrm{M} = \mathrm{constant}$ IMF (indigo patch, thin contour) and a lognormal $M_\mathrm{c}=10\ M_\odot$ and $\sigma =1$ IMF
(blue patch, thick contour), both extending from 10--500 $M_\odot$. We here adopt \citet{Yoon12} stellar evolutionary tracks for non-rotating Pop III stars,
a source-plane $P(>\mu)\propto \mu^{-2}$ magnification distribution with $\mu=1\times 10^2$--$5\times 10^4$, a size-magnification limit (equation~\ref{eq:mu-size}) with $A_0=5\times 10^4$ and a JWST/NIRCam survey with detection limit 30.0 AB mag. The higher upper mass limit adopted here compared to Figure~\ref{HR_bias_PopIII}b does not have any dramatic impact on the $T_\mathrm{eff}$ distribution, but favours the detection of more massive and luminous stars.} 
\label{HR_bias_PopIII_IMFs}
\end{figure}

The predicted distribution of lensed Pop III stars across the HR diagram also depends on the shape and upper mass limit of the IMF, which we demonstrate in Fig.~\ref{HR_bias_PopIII_IMFs} in the case of \citet{Yoon12} tracks and a survey limit of 30 AB mag. Here, we assume that Pop III stars can form up to a mass of 500 $M_\odot$ and show the results for two different top-heavy IMFs: a log-normal IMF \citep[][, with characteristic mass $M_\mathrm{c}=10\ M_\odot$ and $\sigma =1$]{Tumlinson06} and a $\mathrm{d}N/\log\mathrm{M} = \mathrm{constant}$ IMF \citep[e.g.][]{Schauer22}. 

The higher upper mass limit adopted for Pop III stars compared to Fig~\ref{HR_bias_PopIII}b does not have any dramatic effects on the $T_\mathrm{eff}$ distribution of lensed stars across the HR diagram for either the log-normal $M_\mathrm{c}=10\ M_\odot$ IMF, or the $\mathrm{d}N/\log\mathrm{M} = \mathrm{constant}$ IMF. However, the probability density function does extend to higher $L_\mathrm{bol}$ and $M_\mathrm{ZAMS}$, reaching $\approx 500\ M_\odot$ and $\log(L/L_\odot)\approx 7$. If one assumes that the birth of such very massive stars is effectively limited to the Pop III mode of star formation, then the detection of lensed stars in this part of the HR diagram would constitute an indication that Pop III objects have been detected even without detailed metallicity measurements (Section~\ref{sec:PopIII features}). However, ensuring that a given lensed star belongs to this part of the diagram requires that $T_\mathrm{eff}$ is well constrained and that strong upper limits on the magnification can be set to infer a robust lower limit on the luminosity. 

In summary, the identification of candidates for lensed Pop III stars based solely on their position in the HR diagram, without any additional metallicity constraints (Section~\ref{sec:PopIII features}), may be very difficult.

\section{Predicted detection rates for highly magnified Pop III stars}
\label{sec:PopIII_detection_probability}
While Section~\ref{sec:selection function} focused on the {\it relative} probability distribution of stars in the HR diagram of lensed stars\footnote{which provides answers to questions of the form ``if a lensed star is detected at a given redshift, what properties is it most likely to have?''}, we here attempt to estimate the absolute probabilities of detecting Pop III stars at $z\gtrsim 6$ in a JWST survey covering a large number of cluster-lens fields.

When imaging cluster-lens fields, highly magnified stars usually manifest themselves as point sources within lensed arcs, but could in principle also appear as point sources along the macrolensing critical curve of the cluster lens but without any obvious association to an arc. The latter case may arise when the magnification of the star is sufficient to render it detectable, but the host galaxy (which would be subject to a much smaller magnification) remains below the detection threshold. This situation has so far not been seen, but could potentially become increasingly important at the very highest redshifts, where a significant fraction of the cosmic star formation activity may take place in galaxies too faint to be seen through macrolensing even in very deep JWST images 
\citep[e.g.][]{Wu24}. This scenario could also be relevant for Pop III stars forming early on in minihalos, which are not expected to have any association with large galaxies. 

In what follows, we will focus on the the momentary brightness of lensed Pop III stars at high redshifts, and attempt to estimate the probability that a lensed Pop III star appears in a single-epoch imaging campaign covering several cluster-lens fields. The timescale over which stars detected this way will remain observable once discovered -- which depends on the timescale of the crossing of macrolensing caustics and the impact of microlensing -- is outside the scope of the current paper.

Following, \citet{Windhorst18}, we adopt the relation for fold caustics for the macrolensing magnification $\mu_\mathrm{macro}$ as a function of source-plane distance $d_\mathrm{arcsec}$ from the caustic in arcseconds: 
\begin{equation}
\mu_\mathrm{macro} = \frac{B_0}{(d_\mathrm{arcsec})^{1/2}},
\label{eq:mu_B0}
\end{equation}
where $B_0$ is a scaling factor that varies from cluster to cluster, in the $B_0\approx 10$--20 arcsec$^{1/2}$ range for some of the most well-known cluster lenses \citep{Windhorst18}. Strictly speaking, $B_0$ also varies with source redshift, but the dependence is expected to be very weak in the $z\gtrsim 6$ regime studied in this paper, and we therefore treat it as a constant throughout this paper. 

Given an imaging survey with a set of $i$ JWST/NIRCam filters with detection limits $m_{\mathrm{AB limit},\ i}$, one can -- for a star with initial (ZAMS) mass $M$, age $t$ and intrinsic apparent magnitude $m_{M,\ t,\ i}(z)$ -- derive the minimum macrolensing magnification $\mu_\mathrm{detection}$ required to bring the star with mass $M$ and age $t$ above the detection limit in at least one filter as:
\begin{equation}
\mu_\mathrm{detection}(z)=10^{\frac{\min (m_{M,\ t,\ i}(z)-m_{\mathrm{AB limit},\ i}) }{2.5}}.
\label{eq:mu_detection}
\end{equation}
Neglecting microlensing, the maximum distance from the caustic at which such a star may be detected in this survey is then:
\begin{equation}
\max(d_\mathrm{arcsec(z)})=\left( \frac{B_0}{\mu_\mathrm{detection}(z)}\right)^2.
\label{eq:max_dcaustic}
\end{equation}

Given this relation and an assumed source-plane length $L_\mathrm{arcsec}$ of the caustic in arcseconds, we may estimate the angular surface area $\theta^2$ (in arcsec$^2$) over which a given star will attain a magnification equal to or higher than $\mu_\mathrm{detection}(z)$ (at separations $0<d_\mathrm{arcsec}<\max(d_\mathrm{arcsec})$) as 
\begin{equation}
\theta^2\left( \mu\geq \mu_\mathrm{detection}(z) \right)=\max(d_\mathrm{arcsec}(z))L_\mathrm{arcsec}.
\label{eq:A_arcsec2}
\end{equation}

By adopting a certain redshift survey depth $\Delta z$, this angular surface area $\theta^2$ can be converted into a redshift-dependent comoving volume $V_\mathrm{z,\Delta(z)}$ inside which objects will attain $\mu\geq\mu_\mathrm{detection}(z)$:
\begin{equation}
V_\mathrm{z,\Delta(z)} (\mu\geq \mu_\mathrm{detection}(z)) =  V(z,\Delta z)_{\mathrm{arcsec^{-2}}}\ \theta^2(\mu\geq \mu_\mathrm{detection}(z)),
\label{eq:Vz,deltaz}
\end{equation}
where $V(z,\Delta z)_{\mathrm{arcsec^{-2}}}$ is the comoving volume at redshift $z$ that corresponds to depth $\Delta(z)$ per arcsec$^2$. We stress that this treatment only takes macrolensing into account, and neglects the way microlensing is likely to alter the magnification distribution \citep[e.g.][]{Diego19,Palencia23}

\begin{figure*}
\includegraphics[width=\columnwidth]{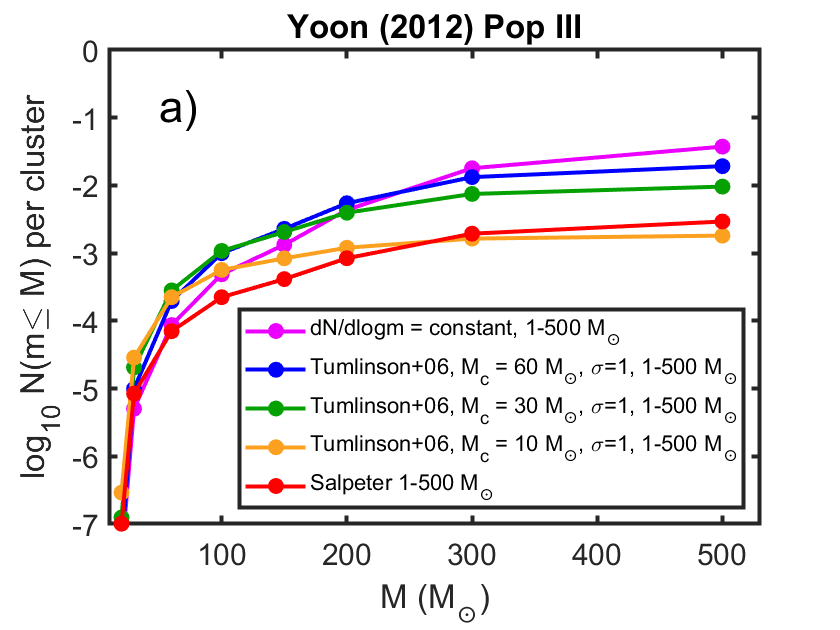}
\includegraphics[width=\columnwidth]{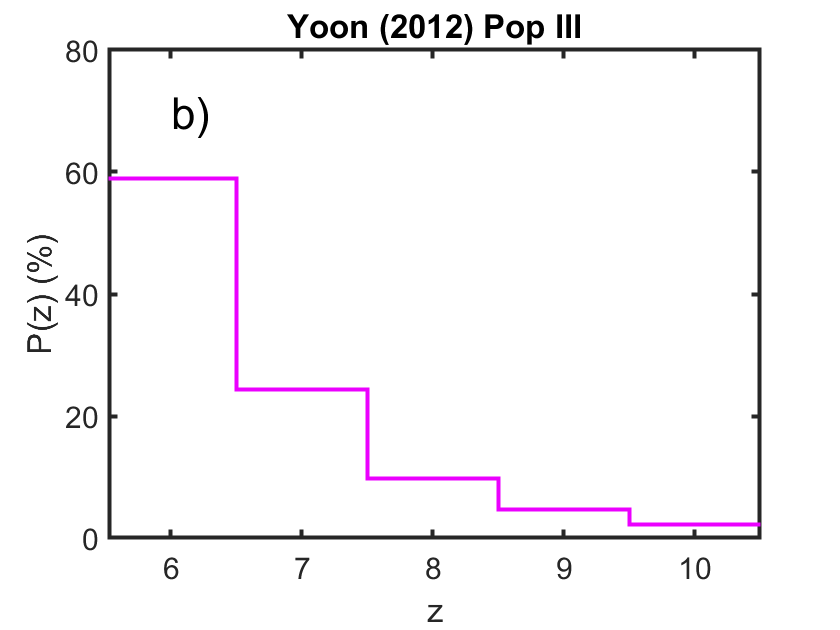}
\caption{{\bf a)} Expected number of detections of lensed Pop III stars at $z=6$--10 in a single cluster lens field observed with JWST/NIRCam to a depth of 28.5 AB mag. The five differently colored lines represent different options for the shape of the Pop III IMF throughout the 1--500 $M_\odot$ range. All models adopt the \citet{Liu20} Pop III SFRD throughout the $z=6$--10 interval, and \citet{Yoon12} stellar evolutionary tracks for non-rotating Pop III stars. Detections of lensed Pop III stars in JWST surveys covering a few tens of cluster-lens fields are only realistic for $\log_{10} N(m\leq M)\gtrsim -2$, which is only possible for the most massive Pop III stars ($M\gtrsim 300\ M_\odot$) and the most top-heavy IMFs (purple, blue and green lines). {\bf b)} The redshift probability distribution for the stars that would make it above the detection threshold in the case of the $\mathrm{d}N/\log\mathrm{M} = \mathrm{constant}$, 1--500 $M_\odot$ IMF (but the distribution is similar for all IMFs used in panel a). The distribution is dominated by detections at $z\approx 6$--7, since many of the lowest-$T_\mathrm{eff}$ stars have been redshifted out of the JWST/NIRCam wavelength range at higher redshifts.}
\label{PopIII_detection_rate}
\end{figure*}

The number of expected detections of lensed Pop III stars in that $\Delta z$ slice can then be derived by estimating the number of luminous Pop III stars within the corresponding volume. Since the magnification required for detection changes dramatically as a function of age of Pop III stars, we sum over $j$ time steps that represent different evolutionary states of the star, each with a duration $\Delta t_j$, and over all redshift slices $z$ for a single cluster:

\begin{equation}
N_\mathrm{detection}(M)\approx \sum_z \sum_j \frac{\mathrm{SFRD}(z) f(M) V_\mathrm{z,\Delta(z)}(\mu\geq \mu_\mathrm{detection}(z)) \Delta t_j}{M}.
\label{eq:Ndetection}
\end{equation}
Here, $N_\mathrm{detection}(M)$ is the expected number of lensed stars of initial mass $M$, $\mathrm{SFRD}(z)$ is the comoving cosmic star formation rate density (in units of $M_\odot$ yr$^{-1}$ cMpc$^{-3}$) of Pop III stars at redshift $z$, and $f(M)$ is the IMF-dependent mass fraction locked up in Pop III stars with initial mass $M$. 

Throughout this paper, we adopt $B_0=15$ and $L_\mathrm{arcsec}=100$ arcsec as fiducial values for strong-lensing clusters. As long as $N_\mathrm{detection}\ll 1$, $N_\mathrm{detection}$ can also be directly interpreted as the approximate probability of detection ($P_\mathrm{detection}\approx N_\mathrm{detection}$) per cluster field surveyed. 

Detecting Pop III stars in JWST surveys of cluster lens fields requires both a relatively high cosmic star formation rate density (SFRD) of Pop III stars at the relevant redshifts, and a top-heavy Pop III IMF. Here, we limit ourselves to $z\gtrsim 6$, since this is the regime that simulations of Pop III stars typically focus on, and for which numerous Pop III SFRD predictions have been published. 

Based on the computational machinery outlined, we find that a SFRD of $\sim 10^{-4}\ M_\odot$ cMpc$^{-3}$ yr$^{-1}$ somewhere in the $z\approx 6$--10 range is required to make detections realistic in the case of very top-heavy IMFs, and that even higher SFRDs are required for the less extreme IMFs. Adopting \citet{Liu20} as our reference SFRD (which reaches a peak of $\approx 10^{-4}\ M_\odot$ cMpc$^{-3}$ yr$^{-1}$ at $z\approx 10$ and drops by a factor of $\approx 3$ by $z=6$) for Pop III stars, we in Fig.~\ref{PopIII_detection_rate}a show the expected detection rate in a 28.5 AB mag JWST/NIRCam survey for five different IMFs covering the 1--500 $M_\odot$ mass range, under the assumption of the \citet{Yoon12} stellar evolutionary tracks for non-rotating Pop III stars. These IMFs include a $\mathrm{d}N/\mathrm{d}M\propto M^{-2.35}$ Salpeter slope IMF, three \citet{Tumlinson06} log-normal IMFs with $\sigma=1$ and characteristic masses $M_\mathrm{c}=10\ M_\odot$, $30\ M_\odot$ and 60 $M_\odot$, plus the \citet{Magg16} $\mathrm{d}N/\mathrm{d}\log(M)=$ constant IMF. Since we are working with discrete stellar evolutionary tracks, each track is in equation~\ref{eq:Ndetection} assigned a mass fraction $f(M)$ derived from the continuous 1--500 $M_\odot$ IMF. The lower and upper limits of the mass interval attached to each track are set by the midpoints in initial mass between adjacent tracks, except for the lowest-mass track in the \citet{Yoon12} set ($M=10\ M_\odot$), for which the assigned interval is 7.5--12.5 $M_\odot$, and the highest-mass track considered ($M=500\ M_\odot$), for which the assigned interval is 400--500 $M_\odot$.

In the following, we consider the $\mu_\mathrm{detection}(z)=100$--$5\times 10^4$ magnification range and adopt a microlensing size-magnification limit with $A_0=5\times 10^4$. The latter limit prevents stars in evolutionary stages where $\mu_\mathrm{detection}(z)$ from Equation~\ref{eq:mu_detection} exceeds $\max(\mu_\mathrm{micro})$ from Equation~\ref{eq:mu-size} to enter the summation when computing $N_\mathrm{detection}(M)$ using Equation~\ref{eq:Ndetection}. However, we note that this only affects the detectability of \citet{Yoon12} Pop III stars with masses $<50 M_\odot$, since the more massive stars can reach the 28.5 AB mag detection limit with much lower magnifications than constrained by this limit. In fact, a 500 $M_\odot$ star at $z=6$ can be rendered detectable at a magnification as low as $\mu\approx 100$ in its brightest phase. At magnifications that low, there may admittedly be observational ambiguities between individual lensed stars and lensed star clusters, as many stars within the same star cluster may experience similar macromagnifications. Hence, our resulting estimate may include cases where the light from several lensed Pop III stars in practice would blend together.

If we take $\log_{10}\mathrm{N_\mathrm{detection}}\gtrsim -2$ as the threshold above which the detection of lensed Pop III stars can be assumed to be realistic (giving a $\gtrsim 10\%$ detection probability for a survey of $\geq 10$ clusters fields), Figure~\ref{PopIII_detection_rate}a suggests that it is only the most massive stars ($\gtrsim 300\ M_\odot$) that have any realistic chance of being detected, and then only for the most top-heavy IMFs considered (\citealt{Tumlinson06} $M_\mathrm{c}=30\ M_\odot$ and 60 $M_\odot$, plus the \citealt{Magg16} $\mathrm{d}N/\mathrm{d}\log(M)=$ constant IMF), where the mass fraction in $\gtrsim 100\ M_\odot$ stars is $\gtrsim 10\%$. 

The most top-heavy IMF considered here ($\mathrm{d}N/\mathrm{d}\log\mathrm{M} = \mathrm{constant}$) produces a total probability for the detection of a Pop III star of $P_\mathrm{detect}\approx N_\mathrm{detection}(m\leq 500\ M_\odot) \approx 10^{-1.4}$. In this scenario, a survey covering $N_\mathrm{clusters}$ cluster-lensing fields would result in a combined detection probability $P_\mathrm{detect,N_\mathrm{clusters}}=1-(1-P_\mathrm{detect})^{N_\mathrm{clusters}}$ across the whole survey. This implies $P_\mathrm{detect,N_\mathrm{clusters}} \approx 33\%$ for a survey covering 10 cluster fields, $\approx 70\%$ for 30 cluster fields and $\approx 87\%$ for 50 clusters. 

$P_\mathrm{detect}$ is lower by $\approx 0.3$, 0.6 and 1.3 dex, respectively, for the \citet{Tumlinson06} $M_\mathrm{c}=60 \ M_\odot$, 30 and 10 $M_\odot$, $\sigma=1$ IMFs and $\approx 1.1$ dex lower for the Salpeter-slope 1--500 $M_\odot$ IMF (please note that the \citealt{Tumlinson06} IMF with $M_\mathrm{c}=10\ M_\odot$ and $\sigma=1$ actually predicts slightly fewer $\gtrsim 300\ M_\odot$ stars compared to the Salpeter-slope IMF extending to 500 $M_\odot$). Even in the most pessimistic case ($M_\mathrm{c}=10\ M_\odot$ and $\sigma=1$), $P_\mathrm{detect,N_\mathrm{clusters}}$ becomes $\approx 10\%$ for an $N_\mathrm{clusters}=50$ survey.

Although the \citet{Liu20} pop III SFRD peaks at $z\approx 10$, we in Figure~\ref{PopIII_detection_rate}b show that the predicted redshift distribution of lensed stars still peaks at $z\approx 6$ (the lowest redshift considered here) and decreases rapidly towards z=10 (the highest redshift considered), primarily due to the progressive loss of the SED peak of the lowest-$T_\mathrm{eff}$ stars (which, we have argued, are often the easiest to detect) from the NIRCam wavelength window as the redshift is increased. For the \citet{Yoon12} set of tracks, this redshift evolution in detection probability only show variations at the level of a few percent among the IMFs considered.

\subsection{Impact of the Pop III star formation rate density evolution}
The redshift evolution of the SFRD of Pop III stars in not well-constrained, and different simulations give rise to radically different SFRD($z$) scenarios -- some attain SFRDs in the $z\approx 6$--10 range lower than that of the \citet{Liu20} model ($\sim 10^{-4}\ M_\odot$ cMpc$^{-3}$ yr$^{-1}$) adopted in Fig.~\ref{PopIII_detection_rate}, and some attain higher values (see \citealt{Klessen23} and \citealt{Venditti23} for recent compilations of simulation results).

To approximately estimate the Pop III IMF and SFRD($z$) limits that would render lensed Pop III stars detectable, we for simplicity here treat the Pop III IMF and the Pop III SFRD($z$) as independent. Naively, one would expect the Pop III IMF to have some impact on the SFRD evolution via feedback and chemical enrichment, but simulations differ significantly in their predictions on this connection. For example, \citet{Maio10} and \citet{Sarmento19} finds that a high characteristic mass of Pop III masses leads to lower Pop III SFRD, whereas \citet{Mebane18} find the opposite trend, and \citet{Pallottini14} find only very small variations in SFRD with different Pop III IMF prescriptions.

In general, keeping the Pop III IMF fixed but varying the SFRD results in a situation in which the $P_\mathrm{detect}$ estimates roughly scale with Pop III SFRD, so that a version of the \citet{Liu20} SFRD scaled up or down by some factor would result in an updated probability $P_\mathrm{detect}$ times this factor. Adopting a constant SFRD($z$) at $3\times 10^{-4}\ M_\odot$ cMpc$^{-3}$ yr$^{-1}$ throughout the $z=6$--10 interval, similar to the Pop III SFRD of \citet{Venditti23}, leads to a redshift distribution similar to that of Fig.~\ref{PopIII_detection_rate}b, but a $P_\mathrm{detect}$ a factor of $\approx 6$ higher than in the \citet{Liu20} case, thereby making even the detection of stars obeying the $M_\mathrm{c}=10\ M_\odot$ and $\sigma=1$ IMF possible ($P_\mathrm{detect,N_\mathrm{clusters}}\approx 40\%$) in a 50-cluster survey. 

On the other hand, if the $z=6$--10 Pop III SFRD is much lower ($10^{-5}\ M_\odot$ cMpc$^{-3}$ yr$^{-1}$), i.e., similar to that derived from the Renaissance void simulations of \citet{Xu16}, then even the most-top heavy IMF we consider here ($\mathrm{d}N/\mathrm{d}\log\mathrm{M} = \mathrm{constant}$) would give only $P_\mathrm{detect,clusters}\approx 30\%$ for a 50-cluster survey.

We note that, if stronger contstraints on the Pop III IMF were available, it may become possible to use the number of JWST detections of any lensed stars (regardless of their Pop III nature) at $z\gtrsim 6$ to set upper limits on the Pop III SFRD at these redshifts, as some SFRDs could predict a higher rate that that observed. Such Pop III IMF constraints may for instance come from the study of second-generation stars in the local Universe \citep[e.g.][]{Koutsouridou23b}.

\begin{figure}
\includegraphics[width=\columnwidth]{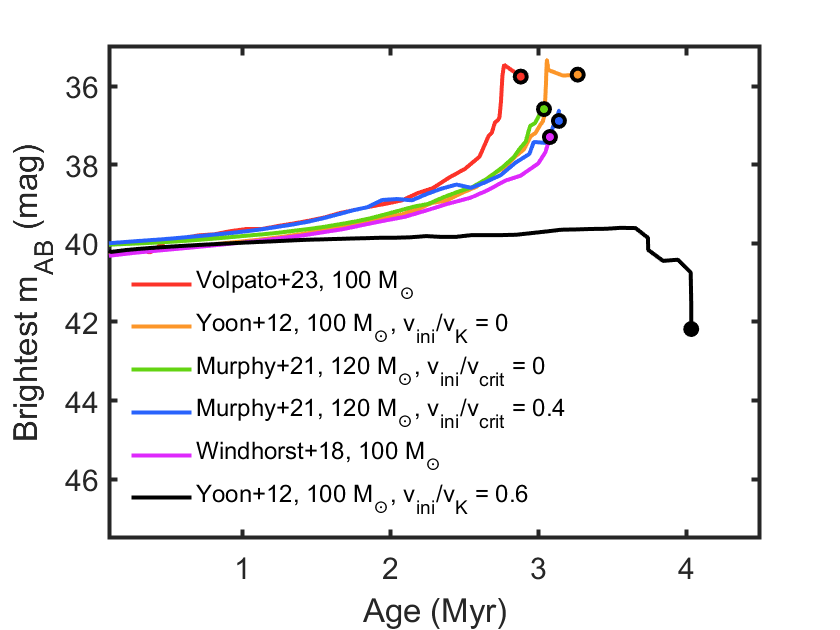}
\caption{Peak brightness in JWST/NIRCam wide-band filters at $z=6$ versus age for 100--120 $M_\odot$ stars based on the stellar evolutionary tracks of \citet{Yoon12}, \citet{Windhorst18}, \citet{Murphy21a} and \citet{Volpato23}. Circles mark the end points along each track. Because the \citet[][red line]{Volpato23} and non-rotating \citet[][orange line]{Yoon12} models for these stars attain much lower temperatures at late stages of evolution ($\approx 4000$--5000 K at their end points) than the other models, they become significantly brighter in the NIRCam filters and therefore more easily detectable. The rotating $v_\mathrm{K}=0.6$ \citet[][black line]{Yoon12} star evolves in the opposite direction, towards increasing $T_\mathrm{eff}$, as it ages and therefore attains a considerably lower peak flux in the JWST/NIRCam bands at an age above $\approx 1$ Myr.}
\label{PopIII_track_comparison}
\end{figure}

\subsection{Impact of late stages of stellar evolution}
\label{late_stages}
As we have argued in Section~\ref{sec:HR_bias}, the post-main sequence evolution of stars has a pronounced impact on the prospects of detecting lensed stars. Since not all Pop III stellar evolutionary tracks show the low-$T_\mathrm{eff}$ extension of the \citet{Yoon12} set, we explore the impact of removing $T_\mathrm{eff}<15000 K$ stars from the calculation. This has the effect of lowering the detection prospects of Fig.~\ref{PopIII_detection_rate}a by more than order of magnitude. Hence, the chances of detecting lensed III stars hinges crucially not just on the IMF, the SFRD but also on the evolutionary paths of these stars in the HR diagram. Because the \citet{Volpato23} stars also evolve to $T_\mathrm{eff}<15 000$ K, we find that the detection prospects at 100-500 $M_\odot$ are very similar to those of the \citet{Yoon12} tracks without rotation. For Pop III evolutionary tracks like those of \citet{Windhorst18}, which at 10--1000 $M_\odot$, never evolve to $T_\mathrm{eff}<30000$ K, and those of \citet{Murphy21a}, where non-rotating stars in the 10--120 $M_\odot$ range never reach $T_\mathrm{eff}<15000$ K, the detection prospects are considerably worse.

\begin{figure*}
\includegraphics[width=\columnwidth]{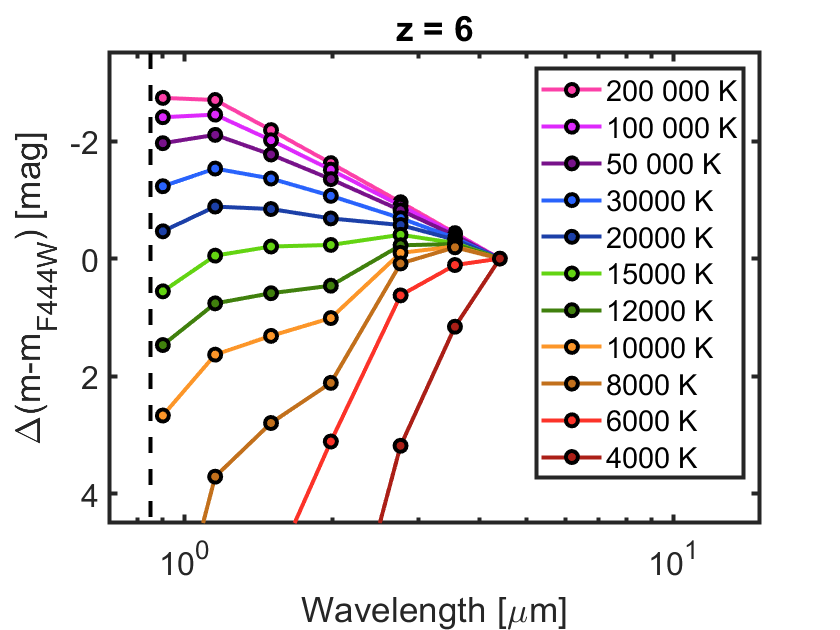}
\includegraphics[width=\columnwidth]{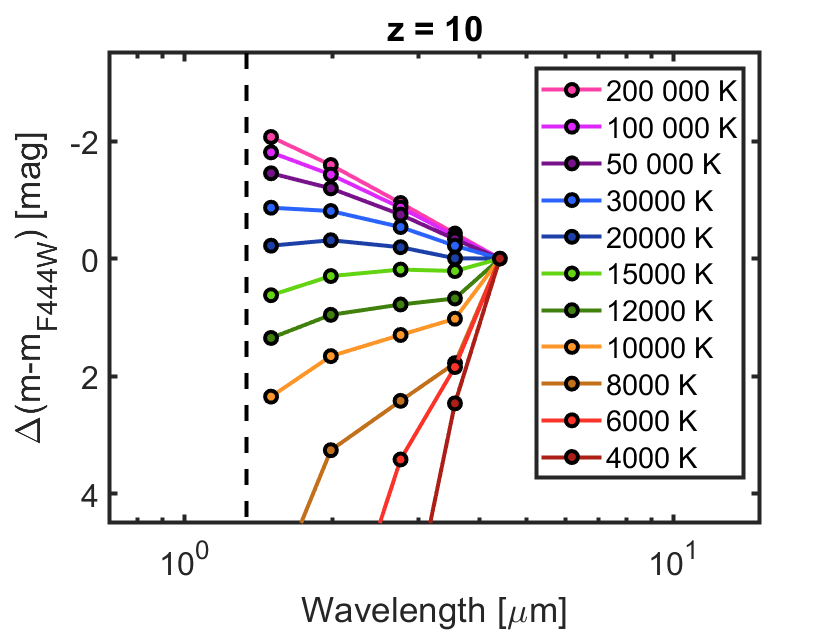}
\caption{NIRCam colour differences between Pop III stars at different temperatures in JWST wide-band filters at $z=6$ ({\bf left}) and $z=10$ ({\bf right}). The photometric fluxes from Muspelheim SEDs for Pop III stars at $T_\mathrm{eff}=4\times 10^3$--$2\times 10^5$ K have here been scaled to the same F444W flux. At each $T_\mathrm{eff}$, the SEDs are based on the lowest $\log(g)$ allowed by the stellar atmosphere grid. The vertical dashed line indicates the position of the Ly$\alpha$ break, to the left of which no detections are possible due to absorption in the neutral intergalactic medium. As seen, colours differences are small at high $T_\mathrm{eff}$ but become progressively larger at low $T_\mathrm{eff}$. We stress that in the case of the reddest (lowest $T_\mathrm{eff}$) SEDs, detections in more than a couple of NIRCam filter are unlikely, since this would require the peak flux to appear many magnitudes above the survey detection limit.}
\label{fig_Teff_col_signatures}
\end{figure*}

In Figure~\ref{PopIII_track_comparison}, we illustrate this by the exploring the brightest JWST/NIRCam broadband magnitudes attained at $z=6$ as a function of age for these four sets of tracks at 100--120 $M_\odot$, and also show the effects of rotation on the \citet{Yoon12} and \citet{Murphy21a} versions of these stars. The \citet{Windhorst18} and the \citet{Murphy21a} Pop III stars attain their peak brightness at levels $\approx 1$--1.5 magnitudes fainter than those of the \citet{Volpato23} and non-rotating \citet{Yoon12} stars and hence require magnifications $\approx 2.5$--4 times higher to be detected (with corresponding lensing probabilities a factor of $\approx 6$--16 lower). The effect of rotation on these results are small for the \citet{Murphy21a} stars at this mass. However, the \citet{Yoon12} 100 $M_\odot$ models with $v_K\geq 0.3$ behave entirely differently, as these have lifetimes extended by up to $\approx 1$ Myr and evolve towards higher $T_\mathrm{eff}$ when they age, eventually reaching $\approx 250 000$ K. This makes these model stars significantly fainter than the others in the NIRCam filters at $z=6$ after the first million years of evolution.

\subsection{Impact of the survey detection threshold}
The use of deeper JWST exposures than assumed in Fig.~\ref{PopIII_detection_rate} (28.5 AB mag) increases the probability for detecting lensed Pop III stars per cluster field, thereby allowing for deeper surveys covering a smaller number of clusters to also reach  substantial probabilities for detections of Pop III stars. For example, pushing the survey depth to 30.0 AB mag allows the two top-heaviest IMFs to reach $P_\mathrm{detect}\gtrsim 10\%$ per cluster field for $\gtrsim 300\ M_\odot$ Pop III stars. In this case, the calculation does however become significantly affected by the requirement that $\mu\geq 100$, since Pop III stars at $\gtrsim 150\ M_\odot$ may in fact reach the detection limit at even lower magnification. It should be noted, however, that lensed Pop III stars that appear at brightness levels as faint as $\approx 29$--30 AB mag would be extremely difficult to characterize in detail using the photometric SED, and perhaps impossible to study spectroscopically throughout the foreseeable future (Section~\ref{sec:PopIII features}).
For reference, the PEARLS program reached detection limits of $\approx 28.5$--29.0 AB mag across several cluster-lens fields using the 7 JWST/NIRCam filters considered here with a total exposure time of 3--4 hours \citep{Windhorst23}. A corresponding programme that reaches 30.0--30.5 AB mag would require $\approx 60$ hours.

While surveys covering several filters are required for a photometric characterization of the SED, some filters are more efficient in catching lensed, high-$z$ Pop III stars than others. Given stellar evolutionary tracks like those of \citet{Yoon12} and \citet{Volpato23}, which predict evolution to $T_\mathrm{eff}\lesssim 6000$ K, the $F444W$ filter provides the largest fraction of detections, since this is where these low-$T_\mathrm{eff}$ stars are the brightest in the $z=6$--10 range. However, the filter that provides the second largest fraction of detections gradually shifts across the NIRCam wavelength range from F356W at $z=6$ to F150W at $z=10$, as an increasing fraction of the lowest-$T_\mathrm{eff}$ stars shift out of NIRCam detection range at the higher redshifts, and higher-$T_\mathrm{eff}$ stars with bluer SEDs start to contribute more to the total detecetion rate, similar to what is seen in Figure~\ref{HR_bias_low_metallicity_z1to10}c and d for metal-enriched stars.

\section{The observational signatures of lensed Population III stars}
\label{sec:PopIII features}
\subsection{Photospheric signatures}
\label{subsec:photospheric signatures}
In the case where the light from a lensed, high-redshift star is dominated by its photosphere, its JWST broadband-filter SEDs is expected to contain relatively little information on the nature of the object. While the rest-frame shape of the Muspelheim SEDs do show some dependence on $T_\mathrm{eff}$, $\log(g)$ and chemical composition \footnote{Please note that in the present version of Muspelheim, the effects of rotation and mass-loss on the model spectra, beyond those imprinted in the $L_\mathrm{bol}$, $T_\mathrm{eff}$, $\log(g)$ evolution, are ignored.}, the effects of $T_\mathrm{eff}$ are the dominant ones. Once $T_\mathrm{eff}$ and redshift has been estimated from the observed shape of a lensed-star SED, the observed flux levels of the SED can be converted into a constraint on $\mu L_\mathrm{bol}$, which will allow the object to be placed into the HR diagram, although with substantial error bars coming from uncertainties on $\mu$, and potentially also from $\mathrm{Teff}$ (see below).  

\begin{figure*}
\includegraphics[width=\columnwidth]{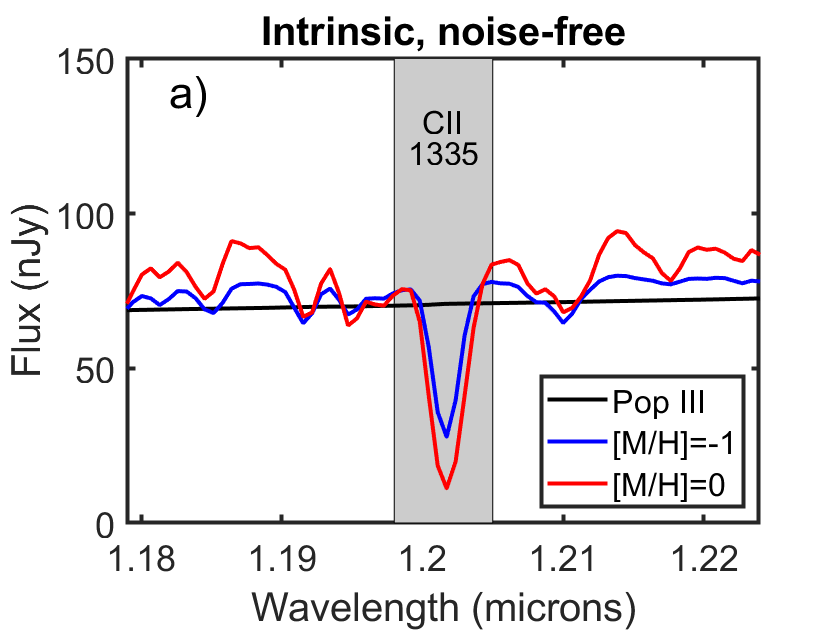}
\includegraphics[width=\columnwidth]{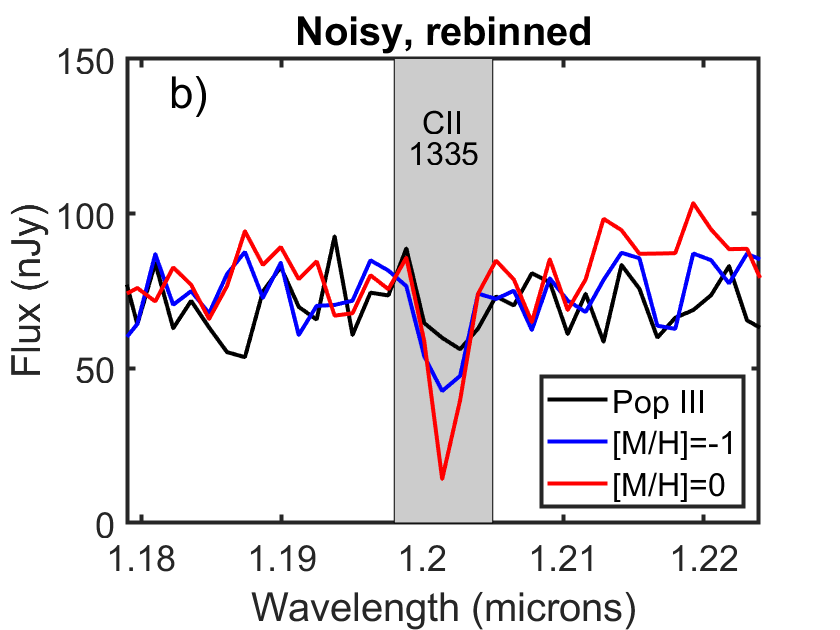}
\caption{Simulated JWST/NIRSpec spectrum around the CII 1335 feature of a $z=8$, $T_\mathrm{eff}=15000$ K, $\log(g)=1.75$ star gravitationally magnified to 26.5 AB mag in the JWST/NIRCam F150W filter. The differently lines correspond to stellar atmosphere spectra at different metallicities from our metal-free (Pop III) TLUSTY grid (black line)  and the TLUSTY B-star grid by \citet{Lanz07} at $Z=0.1\ Z_\odot$ (blue) and $Z=Z_\odot$ (red). {\bf a)} Noise-free, intrinsic spectrum at the resolution of the JWST/NIRSpec G140M disperser and the F100LP filter.
{\bf b)} Mock spectrum with noise level corresponding to 50 hours of single-slit JWST/NIRSpec G140M/F100LP observations based on v.2.2 of the JWST exposure time calculator, after rebinning over 2 adjacent wavelength bins to reduce the noise. Even these very deep spectroscopic observations of an exceptionally bright $z=8$ lensed star only allow for a low-confidence detection of the CII 1335 line (one of the strongest UV lines at 15000 K; \citealt{Leitherer10}) and weak metallicity constraints (at best $Z\lesssim 0.1 Z_\odot$ in the case of a non-detection of the CII 1335 feature).}
\label{fig_spectroscopic_metallicity}
\end{figure*}

The level at which $T_\mathrm{eff}$ can be determined from the SED of a lensed star at high redshifts depends on its intrinsic position in the HR diagram. In Figure~\ref{fig_Teff_col_signatures}, we demonstrate how the JWST/NIRCam colours depend on $T_\mathrm{eff}$ in the case of lensed stars at $z=6$ and $z=10$. In general, the prospects of setting meaningful constraints are optimized when $T_\mathrm{eff}$ is in the range $\approx 8000$--20000 K (F-type to B-type), since this is the range where the SED develops a prominent Balmer/4000\AA{} break -- a feature that cannot easily be mimicked by dust attenuation along the line of sight. At higher temperatures, the coarse spectrum traced by the JWST broadband fluxes appears smooth and blackbody-like. The overall slope of this spectrum becomes increasingly steep in the blueward direction as $T_\mathrm{eff}$ is increased, but this slope evolution saturates at $T_\mathrm{eff}\gtrsim 50000$ K which makes it very challenging to determine higher temperature accurately. Any dust attenuation that remains uncorrected for will also act to make the SED seem to have a lower $T_\mathrm{eff}$. In the absence of other dust constraints, the $T_\mathrm{eff}$ derived from a featureless blue slope of a lensed star therefore merely serves as a lower limit on the actual $T_\mathrm{eff}$ of the object. At $T_\mathrm{eff} < 8000$ K, the SED once again turns almost featureless, with a pronounced red peak towards the longest NIRCam wavelengths. In this case, it becomes unrealistic to determine the redshift from the Ly$\alpha$ limit of the star itself, since the flux directly longward of this edge will be much too low to be detected. Provided that the redshift can be obtained from the spectrum or photometric SED of a gravitational arc to which the star belongs, or from the location of the critical curve in the cluster, the temperature may still be fairly accurately constrained at $\approx 6$. At $z\approx 10$, this would require a situation where the star is magnified to a brightness several magnitudes above the detection level in the F444W filter, since this will otherwise be the only NIRCam filter (among those considered here) in which a $T_\mathrm{eff} < 8000$ K star can be detected. 

The analysis of a $T_\mathrm{eff}<8000$ K star is less sensitive to interstellar dust attenuation and reddening along the line of sight than for a high-$T_\mathrm{eff}$ star, since the peak brightness falls in the rest-frame optical and not the rest-frame ultraviolet. However, circumstellar dust produced by the stars themselves may still complicate the analysis.

While metallicity does have a slight impact on the shape of photometric SEDs of lensed, high-redshift stars in the rest-frame UV where metal absorption bands can cause a slight reduction of flux longward of the Ly$\alpha$, these changes are typically small and difficult to distinguish from effects of dust attenuation (further discussed in Section~\ref{sec:discussion}). The alternative is to attempt to constrain the metallicity of lensed stars through spectroscopy. 

Stars with $T_\mathrm{eff}$ around 15000 K are particularly interesting in this context, both because candidate Pop III stars may potentially be detected at these temperatures (Figure~\ref{HR_bias_PopIII_IMFs}; Section~\ref{sec:PopIII_detection_probability}), because such star remain relatively bright in the part of the rest-frame UV shortward of the Balmer break and because metals imprint a number of absorption features at rest-frame wavelengths $\approx 1216$--2000 \AA{} that can be probed both by JWST/NIRSpec (in principle up to $z\approx 25$) and by future near-IR spectrographs on ground-based ELTs (up to $z\approx 11$, but potentially with higher sensitivity than JWST).

However, absorption line studies of lensed, high-redshift stars are very challenging in the JWST era, due to relatively low continuum signal-to-noise ratios that one can achieve even for exceptionally bright lensed stars.

Some of the most interesting UV absorption features, in the sense of being relatively strong and hence potentially detectable even at lower continuum signal-to-noise ratios ($S/N$) than usually required for absorption-line studies ($S/N\gtrsim 30$), include the SiII 1260 \AA{} line, the CII 1335 \AA{} line, the SiIV 1393, 1403 \AA{} lines and the FeIII 1893 \AA{} for a supergiant at  $T_\mathrm{eff}\approx 15000$ K \citep{Leitherer10}. In Figure~\ref{fig_spectroscopic_metallicity}, we show mock JWST/NIRSpec, $R\sim 1000$ spectra corresponding to 50 h observations centered on the CII 1335 \AA{} feature in the case of a $z=8$, $T_\mathrm{eff}=15000$ K star lensed to 26.5 mag in the JWST/NIRCam F150W filter (this gives a peak brightness in the NIRCam bands of $\approx 26$ AB mag, which is exceptionally bright; $\approx 1$ mag brighter than the $z\approx 6$ lensed star Earendel, \citealt{Welch22b}). Even for these optimistic conditions (super-deep observations of an unusually bright $z=8$ star), metal lines can only be measured at low confidence. Hence, demonstrating the absence of such lines, which would indicate a primordial chemical composition, will be very difficult. Based on the mock observations presented in Figure~\ref{fig_spectroscopic_metallicity}, it would seem possible to constrain the metallicity to $Z<0.1 Z_\odot$ based on the non-detection of the CII 1335 feature feature, but the distinction of a Pop III star from a metal-poor star (e.g. $Z\sim 0.01\ Z_\odot$) would remain difficult in the JWST era. These mock observations are based on TLUSTY spectra which neglect the effect of winds on the UV lines. However, a similar calculation for the CII 1335 features using the PoWR models with various levels of mass-loss \citep{Hainich19} shows that while winds may alter the profile of the line and present a more complicated metallicity dependence with line strength, this does not 
make the line more easily detectable in any radical way. To improve the constraints, data on several metal absorption features would need to be combined. Alternatively, more sensitive spectroscopic data obtained with a future ground-based ELT in the near-IR may also improve the situation.

\subsection{Nebular signatures}
\label{subsec:nebular_signatures}
Pop III stars with $T_\mathrm{eff}\gtrsim 30000$ K produce enough ionizing photons to produce an emission nebula in the surrounding gas, which in principle may be intrinsically brighter than the star itself at the rest wavelengths probed by JWST at high redshifts \citep[e.g.][]{Rydberg13}. This nebula is formed as the surrounding gas absorbs the Lyman continuum radiation at rest wavelengths $<912$ \AA{} and transforms it into nebular continuum and emission lines at $>912$ \AA{}, with a brightness that strongly depends on $T_\mathrm{eff}$ -- for a $T_\mathrm{eff} = 30000$ K Pop III star, the Lyman continuum makes up $\lesssim 10\%$ of the bolometric luminosity of the star, but this fraction rises to $>90\%$ at $T_\mathrm{eff} = 100 000$ K.    

Under the assumption that an ionization-bounded nebula is formed close to the star, we in Fig.~\ref{brightest_ABmag_star_neb_fig} plot the brightest intrinsic AB magnitude attained in the JWST/NIRCam wide-band filters for 60--1000 $M_\odot$, non-rotating Pop III stars from the \citet{Yoon12} set at an age of $\sim 10^5$ yr, along with the brightest JWST/NIRCam AB magnitudes of their associated nebulae, at $z=6$, 10 and 15. Here, we have adopted the Pop III nebular spectrum of a 120 $M_\odot$ Pop III ZAMS star derived by \citet{Rydberg13} for the ZAMS parameters from \citet{Schaerer02}, and scaled it to the LyC luminosities of Pop III stars from our sets. As seen in Fig.~\ref{brightest_ABmag_star_neb_fig}, Pop III nebulae may intrinsically be up to $\approx 3$ magnitudes brighter than the stars that power them in the NIRCam broadband filters considered here. Even so, any potential impact of this nebular emission on observations of lensed high-$z$ stars critically depends on the differential magnification across the star and its nebula, as well as the time scale over which the nebula can be expected to remain compact -- issues that \citet{Rydberg13} did not address in detail.

\subsubsection{Differential magnification of Pop III stars and their HII-regions}
\label{sec:differential magnification}
Because HII-regions are much larger (pc-sized) than stars ($\lesssim 100\ R_\odot$, i.e., $\lesssim 5\times 10^{-4}$ pc, for $\lesssim 1000\ M_\odot$ Pop III stars at $T_\mathrm{eff}\gtrsim 30000$ K; \citealt{Windhorst18}), they cannot attain as large macrolensing magnifications and can furthermore not be significantly magnified by stellar microlensing. Under the simplifying assumption of a constant-density HII-region in which ionization is balanced by recombination, the size of the nebula may be estimated from the Str\"{o}mgren radius ($R_\mathrm{str}$):
\begin{equation}
R_\mathrm{str} = \left( \frac{3Q_\mathrm{tot}}{4\pi \alpha_\mathrm{B}\ n_\mathrm{H}^2}\right) ^{1/3},
\label{eq:Rstr}
\end{equation}
where $Q_\mathrm{tot}$ is the production rate of LyC photons per second, $\alpha_\mathrm{B}$ is the case B recombination rate and $n_\mathrm{H}$ is the hydrogen number density.

\begin{figure}
\includegraphics[width=\columnwidth]{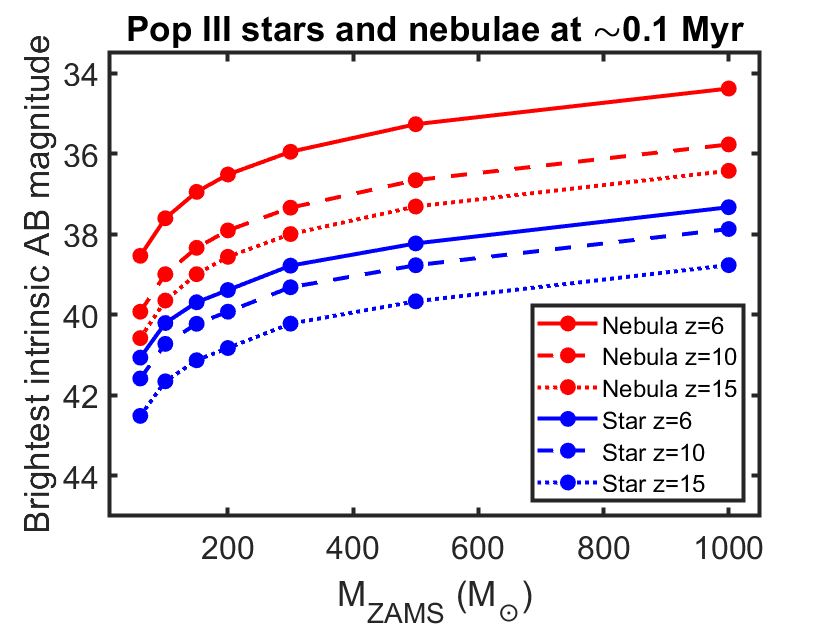}
\caption{The brightest intrinsic (unlensed) JWST/NIRCam wide-band filter fluxes attained by the \citet{Yoon12} non-rotating Pop III stars (blue lines) and their surrounding HII-regions (red lines) as a function of ZAMS mass, at an age of $10^5$ yr. Solid, dashed and dotted lines represent redshifts $z=6$, 10 and 15, respectively. As seen, unlensed Pop III nebulae intrinsically reach peak JWST/NIRCam fluxes up to $\approx 2$--3 magnitudes brighter than those of their Pop III stars at these redshifts. Please note that the stellar and nebular SEDs typically do not attain their brightest magnitudes in the same NIRCam filters, since the shapes of these two SEDs differ substantially.}
\label{brightest_ABmag_star_neb_fig}
\end{figure}

If one adopts an ISM density close to the Pop III star of $n_{\mathrm(H)}=1000$ cm$^{-3}$ (often considered a standard value for O star nebulae), and the initial $Q_\mathrm{tot}$ values for non-rotating 60--1000 $M_\odot$ from \citet{Yoon12}, the radii of the Str\"{o}mgren sphere around these stars become $\approx 1$--4 pc. 

Here, we approximate the maximum macrolensing magnification as function of source size as:
\begin{equation}
\max(\mu_\mathrm{macro})\approx 90 B_0 \ R_\mathrm{S,pc}^{-1/2} \ (D_S/1 \mathrm{Gpc})^{1/2},
\label{mumax_macro}
\end{equation}
derived from eq.~\ref{eq:mu_B0} at the distance from the caustic at which a circular source attains its maximum magnification \citep[$0.65\ R_\mathrm{S}$;][]{Miralda-Escude91}. 

This limits the maximum magnification for nebulae with radii $\approx 1$--4 pc at $z=6$--20 to $\max(\mu_\mathrm{macro})\approx$ 500--1500 in the case of $B_0=15$, i.e., more than an order of magnitude less than what lensed stars may attain. For a JWST/NIRCam photometric detection limit of $\approx 29$ AB mag, the faintest Pop III nebula detectable on its own would (without the additional boost of the star) therefore need to reach an intrinsic brightness $\lesssim 35.7$--36.9 AB mag.

From Fig.~\ref{brightest_ABmag_star_neb_fig}, we see that this limit is, at $z\approx 6$, only met by the nebula of the most massive Pop III stars ($\gtrsim 150 \ M_\odot$). At $z>6$, the Pop III nebulae quickly become fainter because the brightest spectral feature, the H$\alpha$ emission line, redshifts out of JWST/NIRCam range, and at $z=10$, no lensed nebulae of $M \lesssim 500 \ M_\odot$ Pop III stars would be detectable on their own in a survey reaching 29 AB mag.

\begin{figure*}
\includegraphics[width=\columnwidth]{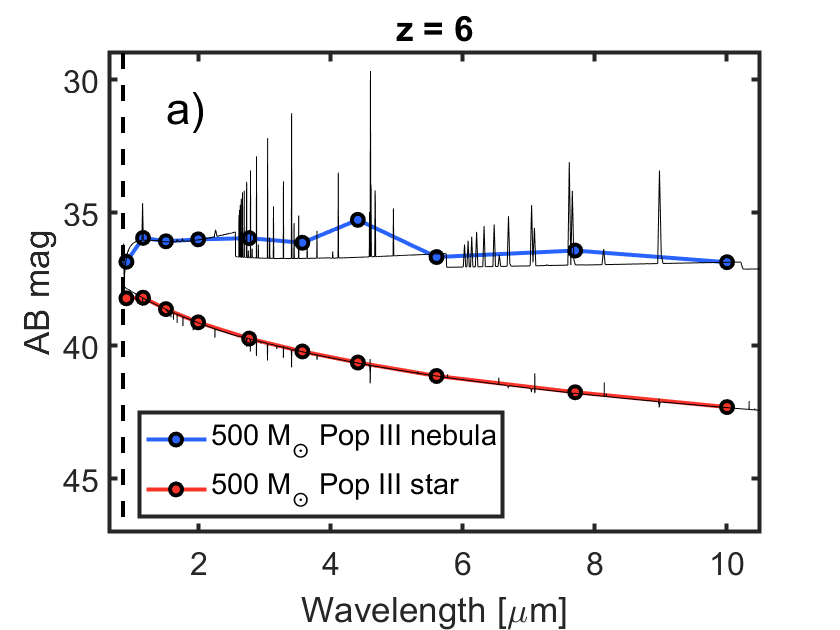}
\includegraphics[width=\columnwidth]{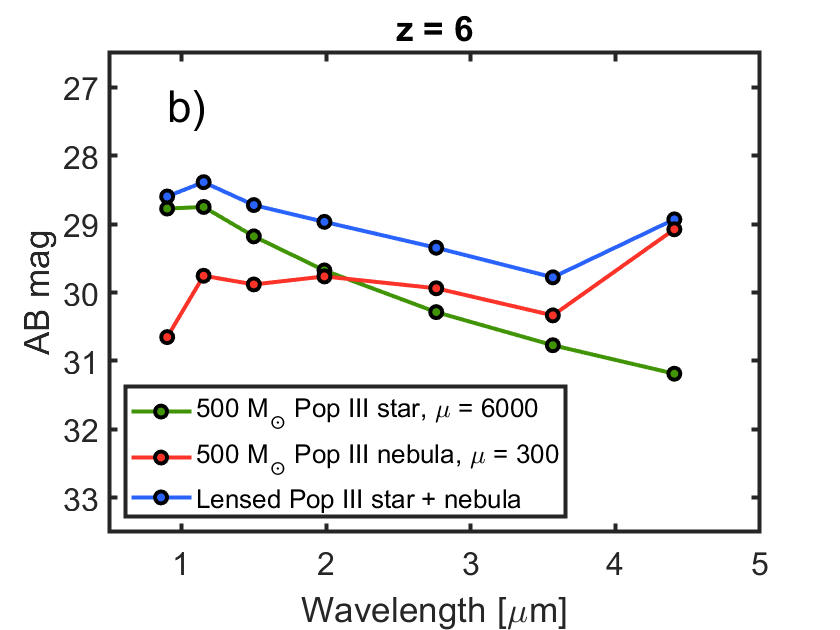}
\caption{Spectral energy distributions of Pop III nebulae. 
{\bf a)} Unlensed spectral energy distributions of a 0.1 Myr old, 500 $M_\odot$ non-rotating \citet{Yoon12} Pop III star (red line) and its surrounding HII-region (blue line) in JWST/NIRCam and MIRI wide filters out to 10 $\mu$m at $z=6$, based on the \citet{Rydberg13} models for the nebulae of Pop III stars. The black solid line indicates the underlying spectra on which the photometric SEDs of the star and its HII-region are based. The dashed vertical line marks the position of the Ly$\alpha$ limit. {\bf b)} The gravitationally lensed versions of the SEDs shown in a), assuming 
$\mu=6000$ for the star (blue line) and $\mu=300$ for its HII-region (red line). The green line represents the SED of the compound object (differentially magnified star + nebula). At these magnifications, either the lensed star, the lensed nebula, or the differentially magnified compound object could potentially be detected by JWST in deep exposures.}
\label{nebular_SED_fig}
\end{figure*}

In Figure~\ref{nebular_SED_fig}a, we show the intrinsic SEDs of a 0.1 Myr old, 500 $M_\odot$ Pop III star from \citet{Yoon12}, and its associated nebula, at $z=6$. In this case, the nebula is intrinsically already $\approx 2$ mag brighter than the star at $\approx 1\ \mu$m, and progressively becomes more dominant at longer wavelengths due to the different slopes of the SEDs.

In Figure~\ref{nebular_SED_fig}b, we show the differentially magnified SED of a $z=6$, 0.1 Myr old, 500 $M_\odot$ star at $\mu=6000$ and its nebula at $\mu=300$. In this case, parts of the JWST/NIRCam SED of the compound object can be detected above a $\approx 29$ AB mag detection threshold at both $\approx 1-2\ \mu$m and at $\approx 4.4\ \mu$m, with the overall shape of the SED having significant contributions from both the star and its nebula (including a notable upturn in F444W due to the nebular H$\alpha$ emission line). If such situations were common, this would have interesting consequences for the prospects of constraining the metallicity of a lensed star, since strong emission lines from hydrogen and helium could be imprinted on the spectrum, whereas metal lines may be absent.

To illustrate this, we in Figure~\ref{etc_fig_neb}, show the simulated JWST/NIRSpec R=1000 spectrum that corresponds to the differentially magnified Pop III star + nebula featured in Fig.\ref{nebular_SED_fig}b. Both the H$\alpha$ and H$\beta$ emission lines are clearly visible, whereas the typically prominent [OIII]4959,5007 lines are absent. Upper limits on these oxygen lines should make it possible to set considerably stronger upper limits on the metallicity compared to the case for absorption line spectroscopy of the Pop III photosphere. We note that it would be considerably more challenging to detect the HeII 1640 emission line, which is often highlighted as a strong Pop III indicator, due to its relatively low intrinsic flux. 

\subsubsection{Temporal evolution of the HII-regions around Pop III stars}
While the analysis presented in Section~\ref{sec:differential magnification} suggests that nebular features in principle could be imprinted in the spectra of lensed $M\gtrsim 150\ M_\odot$ Pop III stars, the {\it probability} of detecting such features critically depends on the timescales over which the HII-region around Pop III stars can remain sufficiently compact (pc-scale) to allow magnification factors of several hundreds or more. 

Even though the gas densities required to keep the Str\"{o}mgren sphere contained at parsec scales ($n_\mathrm{H}\gtrsim 1000$ cm$^{-3}$) are likely to be retained in the vicinity of the star shortly after formation, the hot photoionized HII region that forms around the star will be overpressured compared to the neutral, cooler surroundings and expand. Models for the density evolution of the ionized gas in the HII region around both Pop III stars in $\sim 10^6\ M_\odot$ minihalos \citep[e.g.][]{Whalen04,Alvarez06,Sibony22} and high-mass, metal-enriched stars in more normal HII regions \citep[e.g.][]{Mellema06,Geen21} all suggest that the HII region may expand beyond $>1$ pc already at $\lesssim 10^5$ yr, which makes the timescale for detection perilously short in the lensing scenario considered here.

In Appendix~\ref{sec:Probability Pop III nebula}, we present a simple toy calculation that suggests that if the HII-region around Pop III stars indeed only remains compact (pc-sized) for $\approx 10^5$ yr, then the probability for such HII-regions to be magnified to $\approx 29$ AB mag (as in Fig.~\ref{nebular_SED_fig}b and \ref{etc_fig_neb}) is $N_\mathrm{detect}< 10^{-3}$ per cluster-lens field, which is too low for such effects to realistically be detected with JWST. However, the situation could potentially improve if the expansion of the HII-region is hampered by the gravitational field of a more host massive halo \citep{Kitayama04}, as could be the case if Pop III star formation largely shifts to $\gtrsim 10^7$--$10^8\ M_\odot$ HI-cooling halos at $z\lesssim 15$, or by the gravitational field of the star itself \citep{Jaura22}. Moreover, in the case of significant mass loss from a Pop III star (e.g. due to rotation), the wind/outflow may also imprint emission lines on the spectrum of a lensed star.

\section{Discussion}
\label{sec:discussion}
\subsection{Comparison with other studies}
\citet{Rydberg13} explored the prospects of detecting highly magnified high-redshift Pop III stars with very deep JWST imaging of a single cluster-lens field and came to the conclusion that detections would be highly unlikely, unless essentially {\it all} Pop III stars formed at very high masses $M\gtrsim 300\ M_\odot$ (and only when considering a maximal flux boost due to nebular emission). However, \citet{Rydberg13} did not consider Pop III stars beyond the main sequence and therefore did not include the substantial boost in detection rates provided by potential low-$T_\mathrm{eff}$ states of evolution. Also, the range of Pop III SFRDs explored at $z=6$--10 was also significantly lower than that predicted by some recent simulations and models e.g., \citet{Jaacks19,Liu20,Venditti23,Chantavat23}.

The prospects of detecting lensed Pop III stars was revisited by \citet{Windhorst18}, who did consider post-main sequence of Pop III stars, although not evolution to $T_\mathrm{eff}$ states as low as those explored in the current paper. Based on the assumption that the cosmic infrared background receives a non-negligible contribution from high-redshift Pop III stars, \citet{Windhorst18} argue that there is significant chance of detecting highly magnified Pop III stars or, alternatively, the accretion disks around the black holes they leave behind, through caustic-crossing events with JWST if a handful of strong-lensing galaxy clusters are monitored over a few years. However, in scenarios where the contribution from these objects lie far below the current upper limits on the diffuse infrared background, the JWST detection prospects for lensed Pop III stars or their remnant black holes become progressively worse.

Recently, \citet{Larkin23} presented an extensive study of the magnfications required to detect 1--1000 $M_\odot$ Pop III stars at $z=3$--17 with JWST, albeit only for the ZAMS and without assessing the probability for such detections in realistic surveys of lensed fields.

\subsection{The effect of dust attenuation}
The results presented in this paper assume that lensed stars are not significantly affected by dust attenuation. However, since the stars detected through gravitational lensing are all likely to be have high-masses and short lifetimes ($\approx 2-30$ Myr) they are likely to be sitting in young, star-forming regions, where the interstellar dust attenuation may be substantial, especially in the case of high-metallicity environments. Moreover, red supergiants may be surrounded by circumstellar dust, as seen in local samples \citep{Massey05,Drout12}. This attenuation of the intrinsic flux of lensed stars has the effect of producing an additional, and likely metallicity-dependent bias affecting where in the HR diagram such stars are likely to be found. To first order, the dust attenuation can potentially be assessed from photometric and/or spectroscopic studies of gravitational arcs in which lensed stars have so far been identified, but it must be kept in mind that the specific star-forming region in which a lensed star is sitting may not be the same as that derived for its host galaxy. In general, the effects of dust attenuation on the observed SED of a lensed star candidate is to render it fainter and redder, thereby -- at fixed estimate of the magnification -- shifting it to what (in the absence of corrections for dust) would appear to be a lower $T_\mathrm{eff}$ and lower $L_\mathrm{bol}$ compared to its intrinsic position.

Since Pop III stars are unlikely to form in dusty surroundings, dust reddening and attenuation is generally expected to be a more severe problem for metal-enriched lensed stars. Even so, Pop III supernovae are expected to produce dust \citep[e.g.][]{Marassi15}, and dust generated by the most massive and short-lived members of a small cluster of Pop III stars could potentially affect the line of sight to more long-lived Pop III members in the same region.  

\begin{figure}
\includegraphics[width=\columnwidth]{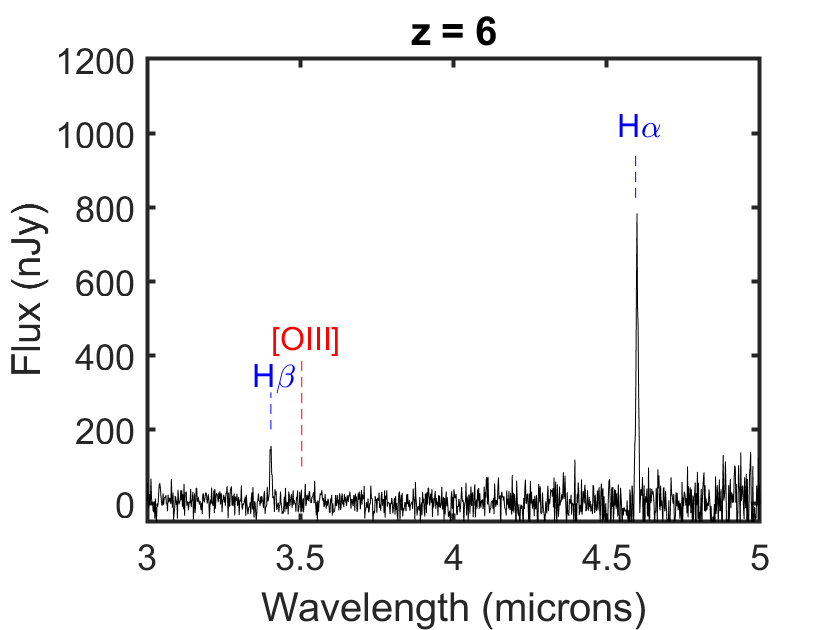}
\caption{Simulated JWST/NIRSpec G395M/F290L fixed-slit spectrum at resolution $R\approx 1000$ after 20 h of observations of the $z=6$, 500 $M_\odot$, 0.1 Myr old differentially magnified star ($\mu=6000$) plus nebula ($\mu=300$) from Figure~\ref{nebular_SED_fig}b. The  H$\alpha$ and H$\beta$ lines from the nebula are clearly detectable (at peak $\mathrm{S/N}\approx 18$ and 7, respectively), whereas the absence of the emission lines [OIII]4959,5007, indicates a very low metallicity.}
\label{etc_fig_neb}
\end{figure}

\subsection{Rotation of Pop III stars}
A significant fraction of Pop III stars may experience rapid rotation \citep{Hirano18}, which is expected to give rise to a number of complicated phenomena such as non-spherical stars, mechanical mass loss and possibly the formation of decretion disks \citep[e.g.][]{Marigo03,Ekstrom08,Yoon12,Murphy21a}. Plane-parallel stellar atmospheres in hydrostatic and radiative equilibrium of the type used here are ill-equipped to model such phenomena, and while such stellar atmosphere models are often used to model the non-ionizing continuum of rotating high-mass stars, the problem becomes more complex when modelling ionizing fluxes, since the outflow of gas from the surface could potentially have a significant impact on the emergent ionizing spectrum \cite[e.g.][]{Schaerer02}.  

\subsection{Binary Pop III stars}
Almost all massive stars are in binary systems \citep[see][for a review]{DeMarco17}, and simulations suggest that Pop III stars may be as well \citep[e.g.][]{Stacy13,Sugamura20}. Aside from having a significant impact on the rotation and evolution of Pop III stars, and hence the overall probability of detecting highly magnified Pop III stars, this also opens up the possibility that what is observed in the case of a Pop III binary located close to a caustic, is the blended contributions from two separate Pop III stars with different properties and potentially slightly different magnifications. This becomes another complication in the analysis of the SEDs of lensed-star candidates, although the two components could potentially be disentangled in the case where $T_\mathrm{eff}$ of the two components are substantially different \citep[as has been suggested to be the case for the $z\approx 6$ object Earendel;][]{Welch22b} or where the radial velocity between the two stars allows for the detection of two sets of absorption lines.  

\subsection{Surface enrichment of Pop III stars}
\label{subsec:surface_enrichment}
Even though Pop III stars form at primordial chemical composition, heavy elements are produced from fusion within these stars. Dredge-ups during late stages of Pop III stellar evolution \citep[e.g.][]{Ekstrom08, Volpato23, Volpato24}, or mixing with fast-rotating Pop III stars \citep{Yoon12,Liu21} would allow nucleosynthetic products to reach the surface. Since late, low-$T_\mathrm{eff}$ stages of evolution are favourable for detection, this leads to an addtional problem when attempting to verify the Pop III nature stars based on spectroscopy. Absorption lines from the stellar surface, or a combination of absorption and emission lines from a wind/outflow may in such cases reveal non-zero CNO levels, which would complicate the identification of the lensed star as Pop III object. 

\subsection{Unsolved problems in the evolution of high-mass stars}
We have throughout this paper stressed that existing evolutionary models differ greatly in their predictions for how Pop III stars move across the HR diagram, and that this has pronounced effects on the detectability of such stars. It must be pointed out, however, that current models for metal-enriched, high-mass stars also suffer from a number of problems, including the lack of observed O-stars close to the predicted ZAMS \citep[e.g.][]{Holgado20} and the observed excess of blue supergiants compared to models \citep[e.g.][]{Bellinger23}. Whatever the resolution to these issues, this is likely to affect the time high-mass stars spend at different $T_\mathrm{eff}$ and hence forecasts of where in the HR diagram lensed stars are the most likely to be found (Figures~\ref{HR_bias_low_metallicity_z2} and ~\ref{HR_bias_low_metallicity_z1to10}). Until such issues have been resolved, caution must be exercised when attempting to interpret  observed distributions of lensed stars across the HR diagram in terms of quantities like the star formation history, metallicity or the IMF of the lensed stellar population.

\section{Conclusions}
\label{sec:summary}
Our results can be summarized as follows:

\begin{itemize}
\item Surveys for highly magnified, high-redshift stars with JWST are subject to strong selection biases related to the JWST wavelength range and detection limits, the lensing probability distribution and size-magnification limits (Section~\ref{sec:HR_bias}). For stars that require the highest magnifications to be rendered detectable, these biases tend to favour the detection of objects in late, lower-$T_\mathrm{eff}$ stages of evolution even though stars only spend a small fraction of their lifetime in such states. For example, based on the $Z\sim 0.1 Z_\odot$ single-star stellar evolutionary tracks used in this paper, $z\approx 3$ stars with initial masses $\lesssim 55\ M_\odot$ found by JWST at $\lesssim 28$ AB magnitudes are only likely to be detected as red supergiants (Figure~ \ref{HR_bias_low_metallicity_z1to10}b).

\item We argue that highly magnified Pop III stars at $z\gtrsim 6$ may be detectable in small numbers with JWST surveys covering a large number ($\approx 10$--50) of cluster-lens fields. However, to make such detections likely, Pop III stars need to obey a number of conditions: The Pop III SFRD at $z\approx 6$--10 needs to be sufficiently high ($\gtrsim 10^{-4}$ M$_\odot$ cMpc$^{-3}$ yr$^{-1}$), the Pop III IMF needs to be very top-heavy and allow a significant mass fraction of the gas converted into Pop III stars to be in $\gtrsim 100\ M_\odot$ objects (mass fraction $\gtrsim 10\%$ for SFRD $\sim 10^{-4}$ M$_\odot$ cMpc$^{-3}$ yr$^{-1}$), and the evolution of Pop III stars must be such that the $M\gtrsim 100\ M_\odot$ stars evolve to temperatures $<15000$ K during their lifetimes (Section~\ref{sec:PopIII_detection_probability}; Figure~\ref{PopIII_detection_rate}).  
Several simulations predict Pop III SFRDs up to an order of magnitude higher than our limit \citep[see e.g.][]{Venditti23}, and the recent stellar evolutionary models by \citet{Volpato23} do predict evolution into low-$T_\mathrm{eff}$ states for  
$\gtrsim 100\ M_\odot$ Pop III stars. Whether the required Pop III IMF is likely to be sufficiently top-heavy to meet our constraints is potentially more controversial, since studies of the abundance patterns of very metal-poor stars in the local Universe do not seem to favour the enrichment signatures that would result from the most top-heavy Pop III IMFs \citep[e.g.][]{Tumlinson06,Koutsouridou23}.

\item Even if the most massive, lensed Pop III stars at $z\gtrsim 6$ were to be lifted above the detection threshold of JWST while close to their ZAMS states at $T_\mathrm{eff}\approx 10^5$ K (Figure~\ref{HR_bias_PopIII}a) -- pinning down $T_\mathrm{eff}$ from JWST photometry will be difficult at $T_\mathrm{eff}\gtrsim 50000$ K due to modest change in spectral slope in this $T_\mathrm{eff}$ regime (Figure~\ref{fig_Teff_col_signatures}). 
 
\item Verifying the Pop III nature of highly-magnified stars at high redshifts using metallicity constraints will be very challenging. In situations where the lensed image is dominated by the stellar photosphere, very deep JWST and/or ELT spectroscopy can be used to target strong metal absorption lines to set an upper limit on the metallicity of the candidate. However, the constraint is unlikely to be better than $Z/Z_\odot<0.1$ given the capabilities of JWST, which makes it difficult to separate a very low-metallicity star from a genuine Pop III star (Section~\ref{subsec:photospheric signatures}; Figure~\ref{fig_spectroscopic_metallicity}). Moreover, Pop III stars may also be self-polluted by metals in the late evolutionary states where many of these stars are likely to be found, thereby complicating the analysis.

\item During early stages of evolution, where a Pop III star may be surrounded by a compact HII region, the differentially magnified nebula could make the object brighter and imprint strong hydrogen  emission lines in the spectrum, without any associated metal emission lines (Figure~\ref{etc_fig_neb}). While this opens up the possibility to detect Pop III stars in their earliest states and probe the metallicity more easily than what is possible using absorption line spectroscopy, the probability of such detections is deemed very low, unless the expanding HII region around the star is confined to $\sim$pc scales for longer than $\sim 10^5$ yr (Section~\ref{subsec:nebular_signatures}).

\item The fact that individual Pop III stars may be rendered detectable by JWST through gravitational lensing in some models has important consequences even in the absence of metallicity constraints on any lensed stars that are found at $z\gtrsim 6$. As JWST continues to observe an increasing number of cluster-lens fields, it should be possible to use the accumulated photometric detections (or even non-detections) of lensed stars at these redshifts to set combined upper limits on parameters that describe the  
SFRD, IMF and evolution of Pop III stars, thereby ruling out the most extreme scenarios. 
\end{itemize}

\section*{Acknowledgements}
EZ acknowledges project grant 2022-03804 from the Swedish Research Council (Vetenskapsr\aa{}det) and has also benefited from a sabbatical at the Swedish Collegium for Advanced Study. EZ and AV acknowledge funding from the Swedish National Space Agency. AN acknowledge funding from the Olle Engkvist Foundation. GV acknowledges support from Padova University, through the research project PRD 2021. GC acknowledges support from the Agence Nationale de la Recherche grant POPSYCLE number ANR-19-CE31-0022.
J.M.D. acknowledges the support of project PID2022-138896NB-C51 (MCIU/AEI/MINECO/FEDER, UE) Ministerio de Ciencia, Investigaci\'on y Universidades. 

\section*{Data Availability}
\label{model_access}
 
Publicly available grids of Muspelheim models are available from: \url{https://www.astro.uu.se/~ez/muspelheim/muspelheim.html}



\bibliographystyle{mnras}
\bibliography{references} 

\begin{thebibliography}{}
\makeatletter
\relax
\def\mn@urlcharsother{\let\do\@makeother \do\$\do\&\do\#\do\^\do\_\do\%\do\~}
\def\mn@doi{\begingroup\mn@urlcharsother \@ifnextchar [ {\mn@doi@}
  {\mn@doi@[]}}
\def\mn@doi@[#1]#2{\def\@tempa{#1}\ifx\@tempa\@empty \href
  {http://dx.doi.org/#2} {doi:#2}\else \href {http://dx.doi.org/#2} {#1}\fi
  \endgroup}
\def\mn@eprint#1#2{\mn@eprint@#1:#2::\@nil}
\def\mn@eprint@arXiv#1{\href {http://arxiv.org/abs/#1} {{\tt arXiv:#1}}}
\def\mn@eprint@dblp#1{\href {http://dblp.uni-trier.de/rec/bibtex/#1.xml}
  {dblp:#1}}
\def\mn@eprint@#1:#2:#3:#4\@nil{\def\@tempa {#1}\def\@tempb {#2}\def\@tempc
  {#3}\ifx \@tempc \@empty \let \@tempc \@tempb \let \@tempb \@tempa \fi \ifx
  \@tempb \@empty \def\@tempb {arXiv}\fi \@ifundefined
  {mn@eprint@\@tempb}{\@tempb:\@tempc}{\expandafter \expandafter \csname
  mn@eprint@\@tempb\endcsname \expandafter{\@tempc}}}

\bibitem[\protect\citeauthoryear{{Alvarez}, {Bromm}  \& {Shapiro}}{{Alvarez}
  et~al.}{2006}]{Alvarez06}
{Alvarez} M.~A.,  {Bromm} V.,   {Shapiro} P.~R.,  2006, \mn@doi [\apj]
  {10.1086/499578}, \href
  {https://ui.adsabs.harvard.edu/abs/2006ApJ...639..621A} {639, 621}

\bibitem[\protect\citeauthoryear{{Bellinger}, {de Mink}, {van Rossem}  \&
  {Justham}}{{Bellinger} et~al.}{2023}]{Bellinger23}
{Bellinger} E.~P.,  {de Mink} S.~E.,  {van Rossem} W.~E.,   {Justham} S.,
  2023, \mn@doi [arXiv e-prints] {10.48550/arXiv.2311.00038}, \href
  {https://ui.adsabs.harvard.edu/abs/2023arXiv231100038B} {p. arXiv:2311.00038}

\bibitem[\protect\citeauthoryear{{Bergemann}}{{Bergemann}}{2014}]{Bergemann14}
{Bergemann} M.,  2014, in , Determination of Atmospheric Parameters of B.
pp 187--205, \mn@doi{10.1007/978-3-319-06956-2_17}

\bibitem[\protect\citeauthoryear{{Broadhurst} et~al.,}{{Broadhurst}
  et~al.}{2024}]{Broadhurst24}
{Broadhurst} T.,  et~al., 2024, \mn@doi [arXiv e-prints]
  {10.48550/arXiv.2405.19422}, \href
  {https://ui.adsabs.harvard.edu/abs/2024arXiv240519422B} {p. arXiv:2405.19422}

\bibitem[\protect\citeauthoryear{{Chantavat}, {Chongchitnan}  \&
  {Silk}}{{Chantavat} et~al.}{2023}]{Chantavat23}
{Chantavat} T.,  {Chongchitnan} S.,   {Silk} J.,  2023, \mn@doi [\mnras]
  {10.1093/mnras/stad1196}, \href
  {https://ui.adsabs.harvard.edu/abs/2023MNRAS.522.3256C} {522, 3256}

\bibitem[\protect\citeauthoryear{{Chen} et~al.,}{{Chen} et~al.}{2022}]{Chen22}
{Chen} W.,  et~al., 2022, \mn@doi [\apjl] {10.3847/2041-8213/ac9585}, \href
  {https://ui.adsabs.harvard.edu/abs/2022ApJ...940L..54C} {940, L54}

\bibitem[\protect\citeauthoryear{{De Marco} \& {Izzard}}{{De Marco} \&
  {Izzard}}{2017}]{DeMarco17}
{De Marco} O.,  {Izzard} R.~G.,  2017, \mn@doi [\pasa] {10.1017/pasa.2016.52},
  \href {https://ui.adsabs.harvard.edu/abs/2017PASA...34....1D} {34, e001}

\bibitem[\protect\citeauthoryear{{Diego}}{{Diego}}{2019}]{Diego19}
{Diego} J.~M.,  2019, \mn@doi [\aap] {10.1051/0004-6361/201833670}, \href
  {https://ui.adsabs.harvard.edu/abs/2019A&A...625A..84D} {625, A84}

\bibitem[\protect\citeauthoryear{{Diego} et~al.,}{{Diego}
  et~al.}{2018}]{Diego18}
{Diego} J.~M.,  et~al., 2018, \mn@doi [\apj] {10.3847/1538-4357/aab617}, \href
  {https://ui.adsabs.harvard.edu/abs/2018ApJ...857...25D} {857, 25}

\bibitem[\protect\citeauthoryear{{Diego} et~al.,}{{Diego}
  et~al.}{2023a}]{Diego22b}
{Diego} J.~M.,  et~al., 2023a, \mn@doi [\aap] {10.1051/0004-6361/202245238},
  \href {https://ui.adsabs.harvard.edu/abs/2023A&A...672A...3D} {672, A3}

\bibitem[\protect\citeauthoryear{{Diego} et~al.,}{{Diego}
  et~al.}{2023b}]{Diego23c}
{Diego} J.~M.,  et~al., 2023b, \mn@doi [\aap] {10.1051/0004-6361/202245238},
  \href {https://ui.adsabs.harvard.edu/abs/2023A&A...672A...3D} {672, A3}

\bibitem[\protect\citeauthoryear{{Diego} et~al.,}{{Diego}
  et~al.}{2023c}]{Diego23b}
{Diego} J.~M.,  et~al., 2023c, \mn@doi [\aap] {10.1051/0004-6361/202347556},
  \href {https://ui.adsabs.harvard.edu/abs/2023A&A...679A..31D} {679, A31}

\bibitem[\protect\citeauthoryear{{Diego} et~al.,}{{Diego}
  et~al.}{2024a}]{Diego24}
{Diego} J.~M.,  et~al., 2024a, \mn@doi [arXiv e-prints]
  {10.48550/arXiv.2404.08033}, \href
  {https://ui.adsabs.harvard.edu/abs/2024arXiv240408033D} {p. arXiv:2404.08033}

\bibitem[\protect\citeauthoryear{{Diego} et~al.,}{{Diego}
  et~al.}{2024b}]{Diego23}
{Diego} J.~M.,  et~al., 2024b, \mn@doi [\aap] {10.1051/0004-6361/202346761},
  \href {https://ui.adsabs.harvard.edu/abs/2024A&A...681A.124D} {681, A124}

\bibitem[\protect\citeauthoryear{{Drout}, {Massey}  \& {Meynet}}{{Drout}
  et~al.}{2012}]{Drout12}
{Drout} M.~R.,  {Massey} P.,   {Meynet} G.,  2012, \mn@doi [\apj]
  {10.1088/0004-637X/750/2/97}, \href
  {https://ui.adsabs.harvard.edu/abs/2012ApJ...750...97D} {750, 97}

\bibitem[\protect\citeauthoryear{{Ekstr{\"o}m}, {Meynet}, {Chiappini},
  {Hirschi}  \& {Maeder}}{{Ekstr{\"o}m} et~al.}{2008}]{Ekstrom08}
{Ekstr{\"o}m} S.,  {Meynet} G.,  {Chiappini} C.,  {Hirschi} R.,   {Maeder} A.,
  2008, \mn@doi [\aap] {10.1051/0004-6361:200809633}, \href
  {https://ui.adsabs.harvard.edu/abs/2008A&A...489..685E} {489, 685}

\bibitem[\protect\citeauthoryear{{Eldridge} \& {Vink}}{{Eldridge} \&
  {Vink}}{2006}]{Eldridge06}
{Eldridge} J.~J.,  {Vink} J.~S.,  2006, \mn@doi [\aap]
  {10.1051/0004-6361:20065001}, \href
  {https://ui.adsabs.harvard.edu/abs/2006A&A...452..295E} {452, 295}

\bibitem[\protect\citeauthoryear{{Fudamoto} et~al.,}{{Fudamoto}
  et~al.}{2024}]{Fudamoto24}
{Fudamoto} Y.,  et~al., 2024, \mn@doi [arXiv e-prints]
  {10.48550/arXiv.2404.08045}, \href
  {https://ui.adsabs.harvard.edu/abs/2024arXiv240408045F} {p. arXiv:2404.08045}

\bibitem[\protect\citeauthoryear{{Furtak} et~al.,}{{Furtak}
  et~al.}{2023}]{Furtak23}
{Furtak} L.~J.,  et~al., 2023, \mn@doi [\mnras] {10.1093/mnrasl/slad135}, \href
  {https://ui.adsabs.harvard.edu/abs/2023MNRAS.tmpL.129F} {}

\bibitem[\protect\citeauthoryear{{Geen}, {Bieri}, {Rosdahl}  \& {de
  Koter}}{{Geen} et~al.}{2021}]{Geen21}
{Geen} S.,  {Bieri} R.,  {Rosdahl} J.,   {de Koter} A.,  2021, \mn@doi [\mnras]
  {10.1093/mnras/staa3705}, \href
  {https://ui.adsabs.harvard.edu/abs/2021MNRAS.501.1352G} {501, 1352}

\bibitem[\protect\citeauthoryear{{Haemmerl{\'e}}, {Woods}, {Klessen}, {Heger}
  \& {Whalen}}{{Haemmerl{\'e}} et~al.}{2018}]{Haemmerle18}
{Haemmerl{\'e}} L.,  {Woods} T.~E.,  {Klessen} R.~S.,  {Heger} A.,   {Whalen}
  D.~J.,  2018, \mn@doi [\mnras] {10.1093/mnras/stx2919}, \href
  {https://ui.adsabs.harvard.edu/abs/2018MNRAS.474.2757H} {474, 2757}

\bibitem[\protect\citeauthoryear{{Hainich}, {Ramachandran}, {Shenar}, {Sander},
  {Todt}, {Gruner}, {Oskinova}  \& {Hamann}}{{Hainich}
  et~al.}{2019}]{Hainich19}
{Hainich} R.,  {Ramachandran} V.,  {Shenar} T.,  {Sander} A.~A.~C.,  {Todt} H.,
   {Gruner} D.,  {Oskinova} L.~M.,   {Hamann} W.~R.,  2019, \mn@doi [\aap]
  {10.1051/0004-6361/201833787}, \href
  {https://ui.adsabs.harvard.edu/abs/2019A&A...621A..85H} {621, A85}

\bibitem[\protect\citeauthoryear{{Hirano} \& {Bromm}}{{Hirano} \&
  {Bromm}}{2018}]{Hirano18}
{Hirano} S.,  {Bromm} V.,  2018, \mn@doi [\mnras] {10.1093/mnras/sty487}, \href
  {https://ui.adsabs.harvard.edu/abs/2018MNRAS.476.3964H} {476, 3964}

\bibitem[\protect\citeauthoryear{{Holgado} et~al.,}{{Holgado}
  et~al.}{2020}]{Holgado20}
{Holgado} G.,  et~al., 2020, \mn@doi [\aap] {10.1051/0004-6361/202037699},
  \href {https://ui.adsabs.harvard.edu/abs/2020A&A...638A.157H} {638, A157}

\bibitem[\protect\citeauthoryear{{Hubeny} \& {Lanz}}{{Hubeny} \&
  {Lanz}}{1995}]{Hubeny95}
{Hubeny} I.,  {Lanz} T.,  1995, \mn@doi [\apj] {10.1086/175226}, \href
  {https://ui.adsabs.harvard.edu/abs/1995ApJ...439..875H} {439, 875}

\bibitem[\protect\citeauthoryear{{Inoue}}{{Inoue}}{2011}]{Inoue11}
{Inoue} A.~K.,  2011, \mn@doi [\mnras] {10.1111/j.1365-2966.2011.18906.x},
  \href {https://ui.adsabs.harvard.edu/abs/2011MNRAS.415.2920I} {415, 2920}

\bibitem[\protect\citeauthoryear{{Jaacks}, {Finkelstein}  \& {Bromm}}{{Jaacks}
  et~al.}{2019}]{Jaacks19}
{Jaacks} J.,  {Finkelstein} S.~L.,   {Bromm} V.,  2019, \mn@doi [\mnras]
  {10.1093/mnras/stz1529}, \href
  {https://ui.adsabs.harvard.edu/abs/2019MNRAS.488.2202J} {488, 2202}

\bibitem[\protect\citeauthoryear{{Jaura}, {Glover}, {Wollenberg}, {Klessen},
  {Geen}  \& {Haemmerl{\'e}}}{{Jaura} et~al.}{2022}]{Jaura22}
{Jaura} O.,  {Glover} S. C.~O.,  {Wollenberg} K. M.~J.,  {Klessen} R.~S.,
  {Geen} S.,   {Haemmerl{\'e}} L.,  2022, \mn@doi [\mnras]
  {10.1093/mnras/stac487}, \href
  {https://ui.adsabs.harvard.edu/abs/2022MNRAS.512..116J} {512, 116}

\bibitem[\protect\citeauthoryear{{Kelly} et~al.,}{{Kelly}
  et~al.}{2018}]{Kelly18}
{Kelly} P.~L.,  et~al., 2018, \mn@doi [Nature Astronomy]
  {10.1038/s41550-018-0430-3}, \href
  {https://ui.adsabs.harvard.edu/abs/2018NatAs...2..334K} {2, 334}

\bibitem[\protect\citeauthoryear{{Kelly} et~al.,}{{Kelly}
  et~al.}{2022}]{Kelly22}
{Kelly} P.~L.,  et~al., 2022, \mn@doi [arXiv e-prints]
  {10.48550/arXiv.2211.02670}, \href
  {https://ui.adsabs.harvard.edu/abs/2022arXiv221102670K} {p. arXiv:2211.02670}

\bibitem[\protect\citeauthoryear{{Kitayama}, {Yoshida}, {Susa}  \&
  {Umemura}}{{Kitayama} et~al.}{2004}]{Kitayama04}
{Kitayama} T.,  {Yoshida} N.,  {Susa} H.,   {Umemura} M.,  2004, \mn@doi [\apj]
  {10.1086/423313}, \href
  {https://ui.adsabs.harvard.edu/abs/2004ApJ...613..631K} {613, 631}

\bibitem[\protect\citeauthoryear{{Klessen} \& {Glover}}{{Klessen} \&
  {Glover}}{2023}]{Klessen23}
{Klessen} R.~S.,  {Glover} S. C.~O.,  2023, \mn@doi [\araa]
  {10.1146/annurev-astro-071221-053453}, \href
  {https://ui.adsabs.harvard.edu/abs/2023ARA&A..61...65K} {61, 65}

\bibitem[\protect\citeauthoryear{{Koutsouridou}, {Salvadori}  \&
  {Sk{\'u}lad{\'o}ttir}}{{Koutsouridou} et~al.}{2023a}]{Koutsouridou23b}
{Koutsouridou} I.,  {Salvadori} S.,   {Sk{\'u}lad{\'o}ttir} {\'A}.,  2023a,
  \mn@doi [arXiv e-prints] {10.48550/arXiv.2312.05309}, \href
  {https://ui.adsabs.harvard.edu/abs/2023arXiv231205309K} {p. arXiv:2312.05309}

\bibitem[\protect\citeauthoryear{{Koutsouridou}, {Salvadori},
  {Sk{\'u}lad{\'o}ttir}, {Rossi}, {Vanni}  \& {Pagnini}}{{Koutsouridou}
  et~al.}{2023b}]{Koutsouridou23}
{Koutsouridou} I.,  {Salvadori} S.,  {Sk{\'u}lad{\'o}ttir} {\'A}.,  {Rossi} M.,
   {Vanni} I.,   {Pagnini} G.,  2023b, \mn@doi [\mnras]
  {10.1093/mnras/stad2304}, \href
  {https://ui.adsabs.harvard.edu/abs/2023MNRAS.525..190K} {525, 190}

\bibitem[\protect\citeauthoryear{{Kroupa}}{{Kroupa}}{2001}]{Kroupa01}
{Kroupa} P.,  2001, \mn@doi [\mnras] {10.1046/j.1365-8711.2001.04022.x}, \href
  {https://ui.adsabs.harvard.edu/abs/2001MNRAS.322..231K} {322, 231}

\bibitem[\protect\citeauthoryear{{Lanz} \& {Hubeny}}{{Lanz} \&
  {Hubeny}}{2003}]{Lanz03}
{Lanz} T.,  {Hubeny} I.,  2003, \mn@doi [\apjs] {10.1086/374373}, \href
  {https://ui.adsabs.harvard.edu/abs/2003ApJS..146..417L} {146, 417}

\bibitem[\protect\citeauthoryear{{Lanz} \& {Hubeny}}{{Lanz} \&
  {Hubeny}}{2007}]{Lanz07}
{Lanz} T.,  {Hubeny} I.,  2007, \mn@doi [\apjs] {10.1086/511270}, \href
  {https://ui.adsabs.harvard.edu/abs/2007ApJS..169...83L} {169, 83}

\bibitem[\protect\citeauthoryear{{Larkin}, {Gerasimov}  \&
  {Burgasser}}{{Larkin} et~al.}{2023}]{Larkin23}
{Larkin} M.~M.,  {Gerasimov} R.,   {Burgasser} A.~J.,  2023, \mn@doi [\aj]
  {10.3847/1538-3881/ac9b43}, \href
  {https://ui.adsabs.harvard.edu/abs/2023AJ....165....2L} {165, 2}

\bibitem[\protect\citeauthoryear{{Leitherer}, {Ortiz Ot{\'a}lvaro}, {Bresolin},
  {Kudritzki}, {Lo Faro}, {Pauldrach}, {Pettini}  \& {Rix}}{{Leitherer}
  et~al.}{2010}]{Leitherer10}
{Leitherer} C.,  {Ortiz Ot{\'a}lvaro} P.~A.,  {Bresolin} F.,  {Kudritzki}
  R.-P.,  {Lo Faro} B.,  {Pauldrach} A. W.~A.,  {Pettini} M.,   {Rix} S.~A.,
  2010, \mn@doi [\apjs] {10.1088/0067-0049/189/2/309}, \href
  {https://ui.adsabs.harvard.edu/abs/2010ApJS..189..309L} {189, 309}

\bibitem[\protect\citeauthoryear{{Lejeune}, {Cuisinier}  \& {Buser}}{{Lejeune}
  et~al.}{1997}]{Lejeune97}
{Lejeune} T.,  {Cuisinier} F.,   {Buser} R.,  1997, \mn@doi [\aaps]
  {10.1051/aas:1997373}, \href
  {https://ui.adsabs.harvard.edu/abs/1997A&AS..125..229L} {125, 229}

\bibitem[\protect\citeauthoryear{{Liu} \& {Bromm}}{{Liu} \&
  {Bromm}}{2020}]{Liu20}
{Liu} B.,  {Bromm} V.,  2020, \mn@doi [\mnras] {10.1093/mnras/staa2143}, \href
  {https://ui.adsabs.harvard.edu/abs/2020MNRAS.497.2839L} {497, 2839}

\bibitem[\protect\citeauthoryear{{Liu}, {Sibony}, {Meynet}  \& {Bromm}}{{Liu}
  et~al.}{2021}]{Liu21}
{Liu} B.,  {Sibony} Y.,  {Meynet} G.,   {Bromm} V.,  2021, \mn@doi [\mnras]
  {10.1093/mnras/stab2057}, \href
  {https://ui.adsabs.harvard.edu/abs/2021MNRAS.506.5247L} {506, 5247}

\bibitem[\protect\citeauthoryear{{Magg}, {Hartwig}, {Glover}, {Klessen}  \&
  {Whalen}}{{Magg} et~al.}{2016}]{Magg16}
{Magg} M.,  {Hartwig} T.,  {Glover} S. C.~O.,  {Klessen} R.~S.,   {Whalen}
  D.~J.,  2016, \mn@doi [\mnras] {10.1093/mnras/stw1882}, \href
  {https://ui.adsabs.harvard.edu/abs/2016MNRAS.462.3591M} {462, 3591}

\bibitem[\protect\citeauthoryear{{Magg}, {Hartwig}, {Agarwal}, {Frebel},
  {Glover}, {Griffen}  \& {Klessen}}{{Magg} et~al.}{2018}]{Magg18}
{Magg} M.,  {Hartwig} T.,  {Agarwal} B.,  {Frebel} A.,  {Glover} S. C.~O.,
  {Griffen} B.~F.,   {Klessen} R.~S.,  2018, \mn@doi [\mnras]
  {10.1093/mnras/stx2729}, \href
  {https://ui.adsabs.harvard.edu/abs/2018MNRAS.473.5308M} {473, 5308}

\bibitem[\protect\citeauthoryear{{Maio}, {Ciardi}, {Dolag}, {Tornatore}  \&
  {Khochfar}}{{Maio} et~al.}{2010}]{Maio10}
{Maio} U.,  {Ciardi} B.,  {Dolag} K.,  {Tornatore} L.,   {Khochfar} S.,  2010,
  \mn@doi [\mnras] {10.1111/j.1365-2966.2010.17003.x}, \href
  {https://ui.adsabs.harvard.edu/abs/2010MNRAS.407.1003M} {407, 1003}

\bibitem[\protect\citeauthoryear{{Marassi}, {Schneider}, {Limongi}, {Chieffi},
  {Bocchio}  \& {Bianchi}}{{Marassi} et~al.}{2015}]{Marassi15}
{Marassi} S.,  {Schneider} R.,  {Limongi} M.,  {Chieffi} A.,  {Bocchio} M.,
  {Bianchi} S.,  2015, \mn@doi [\mnras] {10.1093/mnras/stv2267}, \href
  {https://ui.adsabs.harvard.edu/abs/2015MNRAS.454.4250M} {454, 4250}

\bibitem[\protect\citeauthoryear{{Marigo}, {Chiosi}  \& {Kudritzki}}{{Marigo}
  et~al.}{2003}]{Marigo03}
{Marigo} P.,  {Chiosi} C.,   {Kudritzki} R.~P.,  2003, \mn@doi [\aap]
  {10.1051/0004-6361:20021756}, \href
  {https://ui.adsabs.harvard.edu/abs/2003A&A...399..617M} {399, 617}

\bibitem[\protect\citeauthoryear{{Mas-Ribas}, {Dijkstra}  \&
  {Forero-Romero}}{{Mas-Ribas} et~al.}{2016}]{Mas-Ribas16}
{Mas-Ribas} L.,  {Dijkstra} M.,   {Forero-Romero} J.~E.,  2016, \mn@doi [\apj]
  {10.3847/1538-4357/833/1/65}, \href
  {https://ui.adsabs.harvard.edu/abs/2016ApJ...833...65M} {833, 65}

\bibitem[\protect\citeauthoryear{{Massey}, {Plez}, {Levesque}, {Olsen},
  {Clayton}  \& {Josselin}}{{Massey} et~al.}{2005}]{Massey05}
{Massey} P.,  {Plez} B.,  {Levesque} E.~M.,  {Olsen} K.~A.~G.,  {Clayton}
  G.~C.,   {Josselin} E.,  2005, \mn@doi [\apj] {10.1086/497065}, \href
  {https://ui.adsabs.harvard.edu/abs/2005ApJ...634.1286M} {634, 1286}

\bibitem[\protect\citeauthoryear{{Mebane}, {Mirocha}  \& {Furlanetto}}{{Mebane}
  et~al.}{2018}]{Mebane18}
{Mebane} R.~H.,  {Mirocha} J.,   {Furlanetto} S.~R.,  2018, \mn@doi [\mnras]
  {10.1093/mnras/sty1833}, \href
  {https://ui.adsabs.harvard.edu/abs/2018MNRAS.479.4544M} {479, 4544}

\bibitem[\protect\citeauthoryear{{Meena} et~al.,}{{Meena}
  et~al.}{2023}]{Meena23}
{Meena} A.~K.,  et~al., 2023, \mn@doi [\apjl] {10.3847/2041-8213/acb645}, \href
  {https://ui.adsabs.harvard.edu/abs/2023ApJ...944L...6M} {944, L6}

\bibitem[\protect\citeauthoryear{{Mellema}, {Arthur}, {Henney}, {Iliev}  \&
  {Shapiro}}{{Mellema} et~al.}{2006}]{Mellema06}
{Mellema} G.,  {Arthur} S.~J.,  {Henney} W.~J.,  {Iliev} I.~T.,   {Shapiro}
  P.~R.,  2006, \mn@doi [\apj] {10.1086/505294}, \href
  {https://ui.adsabs.harvard.edu/abs/2006ApJ...647..397M} {647, 397}

\bibitem[\protect\citeauthoryear{{Miralda-Escude}}{{Miralda-Escude}}{1991}]{Miralda-Escude91}
{Miralda-Escude} J.,  1991, \mn@doi [\apj] {10.1086/170486}, \href
  {https://ui.adsabs.harvard.edu/abs/1991ApJ...379...94M} {379, 94}

\bibitem[\protect\citeauthoryear{{Murphy} et~al.,}{{Murphy}
  et~al.}{2021}]{Murphy21a}
{Murphy} L.~J.,  et~al., 2021, \mn@doi [\mnras] {10.1093/mnras/staa3803}, \href
  {https://ui.adsabs.harvard.edu/abs/2021MNRAS.501.2745M} {501, 2745}

\bibitem[\protect\citeauthoryear{{Nakajima} \& {Maiolino}}{{Nakajima} \&
  {Maiolino}}{2022}]{Nakajima22}
{Nakajima} K.,  {Maiolino} R.,  2022, \mn@doi [\mnras]
  {10.1093/mnras/stac1242}, \href
  {https://ui.adsabs.harvard.edu/abs/2022MNRAS.513.5134N} {513, 5134}

\bibitem[\protect\citeauthoryear{{Palencia}, {Diego}, {Kavanagh}  \&
  {Martinez}}{{Palencia} et~al.}{2023}]{Palencia23}
{Palencia} J.~M.,  {Diego} J.~M.,  {Kavanagh} B.~J.,   {Martinez} J.,  2023,
  \mn@doi [arXiv e-prints] {10.48550/arXiv.2307.09505}, \href
  {https://ui.adsabs.harvard.edu/abs/2023arXiv230709505P} {p. arXiv:2307.09505}

\bibitem[\protect\citeauthoryear{{Pallottini}, {Ferrara}, {Gallerani},
  {Salvadori}  \& {D'Odorico}}{{Pallottini} et~al.}{2014}]{Pallottini14}
{Pallottini} A.,  {Ferrara} A.,  {Gallerani} S.,  {Salvadori} S.,   {D'Odorico}
  V.,  2014, \mn@doi [\mnras] {10.1093/mnras/stu451}, \href
  {https://ui.adsabs.harvard.edu/abs/2014MNRAS.440.2498P} {440, 2498}

\bibitem[\protect\citeauthoryear{{Rieke} et~al.,}{{Rieke}
  et~al.}{2023}]{Rieke23}
{Rieke} M.~J.,  et~al., 2023, \mn@doi [\pasp] {10.1088/1538-3873/acac53}, \href
  {https://ui.adsabs.harvard.edu/abs/2023PASP..135b8001R} {135, 028001}

\bibitem[\protect\citeauthoryear{{Rydberg}, {Zackrisson}, {Lundqvist}  \&
  {Scott}}{{Rydberg} et~al.}{2013}]{Rydberg13}
{Rydberg} C.-E.,  {Zackrisson} E.,  {Lundqvist} P.,   {Scott} P.,  2013,
  \mn@doi [\mnras] {10.1093/mnras/sts653}, \href
  {https://ui.adsabs.harvard.edu/abs/2013MNRAS.429.3658R} {429, 3658}

\bibitem[\protect\citeauthoryear{{Sanyal}, {Grassitelli}, {Langer}  \&
  {Bestenlehner}}{{Sanyal} et~al.}{2015}]{Sanyal15}
{Sanyal} D.,  {Grassitelli} L.,  {Langer} N.,   {Bestenlehner} J.~M.,  2015,
  \mn@doi [\aap] {10.1051/0004-6361/201525945}, \href
  {https://ui.adsabs.harvard.edu/abs/2015A&A...580A..20S} {580, A20}

\bibitem[\protect\citeauthoryear{{Sarmento}, {Scannapieco}  \&
  {C{\^o}t{\'e}}}{{Sarmento} et~al.}{2019}]{Sarmento19}
{Sarmento} R.,  {Scannapieco} E.,   {C{\^o}t{\'e}} B.,  2019, \mn@doi [\apj]
  {10.3847/1538-4357/aafa1a}, \href
  {https://ui.adsabs.harvard.edu/abs/2019ApJ...871..206S} {871, 206}

\bibitem[\protect\citeauthoryear{{Schaerer}}{{Schaerer}}{2002}]{Schaerer02}
{Schaerer} D.,  2002, \mn@doi [\aap] {10.1051/0004-6361:20011619}, \href
  {https://ui.adsabs.harvard.edu/abs/2002A&A...382...28S} {382, 28}

\bibitem[\protect\citeauthoryear{{Schaerer}}{{Schaerer}}{2003}]{Schaerer03}
{Schaerer} D.,  2003, \mn@doi [\aap] {10.1051/0004-6361:20021525}, \href
  {https://ui.adsabs.harvard.edu/abs/2003A&A...397..527S} {397, 527}

\bibitem[\protect\citeauthoryear{{Schauer}, {Drory}  \& {Bromm}}{{Schauer}
  et~al.}{2020}]{Schauer20}
{Schauer} A. T.~P.,  {Drory} N.,   {Bromm} V.,  2020, \mn@doi [\apj]
  {10.3847/1538-4357/abbc0b}, \href
  {https://ui.adsabs.harvard.edu/abs/2020ApJ...904..145S} {904, 145}

\bibitem[\protect\citeauthoryear{{Schauer}, {Bromm}, {Drory}  \&
  {Boylan-Kolchin}}{{Schauer} et~al.}{2022}]{Schauer22}
{Schauer} A. T.~P.,  {Bromm} V.,  {Drory} N.,   {Boylan-Kolchin} M.,  2022,
  \mn@doi [\apjl] {10.3847/2041-8213/ac7f9a}, \href
  {https://ui.adsabs.harvard.edu/abs/2022ApJ...934L...6S} {934, L6}

\bibitem[\protect\citeauthoryear{{Schneider}, {Ehlers}  \& {Falco}}{{Schneider}
  et~al.}{1992}]{Schneider92}
{Schneider} P.,  {Ehlers} J.,   {Falco} E.~E.,  1992, {Gravitational Lenses},
  \mn@doi{10.1007/978-3-662-03758-4.
}

\bibitem[\protect\citeauthoryear{{Sibony}, {Liu}, {Simmonds}, {Meynet}  \&
  {Bromm}}{{Sibony} et~al.}{2022}]{Sibony22}
{Sibony} Y.,  {Liu} B.,  {Simmonds} C.,  {Meynet} G.,   {Bromm} V.,  2022,
  \mn@doi [\aap] {10.1051/0004-6361/202244146}, \href
  {https://ui.adsabs.harvard.edu/abs/2022A&A...666A.199S} {666, A199}

\bibitem[\protect\citeauthoryear{{Song}, {Meynet}, {Li}, {Peng}, {Zhang}  \&
  {Zhan}}{{Song} et~al.}{2020}]{Song20}
{Song} H.,  {Meynet} G.,  {Li} Z.,  {Peng} W.,  {Zhang} R.,   {Zhan} Q.,  2020,
  \mn@doi [\apj] {10.3847/1538-4357/ab7993}, \href
  {https://ui.adsabs.harvard.edu/abs/2020ApJ...892...41S} {892, 41}

\bibitem[\protect\citeauthoryear{{Stacy} \& {Bromm}}{{Stacy} \&
  {Bromm}}{2013}]{Stacy13}
{Stacy} A.,  {Bromm} V.,  2013, \mn@doi [\mnras] {10.1093/mnras/stt789}, \href
  {https://ui.adsabs.harvard.edu/abs/2013MNRAS.433.1094S} {433, 1094}

\bibitem[\protect\citeauthoryear{{Sugimura}, {Matsumoto}, {Hosokawa}, {Hirano}
  \& {Omukai}}{{Sugimura} et~al.}{2020}]{Sugamura20}
{Sugimura} K.,  {Matsumoto} T.,  {Hosokawa} T.,  {Hirano} S.,   {Omukai} K.,
  2020, \mn@doi [\apjl] {10.3847/2041-8213/ab7d37}, \href
  {https://ui.adsabs.harvard.edu/abs/2020ApJ...892L..14S} {892, L14}

\bibitem[\protect\citeauthoryear{{Surace} et~al.,}{{Surace}
  et~al.}{2018}]{Surace18}
{Surace} M.,  et~al., 2018, \mn@doi [\apjl] {10.3847/2041-8213/aaf80d}, \href
  {https://ui.adsabs.harvard.edu/abs/2018ApJ...869L..39S} {869, L39}

\bibitem[\protect\citeauthoryear{{Surace}, {Zackrisson}, {Whalen}, {Hartwig},
  {Glover}, {Woods}, {Heger}  \& {Glover}}{{Surace} et~al.}{2019}]{Surace19}
{Surace} M.,  {Zackrisson} E.,  {Whalen} D.~J.,  {Hartwig} T.,  {Glover}
  S.~C.~O.,  {Woods} T.~E.,  {Heger} A.,   {Glover} S.~C.~O.,  2019, \mn@doi
  [\mnras] {10.1093/mnras/stz1956}, \href
  {https://ui.adsabs.harvard.edu/abs/2019MNRAS.488.3995S} {488, 3995}

\bibitem[\protect\citeauthoryear{{Sz{\'e}csi}, {Agrawal}, {W{\"u}nsch}  \&
  {Langer}}{{Sz{\'e}csi} et~al.}{2022}]{Szecsi20}
{Sz{\'e}csi} D.,  {Agrawal} P.,  {W{\"u}nsch} R.,   {Langer} N.,  2022, \mn@doi
  [\aap] {10.1051/0004-6361/202141536}, \href
  {https://ui.adsabs.harvard.edu/abs/2022A&A...658A.125S} {658, A125}

\bibitem[\protect\citeauthoryear{{Trussler} et~al.,}{{Trussler}
  et~al.}{2023}]{Trussler22}
{Trussler} J. A.~A.,  et~al., 2023, \mn@doi [\mnras] {10.1093/mnras/stad2553},
  \href {https://ui.adsabs.harvard.edu/abs/2023MNRAS.525.5328T} {525, 5328}

\bibitem[\protect\citeauthoryear{{Tumlinson}}{{Tumlinson}}{2006}]{Tumlinson06}
{Tumlinson} J.,  2006, \mn@doi [\apj] {10.1086/500383}, \href
  {https://ui.adsabs.harvard.edu/abs/2006ApJ...641....1T} {641, 1}

\bibitem[\protect\citeauthoryear{{Tumlinson} \& {Shull}}{{Tumlinson} \&
  {Shull}}{2000}]{Tumlinson00}
{Tumlinson} J.,  {Shull} J.~M.,  2000, \mn@doi [\apjl] {10.1086/312432}, \href
  {https://ui.adsabs.harvard.edu/abs/2000ApJ...528L..65T} {528, L65}

\bibitem[\protect\citeauthoryear{{Tumlinson}, {Giroux}  \& {Shull}}{{Tumlinson}
  et~al.}{2001}]{Tumlinson01}
{Tumlinson} J.,  {Giroux} M.~L.,   {Shull} J.~M.,  2001, \mn@doi [\apjl]
  {10.1086/319477}, \href
  {https://ui.adsabs.harvard.edu/abs/2001ApJ...550L...1T} {550, L1}

\bibitem[\protect\citeauthoryear{{Tumlinson}, {Shull}  \&
  {Venkatesan}}{{Tumlinson} et~al.}{2003}]{Tumlinson03}
{Tumlinson} J.,  {Shull} J.~M.,   {Venkatesan} A.,  2003, \mn@doi [\apj]
  {10.1086/345737}, \href
  {https://ui.adsabs.harvard.edu/abs/2003ApJ...584..608T} {584, 608}

\bibitem[\protect\citeauthoryear{{Venditti}, {Graziani}, {Schneider},
  {Pentericci}, {Di Cesare}, {Maio}  \& {Omukai}}{{Venditti}
  et~al.}{2023}]{Venditti23}
{Venditti} A.,  {Graziani} L.,  {Schneider} R.,  {Pentericci} L.,  {Di Cesare}
  C.,  {Maio} U.,   {Omukai} K.,  2023, \mn@doi [\mnras]
  {10.1093/mnras/stad1201}, \href
  {https://ui.adsabs.harvard.edu/abs/2023MNRAS.522.3809V} {522, 3809}

\bibitem[\protect\citeauthoryear{{Venumadhav}, {Dai}  \&
  {Miralda-Escud{\'e}}}{{Venumadhav} et~al.}{2017}]{Venumadhav17}
{Venumadhav} T.,  {Dai} L.,   {Miralda-Escud{\'e}} J.,  2017, \mn@doi [\apj]
  {10.3847/1538-4357/aa9575}, \href
  {https://ui.adsabs.harvard.edu/abs/2017ApJ...850...49V} {850, 49}

\bibitem[\protect\citeauthoryear{{Volpato}, {Marigo}, {Costa}, {Bressan},
  {Trabucchi}  \& {Girardi}}{{Volpato} et~al.}{2023}]{Volpato23}
{Volpato} G.,  {Marigo} P.,  {Costa} G.,  {Bressan} A.,  {Trabucchi} M.,
  {Girardi} L.,  2023, \mn@doi [\apj] {10.3847/1538-4357/acac91}, \href
  {https://ui.adsabs.harvard.edu/abs/2023ApJ...944...40V} {944, 40}

\bibitem[\protect\citeauthoryear{{Volpato}, {Marigo}, {Costa}, {Bressan},
  {Trabucchi}, {Girardi}  \& {Addari}}{{Volpato} et~al.}{2024}]{Volpato24}
{Volpato} G.,  {Marigo} P.,  {Costa} G.,  {Bressan} A.,  {Trabucchi} M.,
  {Girardi} L.,   {Addari} F.,  2024, \mn@doi [\apj]
  {10.3847/1538-4357/ad1185}, \href
  {https://ui.adsabs.harvard.edu/abs/2024ApJ...961...89V} {961, 89}

\bibitem[\protect\citeauthoryear{{Wang} et~al.,}{{Wang} et~al.}{2022}]{Wang22}
{Wang} C.,  et~al., 2022, \mn@doi [Nature Astronomy]
  {10.1038/s41550-021-01597-5}, \href
  {https://ui.adsabs.harvard.edu/abs/2022NatAs...6..480W} {6, 480}

\bibitem[\protect\citeauthoryear{{Welch} et~al.,}{{Welch}
  et~al.}{2022a}]{Welch22}
{Welch} B.,  et~al., 2022a, \mn@doi [\nat] {10.1038/s41586-022-04449-y}, \href
  {https://ui.adsabs.harvard.edu/abs/2022Natur.603..815W} {603, 815}

\bibitem[\protect\citeauthoryear{{Welch} et~al.,}{{Welch}
  et~al.}{2022b}]{Welch22b}
{Welch} B.,  et~al., 2022b, \mn@doi [\apjl] {10.3847/2041-8213/ac9d39}, \href
  {https://ui.adsabs.harvard.edu/abs/2022ApJ...940L...1W} {940, L1}

\bibitem[\protect\citeauthoryear{{Whalen}, {Abel}  \& {Norman}}{{Whalen}
  et~al.}{2004}]{Whalen04}
{Whalen} D.,  {Abel} T.,   {Norman} M.~L.,  2004, \mn@doi [\apj]
  {10.1086/421548}, \href
  {https://ui.adsabs.harvard.edu/abs/2004ApJ...610...14W} {610, 14}

\bibitem[\protect\citeauthoryear{{Windhorst} et~al.,}{{Windhorst}
  et~al.}{2018}]{Windhorst18}
{Windhorst} R.~A.,  et~al., 2018, \mn@doi [\apjs] {10.3847/1538-4365/aaa760},
  \href {https://ui.adsabs.harvard.edu/abs/2018ApJS..234...41W} {234, 41}

\bibitem[\protect\citeauthoryear{{Windhorst} et~al.,}{{Windhorst}
  et~al.}{2023}]{Windhorst23}
{Windhorst} R.~A.,  et~al., 2023, \mn@doi [\aj] {10.3847/1538-3881/aca163},
  \href {https://ui.adsabs.harvard.edu/abs/2023AJ....165...13W} {165, 13}

\bibitem[\protect\citeauthoryear{{Wu} \& {Kravtsov}}{{Wu} \&
  {Kravtsov}}{2024}]{Wu24}
{Wu} Z.,  {Kravtsov} A.,  2024, \mn@doi [arXiv e-prints]
  {10.48550/arXiv.2405.08066}, \href
  {https://ui.adsabs.harvard.edu/abs/2024arXiv240508066W} {p. arXiv:2405.08066}

\bibitem[\protect\citeauthoryear{{Xu}, {Ahn}, {Norman}, {Wise}  \&
  {O'Shea}}{{Xu} et~al.}{2016}]{Xu16}
{Xu} H.,  {Ahn} K.,  {Norman} M.~L.,  {Wise} J.~H.,   {O'Shea} B.~W.,  2016,
  \mn@doi [\apjl] {10.3847/2041-8205/832/1/L5}, \href
  {https://ui.adsabs.harvard.edu/abs/2016ApJ...832L...5X} {832, L5}

\bibitem[\protect\citeauthoryear{{Yan}, {Jerabkova}  \& {Kroupa}}{{Yan}
  et~al.}{2023}]{Yan23}
{Yan} Z.,  {Jerabkova} T.,   {Kroupa} P.,  2023, \mn@doi [\aap]
  {10.1051/0004-6361/202244919}, \href
  {https://ui.adsabs.harvard.edu/abs/2023A&A...670A.151Y} {670, A151}

\bibitem[\protect\citeauthoryear{{Yoon}, {Dierks}  \& {Langer}}{{Yoon}
  et~al.}{2012}]{Yoon12}
{Yoon} S.~C.,  {Dierks} A.,   {Langer} N.,  2012, \mn@doi [\aap]
  {10.1051/0004-6361/201117769}, \href
  {https://ui.adsabs.harvard.edu/abs/2012A&A...542A.113Y} {542, A113}

\bibitem[\protect\citeauthoryear{{Zackrisson}, {Rydberg}, {Schaerer},
  {{\"O}stlin}  \& {Tuli}}{{Zackrisson} et~al.}{2011}]{Zackrisson11}
{Zackrisson} E.,  {Rydberg} C.-E.,  {Schaerer} D.,  {{\"O}stlin} G.,   {Tuli}
  M.,  2011, \mn@doi [\apj] {10.1088/0004-637X/740/1/13}, \href
  {https://ui.adsabs.harvard.edu/abs/2011ApJ...740...13Z} {740, 13}

\bibitem[\protect\citeauthoryear{{Zitrin} et~al.,}{{Zitrin}
  et~al.}{2013}]{Zitrin13}
{Zitrin} A.,  et~al., 2013, \mn@doi [\apjl] {10.1088/2041-8205/762/2/L30},
  \href {https://ui.adsabs.harvard.edu/abs/2013ApJ...762L..30Z} {762, L30}

\bibitem[\protect\citeauthoryear{{de Mink}, {Cantiello}, {Langer}, {Pols},
  {Brott}  \& {Yoon}}{{de Mink} et~al.}{2009}]{deMink09}
{de Mink} S.~E.,  {Cantiello} M.,  {Langer} N.,  {Pols} O.~R.,  {Brott} I.,
  {Yoon} S.~C.,  2009, \mn@doi [\aap] {10.1051/0004-6361/200811439}, \href
  {https://ui.adsabs.harvard.edu/abs/2009A&A...497..243D} {497, 243}

\makeatother
\end{thebibliography}


\appendix
\section{Stellar atmosphere grids}
\label{sec:appendix_stellar_atmospheres}
\begin{figure*}
\includegraphics[width=\columnwidth]{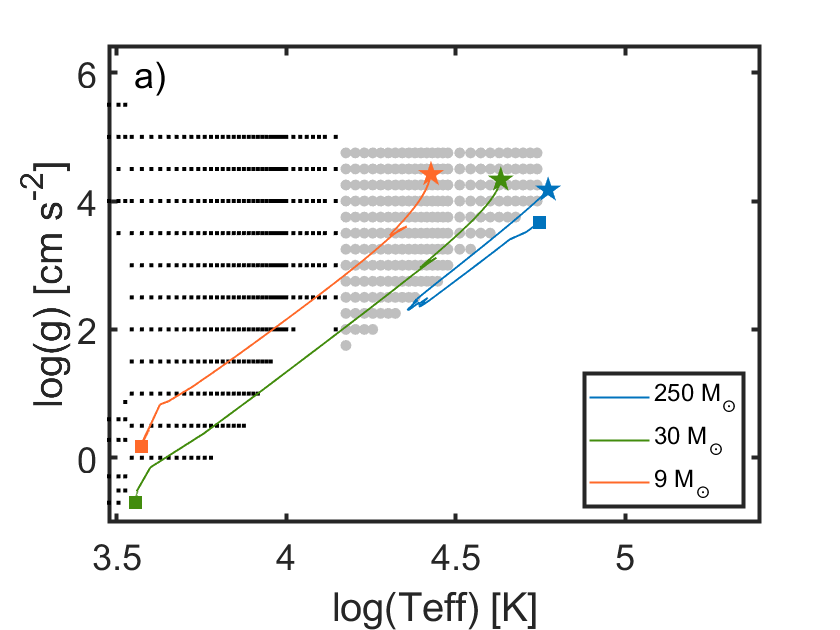}
\includegraphics[width=\columnwidth]{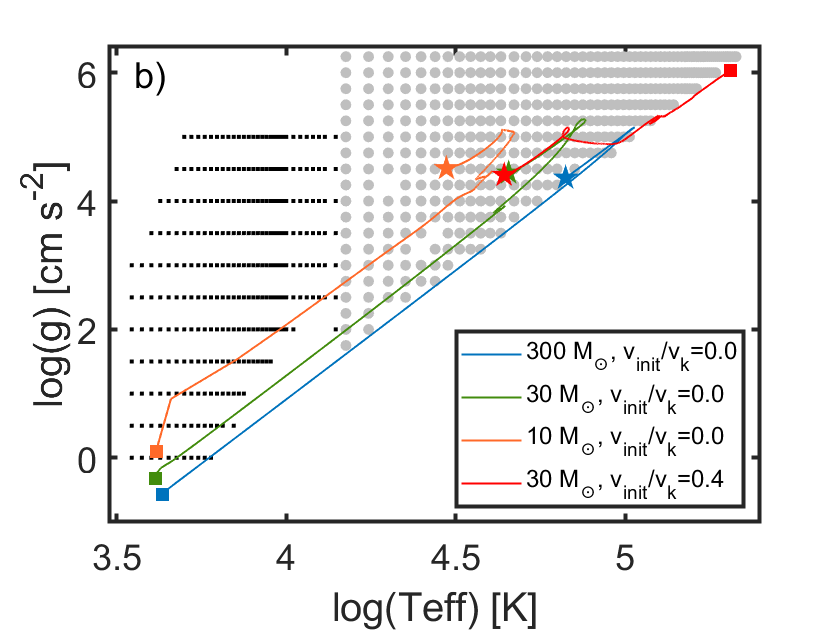}
\caption{$T_\mathrm{eff}$-$\log (g)$ diagram showing selected stellar evolutionary tracks for massive stars overlaid on the corresponding grid of stellar evolutionary spectra. {\bf a)} $Z\approx 0.002$ (SMC) metallicity tracks from \citet{Szecsi20} at 9 $M_\odot$ (orange line), 30 $M_\odot$ (green) and 250 $M_\odot$ (blue) overlaid on a $Z\approx 0.1\ Z_\odot$ stellar atmosphere grid. {\bf b)} Pop III tracks without rotation from \citet{Yoon12} at 10 $M_\odot$ (orange line), 30 $M_\odot$ (green) and 300 $M_\odot$ (blue), along with one 30 $M_\odot$ track with rotation ($v_\mathrm{init}/v_\mathrm{k}=0.4$; red line) overlaid on the $Z\approx 0$ stellar atmosphere grid. Along each evolutionary track in both panels, stars mark the start (youngest age) and squares the end. Grey markers indicate the locations of TLUSTY grid points and black points the positions of \citet{Lejeune97} grid points. As seen in both panels, the $M= 30\ M_\odot$ tracks venture outside the boundaries of these grids during certain points during their evolution, and the $M\approx 250$--$300\ M_\odot$ tracks spend a large fraction of their evolution below the lowest $\log(g)$ boundary of the grids.
.}
\label{fig_atmos_grid}
\end{figure*}

Here we provide further information on the technical procedure for generating SEDs of stars in the Muspelheim models. In Figure~\ref{fig_atmos_grid}, we show selected stellar evolutionary tracks overlaid on the $T_\mathrm{eff}$-$\log (g)$ coverage of the stellar atmosphere grids used at $Z=0.1\ Z_\odot$ (a) and $Z=0$ (b). This illustrates a common problem, in that the most massive stars often venture into a region of parameter space at very low $\log (g)$ where stellar atmosphere models are absent. For the lowest-mass tracks, this happens during a very small fraction of their lifetimes, but at $M\gtrsim 50\ M_\odot$, this fraction grows to $\gtrsim 10\%$, and for the highest-mass tracks, a majority of their evolution occurs at lower $\log(g)$ than formally covered by the grid. The low-$\log (g)$ boundary coarsely traces the Eddington limit for these stars, but since the grids are largely based on a fixed 
$\Delta\log(g)$ step size, the actual limit occurs at $\log (g)$ slightly outside these grids. In the case of the 300 $M_\odot$ track featured in panel \ref{fig_atmos_grid}b, $\gtrsim 90\%$ of the lifetime is spent within one $\Delta\log(g)$ grid step of the current grid boundary, and it is only at the late, short-lived, lower-$T_\mathrm{eff}$ stages where the discrepancy becomes more substantial.  On the other hand, In the case of the 250 $M_\odot$ track in panel a, the majority of the lifetime is spent farther away than this from the grid. As argued by \citet{Sanyal15}, many massive stars are likely to straddle the Eddington limit at their surface. The effects of forcing one-dimensional, hydrostatic atmosphere models onto such stars, which may also be non-spherical due to rotation, will inevitably lead to inaccuracies when it comes to individual spectral features. There is, however, no set of published stellar atmosphere models that can deal with such effects over the large regions of parameter space required by sets of stellar evolutionary tracks of the type used here.   

In cases where data points along a stellar evolutionary tracks fall outside the $\log(g)$ limits of the grid of stellar atmosphere spectra, an interpolation between the two closest $\log(g)$ grid models is performed in $T_\mathrm{eff}$, instead of an attempted extrapolation. In the procedure of interpolating the stellar atmosphere grids, wavelength interpolations are also occasionally required due to differences in wavelength coverage for stellar atmosphere spectra in different parts of these grids. While this is expected not to have any significant impact on broad- and medium-band magnitudes, individual absorption or emission features could get distorted. For this reason, an alternative mode is also implemented, in which the stellar atmosphere spectra that lies the closest in $T_\mathrm{eff}$ and $\log(g)$ is instead selected, without any interpolation. While this ensures that the native spectral sampling is always used and that no spurious spectral features are introduced by the interpolation procedure, this implies that quantities like colours, that depend solely on the shape of the SEDs, evolve in a less smooth, step-wise fashion as as these stars evolve through the $T_\mathrm{eff}$-$\log(g)$ grid. By comparing the results from the two strategies, we have checked that the results presented in this paper are not strongly affected by the wavelength interpolation procedure. 

\section{Relation between brightness and effective temperature}
\label{sec:Teff dependence}
\begin{figure*}
\includegraphics[width=\columnwidth]{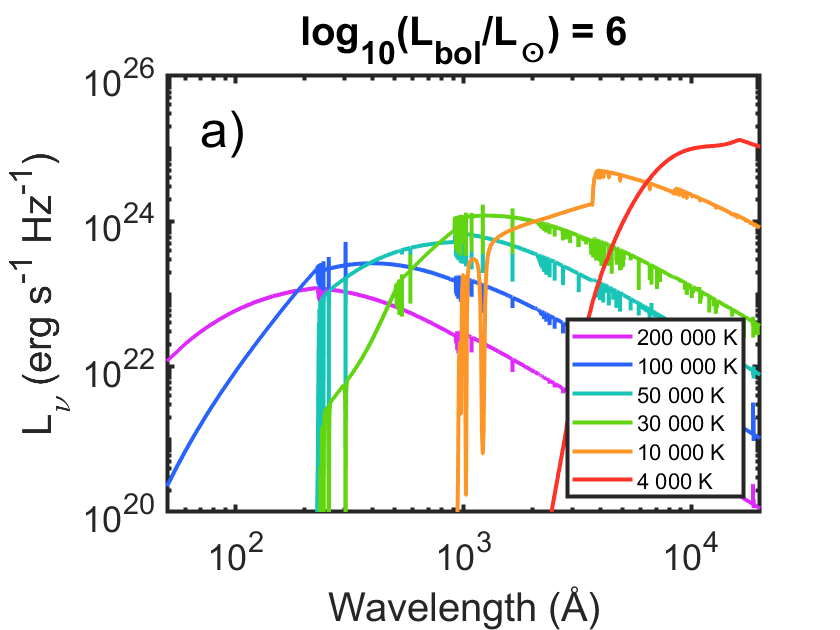}
\includegraphics[width=\columnwidth]{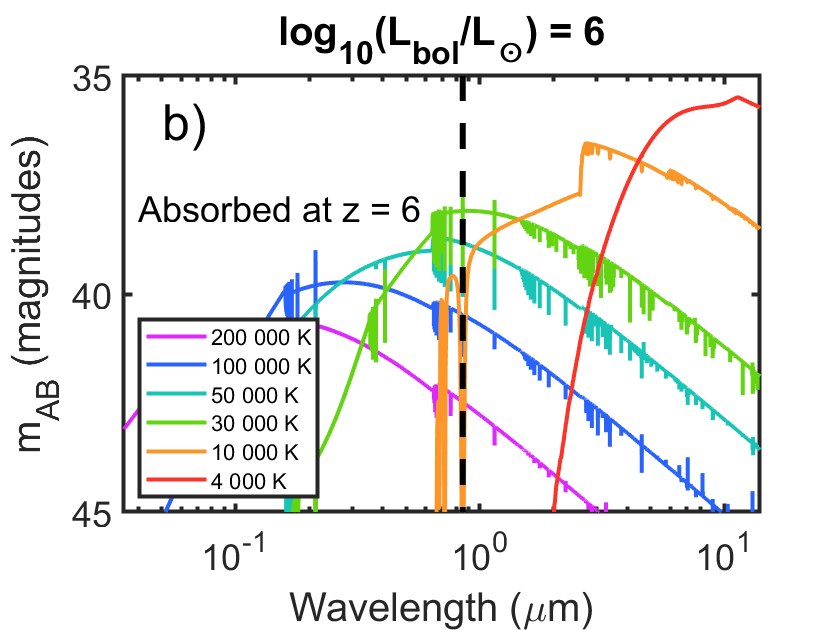}
\caption{The (a) rest-frame monochromatic luminosity $L_\nu$  and (b) observed-frame brightness (AB magnitude) at $z=6$ for a star at different $T_\mathrm{eff}$ but fixed bolometric luminosity $L_\mathrm{bol}=10^6\ L_\odot$. As $T_\mathrm{eff}$ decreases, the peak $L_\nu$ and observed brightness increase substantially. In panel b, the vertical dashed line marks the $z=6$ Ly$\alpha$ limit, to the left of which the radiation is expected to be largely absorbed by the neutral IGM. Hence, the part accessible to JWST lies to the right of this line.}
\label{appA_fig}
\end{figure*}

Here we provide further explanations for why observations of lensed stars will favour the detection of stars in relatively short-lived low-$T_\mathrm{eff}$, states over the more long-lived high-$T_\mathrm{eff}$ states associated with the stellar main sequence.
In Figure~\ref{appA_fig}, we show the rest-frame evolution in monochromatic luminosity $L_\nu$ and observed-frame evolution of $f_\nu$ (expressed in AB magnitudes) of the SED of an artificial star at $z=6$ that is assumed to evolve from $T_\mathrm{eff}=200 000$ K to 4000 K
at constant $L_\mathrm{bol}=10^6\ L_\odot$. While realistic stars admittedly do not evolve at fixed $L_\mathrm{bol}$, this is not far from the truth of $M\gtrsim 100 \ M_\odot$ Pop III stars, which are often predicted to evolve almost horizontally in the HR diagram \citep{Windhorst18,Volpato23}. 
Figure~\ref{appA_fig}a indicates a factor of $\sim 100$ difference in peak $L_\nu$ when  $T_\mathrm{eff}$ goes from high to low. However, the full wavelength range is not directly observable (at any redshift) due to foreground gas absorption at $<912$ \AA{} (neutral ISM) or $<1216$ \AA{} (neutral IGM at $z\gtrsim 6$) and the wavelength coverage of the most sensitive JWST instruments also limits the relevant range. Figure~\ref{appA_fig}b shows that there is a $\approx 5.5$ magnitude difference at $z=6$ in peak brightness in the $\approx1$--5 $\mu$m window available for JWST NIRCam or NIRSpec observations, between the highest and lowest $T_\mathrm{eff}$ stars plotted.

As a toy example of how these effects convert into detection probabilities, consider a scenario where this artificial star only has two significant $T_\mathrm{eff}$ stages during its evolution across the HR diagram -- one long-lived $10^5$ K state and one short-lived $10^4$ K state. In Figure~\ref{appA_fig}b, we see that the latter state could be $\approx 4$ mag brighter in the JWST window at $z=6$. Since both of these stars would intrinsically lie far below the detection threshold of JWST, a substantial magnification boost is required to bring them into the detectable range. Assuming that the magnification probability distribution follows $P(>\mu)\propto \mu^{-2}$, the magnification required for the $10^4$ K state would be $\approx 40$ times lower than for the $10^5$ K state, which would make it $40^2=1600$ times more likely to occur. Hence, even if the star only spends 1\%  of its lifetime in the short-lived $10^4$ K state, it would still be a factor of $\approx 16$ times more likely to be found in this state rather than in the more long-lived $10^5$ K state. 

\section{Probability of detecting a lensed Pop III nebula}
\label{sec:Probability Pop III nebula}
Here, we present a rough estimate of the probability to detect a Pop III nebula lensed above the detection limit of a 29 AB mag JWST/NIRCam detection limit, arguing that this probability is no more than $\sim 10^{-3}$ per cluster field if the HII-region remains confined to parsec scales for time scales of $\sim 10^5$ yr.

\begin{figure}
\includegraphics[width=\columnwidth]{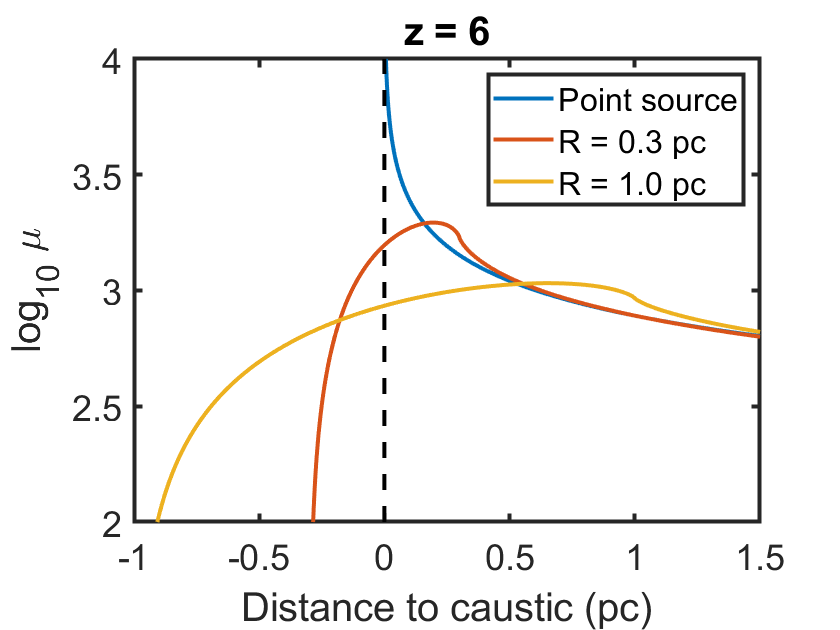}
\caption{Macrolensing magnification $\mu$ in the absence of microlenses as a function of distance from the cluster caustic (position marked by the dashed vertical line) at $z=6$ for a point source (blue line, also suitable for the photosphere of a Pop III star within the range of magnifications plotted), and for sizes more characteristic of HII-regions around Pop III regions (red and yellow lines, for 0.3 and 1.0 pc respectively). The situation shown corresponds to a position along the cluster caustic at which $B_0\approx 10$ (eq.~\ref{eq:mu_B0}).}
\label{lightcurve}
\end{figure}

In Figure~\ref{lightcurve}, we use the inverse ray-shooting code of \citet{Meena23} to plot the $z=6$ macrolensing magnification for different source sizes as a function of distance from the caustic in the MACS J0416.1-2403 galaxy cluster based on the mass model from \citet{Zitrin13}. The behaviour of circular sources with two different sizes relevant for Pop III nebulae ($R_\mathrm{S}=0.3$ pc and 1 pc), are compared to the point-source approximation (eq.~\ref{eq:mu_B0}, with $B_0\approx 10$ at the chosen position along the critical curve) which, within the range of magnifications plotted, illustrates the magnification curve of a stellar-sized source.
 
 While the magnification continues to increase to $\mu\gtrsim 10^4$ as the stellar-sized source approaches the caustic, the two nebular-sized sources reach their maximum magnifications at a distance of $0.65R_\mathrm{S}$ \citep{Miralda-Escude91}. Moreover, while the stellar-sized source becomes unobservable at the point of crossing the caustic, the larger sources can retain a relatively high magnification even when the midpoint has passed slightly beyond the caustic.

At a distance $\approx 1.5$ pc from the caustic, where the macrolensing magnification in the situation plotted is $\mu\approx 600$, a young Pop III star would -- owing to its lower intrinsic flux (Figure~\ref{brightest_ABmag_star_neb_fig}) -- typically not be rendered detectable by JWST from macrolensing alone, whereas an ionization-bounded nebula around a $\gtrsim 200\ M_\odot$ Pop III star at $z=6$ would, by being boosted to $\gtrsim 29.5$ AB mag (Figure~\ref{brightest_ABmag_star_neb_fig}). In the absence of an additional microlensing boost to the flux of the star, this would correspond to a case where the observed SED would be nebular dominated. If the star, on the other hand, is highly microlensed at this position, both components could contribute significantly to the SED.

To provide a rough estimate of the probability to detect the lensed Pop III nebula of a single star, we here consider the very optimistic case where all Pop III stars form with a mass of $500 M_\odot$. For the \citet{Yoon12} non-rotating tracks, we find that a 500 $M_\odot$ star at age $10^5$ yr would produce an ionizing flux $Q_\mathrm{tot}\approx 7.4\times 10^{50}$ s$^{-1}$ and a nebula that reaches a peak JWST/NIRCam flux corresponding to $\approx 35.3$ AB mag at $z=6$. To reach a NIRCam detection limit of 29 AB mag, we therefore need a magnification of $\mu\approx 330$, which according to eq.~\ref{mumax_macro} can be achieved for source sizes with radii up to $\approx 20$ pc at this redshift. To contain the Str\"{o}mgren radius of the HII region within this radius would require a gas density $n_\mathrm{H}\geq 50$ cm$^{-3}$ (eq.~\ref{eq:Rstr}), which we assume can be retained for $t_{HII}\sim 10^5$ yr.

Using eq.~\ref{eq:max_dcaustic} with $B_0=15$ implies that we can reach $\mu=330$ at a maximum separation from the caustic of $\max(d_\mathrm{arcsec})\approx 2\times 10^{-3}$ arcsec. If we assume that the magnification is retained at this level or higher from this location and all the way up to the caustic (see figure~\ref{lightcurve}), and moreover adopt a length for the cluster caustic of $L_\mathrm{arcsec}=100\arcsec$, we get a source-plane area (eq.~\ref{eq:A_arcsec2}) $\theta^2 \approx 0.2$ arcsec$^2$ over which the nebula may be rendered detectable in a single cluster-lens field. For $\Delta(z)=1$ at $z=6$, this corresponds to a volume of $V_\mathrm{z,\Delta(z)}\approx 0.13$ cMpc$^{-3}$ (eq.~\ref{eq:Vz,deltaz}). Adopting the \citet{Liu20} Pop III SFRD of $\approx 4\times 10^{-5}\ M_\odot$ yr$^{-1}$ cMpc$^{-3}$ at $z=6$ and using eq.~\ref{eq:Ndetection} (with $f(M)=1$ and $\Delta t_j=t_\mathrm{HII}=10^5$ yr), we find that only $N_\mathrm{detect}\approx 10^{-3}$ Pop III nebulae are expected to be detected per cluster field. 

Hence, unless the HII-region can remain sufficiently compact to attain magnifications in the hundreds for significantly longer than $\sim 10^5$ years, it is unlikely to be detectable. This result is strictly valid only for the volume spanning $\Delta(z)=1$ at $z=6$, but higher redshifts contribute relatively little to the $N_\mathrm{detect}$ estimate since H$\alpha$ gets redshifted out of the NIRCam window at $z\geq 6.9$, thereby rendering the Pop III nebula intrinsically much fainter (see Figure \ref{brightest_ABmag_star_neb_fig}).

\end{document}